\documentclass[aps,pra,onecolumn,superscriptaddress,11pt]{revtex4}
\usepackage{epstopdf}
\usepackage{amsmath}
\usepackage{amssymb}
\usepackage{amsbsy}
\usepackage{makeidx}
\usepackage{float}
\usepackage{graphicx}
\usepackage{amsfonts}
\usepackage{placeins}
\usepackage[caption = false]{subfig}
\usepackage{wasysym}
\usepackage{comment}
\usepackage[toc,page]{appendix}
\usepackage{color}
\begin{document}
\title{Traveling Wave Solutions in a Chain of Periodically Forced Coupled Nonlinear Oscillators}
\author{M. Duanmu}
\affiliation{Department of Mathematics and Statistics, University of Massachusetts, Amherst MA 01003-4515, USA}
\author{N. Whitaker}
\affiliation{Department of Mathematics and Statistics, University of Massachusetts, Amherst MA 01003-4515, USA}
\author{ P. G.  Kevrekidis}
\affiliation{Department of Mathematics and Statistics, University of Massachusetts, Amherst MA 01003-4515, USA}
\affiliation{Center for Nonlinear Studies and Theoretical Division, Los Alamos National Laboratory, Los Alamos, New Mexico 87545, USA}
\author{A. Vainchtein}
\affiliation{Department of Mathematics, University of
Pittsburgh, Pittsburgh, PA 15260, USA}
\author{J. Rubin}
\affiliation{Department of Mathematics and Center for the Neural Basis of Cognition, University of
Pittsburgh, Pittsburgh, PA 15260, USA}
\begin{abstract}
Motivated by earlier studies of artificial perceptions of light called
phosphenes, we analyze traveling wave solutions in a chain of periodically forced
coupled nonlinear oscillators modeling this phenomenon. We examine
the discrete model problem in its co-traveling frame and systematically
obtain the corresponding traveling waves in one spatial dimension.
Direct numerical simulations as well as linear stability analysis
are employed  to reveal the parameter regions where the traveling
waves are stable, and these waves are, in turn, connected to the standing waves analyzed
in earlier work. We also consider a two-dimensional extension of the model and
demonstrate the robust evolution and stability of planar fronts and
annihilation of radial ones.
Finally, we show that solutions that initially feature two symmetric fronts with bulged centers evolve in
qualitative agreement with experimental observations of phosphenes.
\end{abstract}

\maketitle

\section{Introduction}

Electrical stimulation of
the retina can produce artificial perceptions of luminance changes called phosphenes,
which may also arise in early stages of retinal or visual disease \cite{cervetto2007}.
The induction of phosphenes is being used to help restore vision or develop visual aids
for patients with severely compromised vision \cite{zrenner2002,tehovnik2013}, and an understanding of
how phosphenes arise and behave could contribute to such efforts.
In a detailed experimental study \cite{carp}, Carpenter explored electrically induced
phosphenes in human subjects. Each subject's eyes were immersed in a saline
bath to which an  alternating current was applied.  When a dark object was passed through a subject's visual field in the presence of such stimulation,
visual perceptions of line phosphenes occurred in its wake. The lines were
observed to move and
interact but
never cross. This work  suggested the presence of a bistability of activity states in the system, with the moving lines representing boundaries between sets of cells in different activity regimes.

Drover and Ermentrout~\cite{drover} developed a
one-dimensional model providing a simple representation of the
phosphenes in Carpenter's experiments and their motion. In their work, a chain of excitable neurons  was driven
by a spatially uniform periodic stimulus at a frequency higher than
the cells could follow, inducing a
$1:2$ phase locking with the stimulus.
In the absence of coupling, neurons could fire on
even or odd cycles of the stimulus, resulting in an intrinsic bistability for
the forced
system. Sufficiently strong coupling, even if directionally unbiased, elicited unidirectional traveling waves in which cells were recruited to switch phase.
Large amplitude forcing of an excitable system is a difficult
problem to tackle
analytically. In view of that difficulty, the more recent analysis of~\cite{parks}
assumed that each neuron is intrinsically oscillatory and characterized
by a state evolving at half the frequency of an applied forcing signal.
Using multiple time scale expansion and the
Fredholm alternative, the authors of~\cite{parks} derived a
more analytically tractable
{\it effective model} for the time evolution of the neurons'  phases.
This reduced model, which is quite general and particularly interesting
in its own right, will be the focus of the present work.

In~\cite{parks}, a detailed numerical existence
and stability analysis was done using XPPAUT~\cite{ermen} for a finite chain
of coupled phase oscillators. Certain interesting
bifurcation phenomena were identified including saddle-node and
pitchfork bifurcations in a two-dimensional parameter space
characterizing the strength and asymmetry of coupling between nearest-neighbor oscillators. Beyond the critical points involved in
these bifurcations, direct numerical simulations identified traveling waves
that are strongly reminiscent of the ``recruitment waves'' obtained in the
original chain of forced neural oscillators.
It is exactly these traveling waves that we systematically obtain and analyze
in the present manuscript, by a combination of numerical and, whenever possible, semi-analytical techniques.

Specifically, we consider the exact traveling
wave problem, which takes the form of an advance-delay differential ordinary equation in
the co-traveling frame of such waves and which we introduce along with
the mathematical formulation of the problem
in Section 2.  In Section 3, we proceed to solve
this problem numerically, identifying exact (up to numerical error) traveling
and standing (zero-speed) waves within the
full two-dimensional parameter space used in~\cite{parks}.
We explore the stability of these waves in two complementary ways. On the one
hand, we consider the traveling waves as steady states of the associated
advance-delay partial differential equation (PDE). On the other
hand, we examine them via direct numerical simulations of the original system
of ordinary differential equations (ODEs) for the coupled oscillators with the initial condition ``distilled''
on the lattice from the obtained traveling wave solution.
In~\cite{parks}, this problem was studied in a limited range for the parameter $\mu$ measuring the asymmetry of the nearest-neighbor coupling function.
Here, our analysis extends to all values of $\mu$. We show that traveling waves are stable in certain parameter regions that are periodic in $\mu$ and are
located above a certain curve, below which there exist stable standing waves. The stability regions alternate along the direction of $\mu$ with regions
where the waves are unstable due to the instability of the background state as well as a frontal instability. Simulations of the lattice system initialized by an unstable traveling wave
show that the frontal instability results in formation of two fronts that propagate in the opposite directions with the same speed as the initial wave.

In the final part of Section 3, we extend the lattice model to a two-dimensional setting and show that the planar fronts obtained from the one-dimensional traveling wave are very robust
even with a local initial distortion, unlike radial fronts, which
are eventually annihilated in the dynamical evolution. Finally, we consider
the evolution of two symmetric fronts with initially bulged centers and show that the resulting dynamics is in agreement with Carpenter's findings, based on observations of phosphenes, that lines form loops instead of crossing through each other and that a line does not break apart unless it meets another line.
This work adds support to the idea that a phase model may provide a useful framework for studying phosphenes and also raises interesting mathematical issues about the relation between advance-delay PDE and lattice ODE system solutions, and we conclude with a brief discussion of these points in Section 4.

\section{Discrete model and the traveling wave equation}

Consider a chain of $N$ periodically forced oscillators governed by the reduced spatially discrete model of stimulated retinal cells derived in \cite{parks}:
\begin{equation}
\begin{split}
\dot{\theta}_{-n} &= k H(\theta_{-n+1}-\theta_{-n})+f(\theta_{-n}) \\
\dot{\theta}_j &= k [H(\theta_{j-1}-\theta_j)+H(\theta_{j+1}-\theta_j)]+f(\theta_j), \; \;
\mbox{for} \; \;  j=-n+1,\dots,N-n-2\\
\dot{\theta}_{N-n-1} &= k H(\theta_{N-n-2}-\theta_{N-n-1})+f(\theta_{N-n-1}).
\end{split}
\label{ee1}
\end{equation}
Here $\theta_j(t)$ is the slowly evolving phase of each neuron, $f(\theta)$ is a $\pi$-periodic forcing or locking function, and $H(\theta)$ is a $2 \pi$-periodic function characterizing the coupling of the nearest neighbors and multiplied by the coupling constant $k>0$. The periodicities of $f(\theta)$ and $H(\theta)$ represent the $1:2$ frequency locking
present in the system and can be easily generalized to different types of frequency locking. In \cite{parks} the number of oscillators was set to be even, $N=2n$, but here we also allow it to be odd, $N=2n+1$.

\begin{figure}[tbp!]
\begin{center}
{\includegraphics[width=0.6\textwidth]{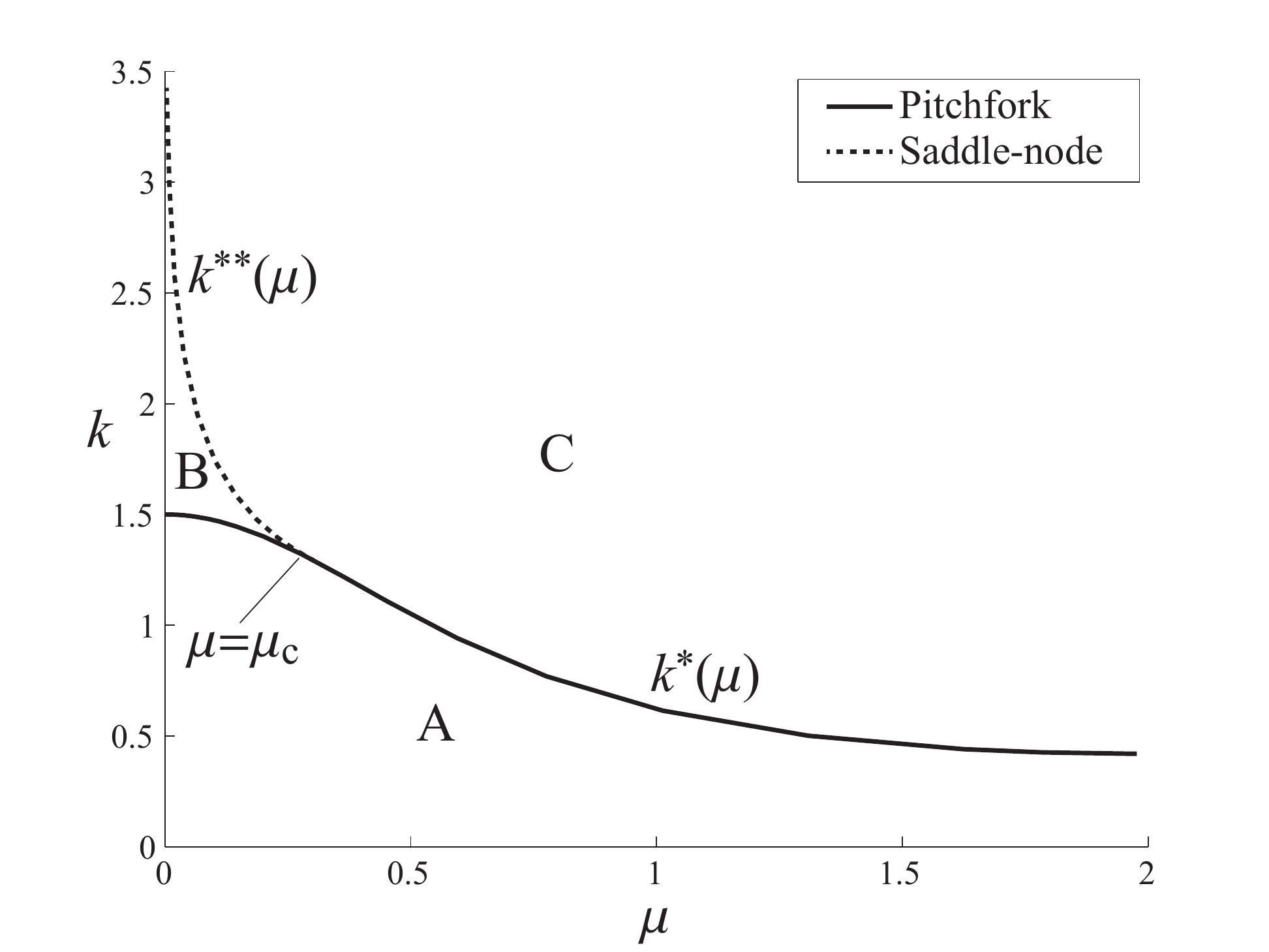}}
\end{center}
\caption{Bifurcation diagram for stationary solutions of the system \eqref{ee1} with $N=2n=10$, as obtained in \cite{parks}. The primary split state equilibria are stable in region A and destabilize via a pitchfork bifurcation (solid curve, $k=k^*(\mu)$). At $\mu<\mu_c$ the bifurcation is supercritical and gives rise to a pair of secondary split state equilibria that are stable in region B and disappear through a saddle-node bifurcation (dashed curve, $k=k^{**}(\mu)$, $\mu<\mu_c$). }
\label{f00}
\end{figure}

In what follows, we consider prototypical examples of the two functions proposed in \cite{parks},
\[
H(\theta)=\sin(\theta+\mu)-\sin(\mu), \quad f(\theta)=-\sin(2\theta),
\]
where the parameter $\mu$ measures the asymmetry of the coupling function $H(\theta)$.
We can identify 0 with $2\pi$ such that $\theta_j \in [0,2\pi)$ and then the firing of the $j$th neuron corresponds to $\theta_j$ crossing through some distinguished value, taken for some oscillator models, for example, to be $\theta_j=\pi$.
It is not hard to see that in the case of $\mu=0$, when $H(\theta)=\sin\theta$ is odd, there exist equilibrium split state (antiphase) solutions given by
\[
(\theta_{-n}, \dots, \theta_{-1}, \theta_{0}, \dots, \theta_{N-n-1}) = (0, \dots, 0,\pi, \dots, \pi)  \quad \text{and} \quad (\pi, \dots, \pi,0, \dots, 0).
\]
If $N=2n$, these solutions have equal numbers of oscillators firing in each cycle. When $\mu$ is nonzero, such piecewise constant split states no longer exist. However, there are single-front equilibrium split states that are close to the above antiphase state but have boundary layers near the front. In \cite{parks}, such steady state solutions of
(\ref{ee1}) are obtained numerically by varying $\mu$ and $k$. A prototypical example of the resulting diagram in the $(\mu,k)$ plane is shown in Fig.~\ref{f00}, adapted from Fig.~4
in \cite{parks}. There we observe that the primary split state solution is stable in a region of small enough
$k$ (region A) but becomes unstable through a
pitchfork bifurcation at $k=k^*(\mu)$ shown by the solid curve. This bifurcation is subcritical for $\mu$ above a certain threshold, $\mu>\mu_c$. For $\mu<\mu_c$, the bifurcation is
supercritical and gives rise to two other, secondary, stable nonsymmetric split states that exist in region B. These states, in turn, disappear through a saddle-node bifurcation at $k=k^{**}(\mu)$ (dashed curve), where $k^{**}(\mu)>k^*(\mu)$.  When exploring the dynamical byproducts of these instabilities, the
authors of~\cite{parks} identified a spontaneous emergence of traveling
waves in region C. This motivates us to seek (numerically) exact traveling wave solutions for this problem.

To this end, we consider the infinite chain of oscillators
\begin{align}
\dot{\theta}_j = k [H(\theta_{j-1}-\theta_j)+H(\theta_{j+1}-\theta_j)]+f(\theta_j), \quad j \in \mathbb{Z}. \label{tw0}
\end{align}
We start by seeking a solution of this system in the form $\theta_j(t)=\Theta(z,\tau)$,
where $z = j-ct$ is a traveling wave coordinate with wave velocity $c$ and $\tau=t$, such that $\Theta \rightarrow 0$ and
$\Theta \rightarrow \pi$ when $z \rightarrow -\infty$ and $\infty$, respectively, or vice versa.
Substitution of this ansatz into (\ref{tw0}) leads to the partial differential advance-delay equation (co-traveling frame PDE) of the form
\begin{align}
\Theta_{\tau} - c\Theta_z=& k [H(\Theta(z+1,\tau)-\Theta(z,\tau))+H(\Theta(z-1,\tau)-\Theta(z,\tau))]+f(\Theta(z,\tau)). \label{tw1}
\end{align}
\emph{Traveling wave} solutions of \eqref{tw0},
\begin{equation}
\theta_j(t)=\phi(z), \quad z=j-ct,
\label{TWansatz}
\end{equation}
are stationary solutions of \eqref{tw1} and satisfy the advance-delay ordinary differential equation
\begin{align}
-c \phi'(z) =k[ H(\phi(z+1)-\phi(z))+H(\phi(z-1)-\phi(z))]+f(\phi(z)). \label{tw11}
\end{align}
Since every translate of the traveling
solution is also a solution, the additional pinning condition $\phi(0)=\pi/2$ is
 imposed on the nonlinear system in order to fix (i.e., pin)
the traveling wave. This condition uniquely identifies the solution, as well
as the corresponding velocity $c$. Note that the traveling wave ansatz \eqref{TWansatz} implies that
$\theta_j(t)=\theta_{j+1}(t+1/c)$ for any integer $j$, meaning that the
phase value $\theta_j(t)$ is periodic (modulo shifts) with a period of $1/c$.

\section{Existence, Stability and Dynamics}

\subsection{Examples of traveling wave solutions}

We start by considering solutions of the traveling wave Eq. \eqref{tw11} satisfying
$\phi(0)=\pi/2$, $\phi(z) \rightarrow \pi$ as $z \rightarrow \infty$, $\phi(z) \rightarrow 0$
as $z \rightarrow -\infty$. Eq.~\eqref{tw11} is discretized
in the interval $[-25,25]$
with a forward difference approximation of the derivative in the left hand side
(for comparison purposes, a centered difference scheme was also used). The
middle grid point is fixed to satisfy the pinning
condition $\phi(0)=\pi/2$, and the speed $c$ is added as a variable.
We solve the
resulting system using Newton's method for the traveling wave $\phi(z)$
and its velocity $c$, thus identifying numerically exact (at least up to a prescribed tolerance of $10^{-12}$ in the discretized system) solutions.

\begin{figure}[!htb]
\begin{center}

{\subfloat[]{\includegraphics[width=0.4\textwidth]{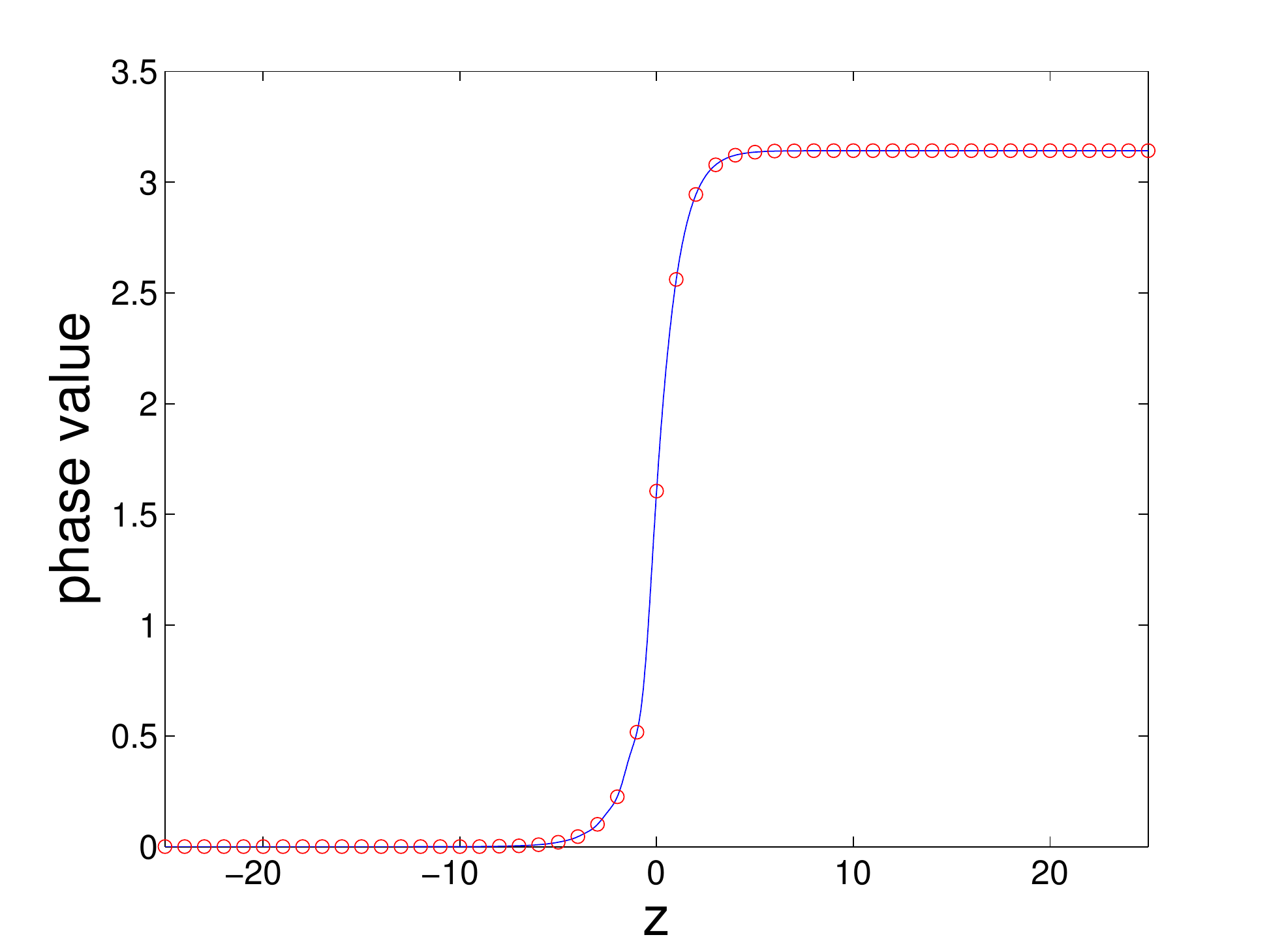}}}
{\subfloat[]{\includegraphics[width=0.4\textwidth]{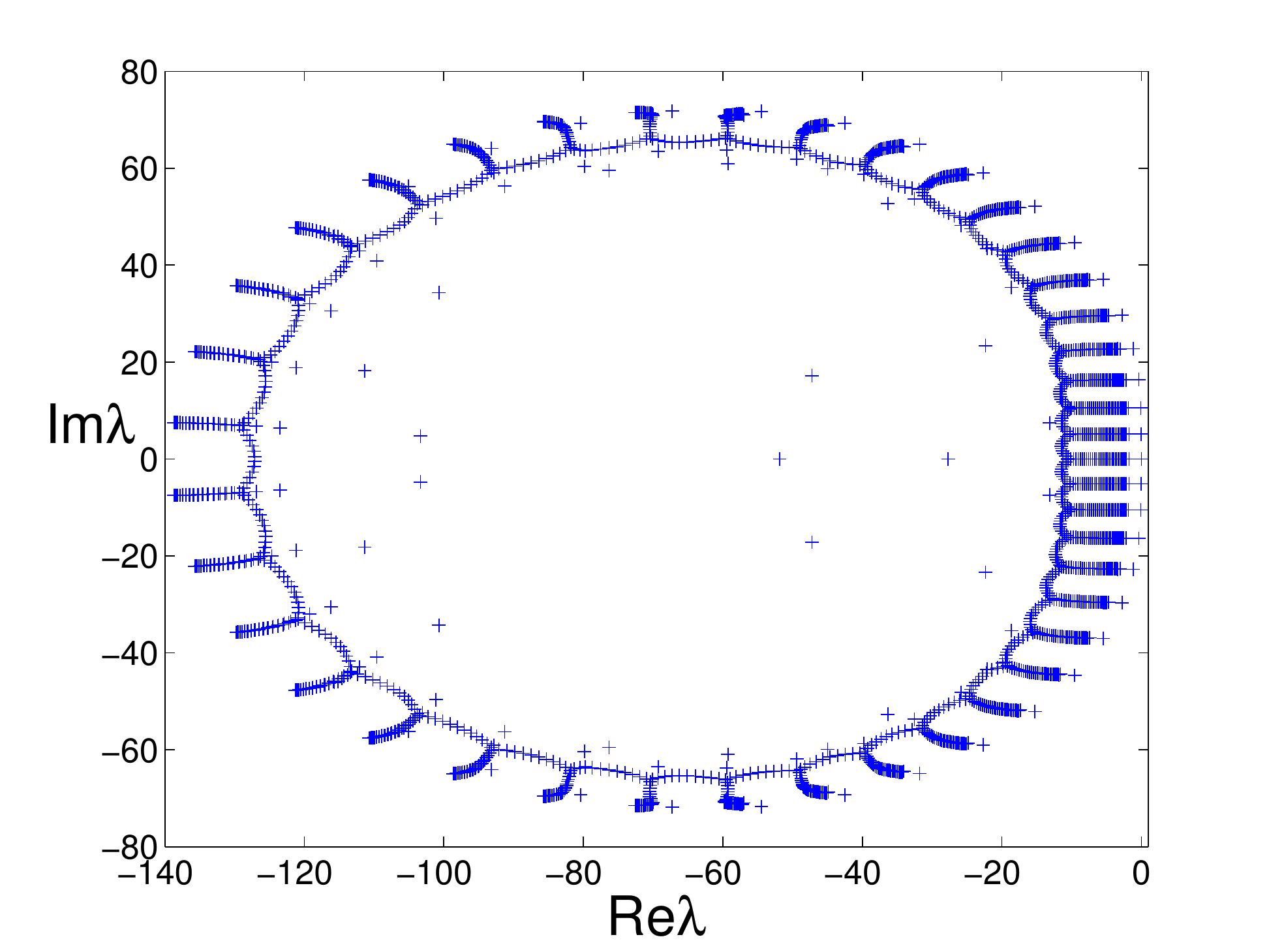}}}
{\subfloat[]{\includegraphics[width=0.4\textwidth]{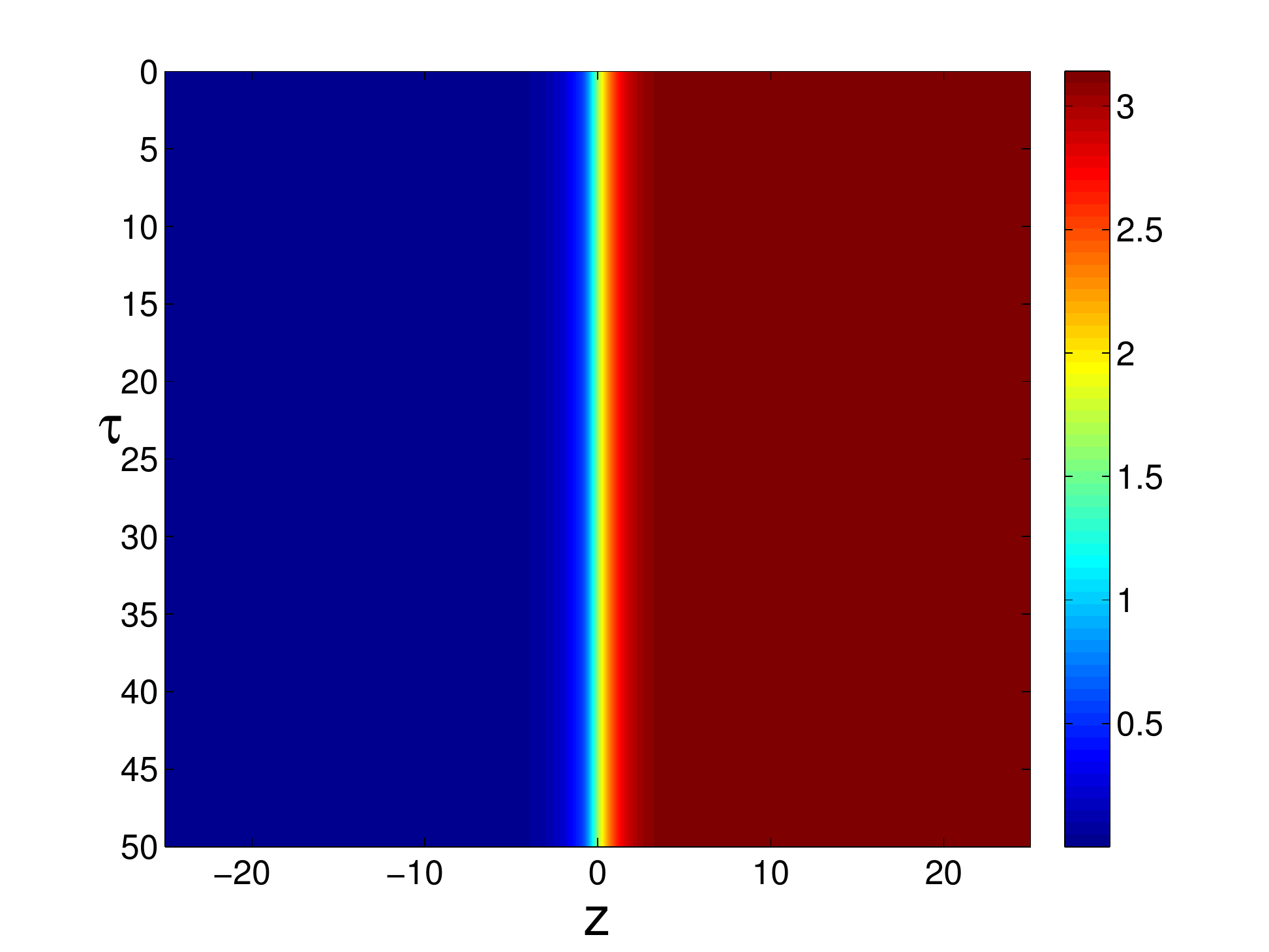}}}
{\subfloat[]{\includegraphics[width=0.4\textwidth]{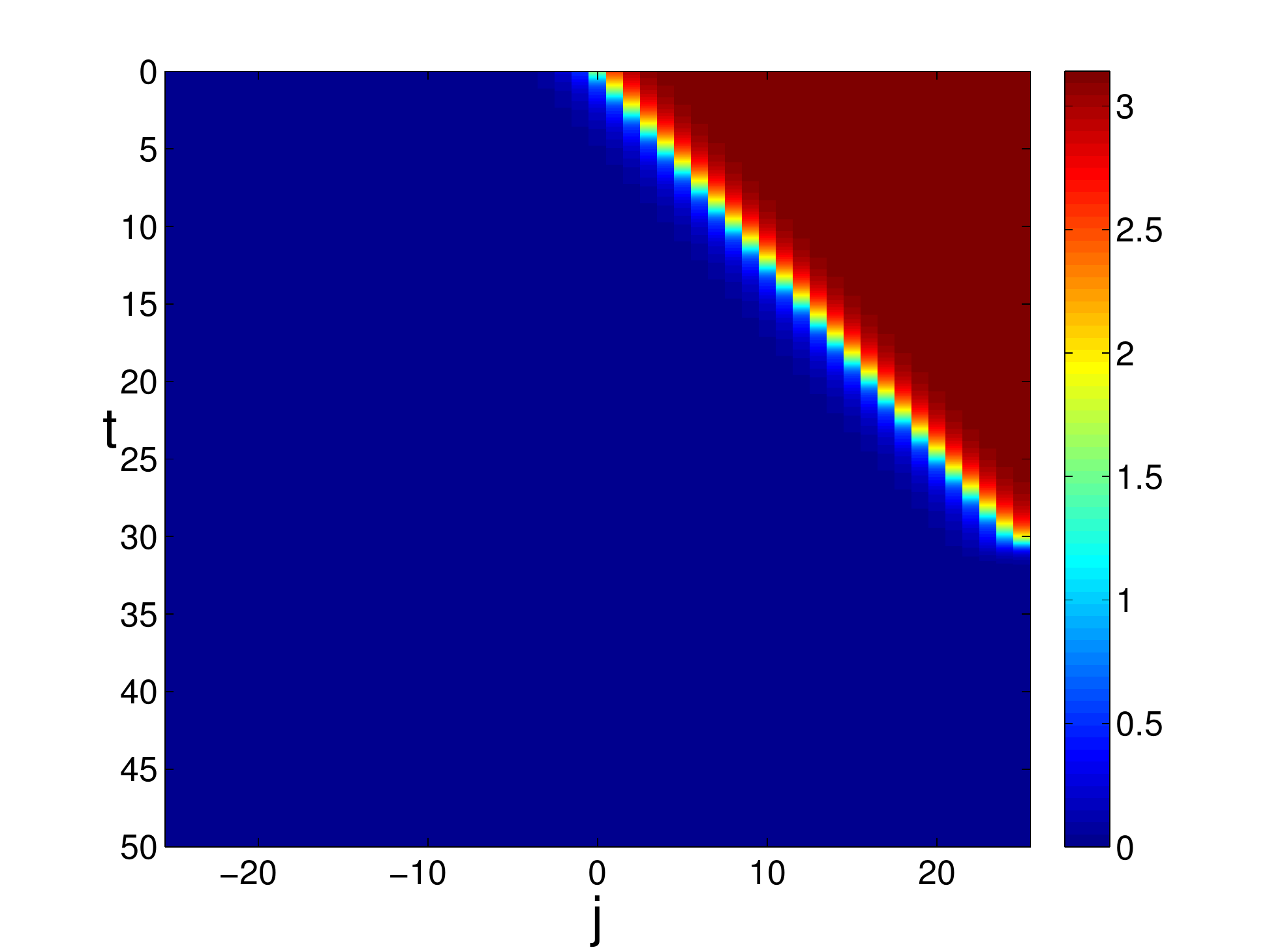}}}
\end{center}
\caption{(a) Traveling wave solution (solid curve) $\phi(z)$ of \eqref{tw11} at $k=2.25$ and $\mu=0.5$, yielding $c=0.8123$, and the initial condition (circles) $\theta_j(0)=\phi(j)$ on the lattice for the ODE simulations shown in (d). (b) Eigenvalues of the Jacobian of the linearization of the co-traveling frame PDE \eqref{tw1} around the stationary solution $\phi(z)$. (c) The space-time evolution of the solution of PDE \eqref{tw1} with the stationary state as the initial condition, $\Theta(z,0)=\phi(z)$. (d) The space-time evolution of the solution of ODE \eqref{ee1} with initial condition $\theta_j(0)=\phi(j)$ shown by circles in (a).}
\label{f200}
\end{figure}

\begin{figure}[!htb]
\begin{center}

 {\subfloat[]{\includegraphics[width=0.4\textwidth]{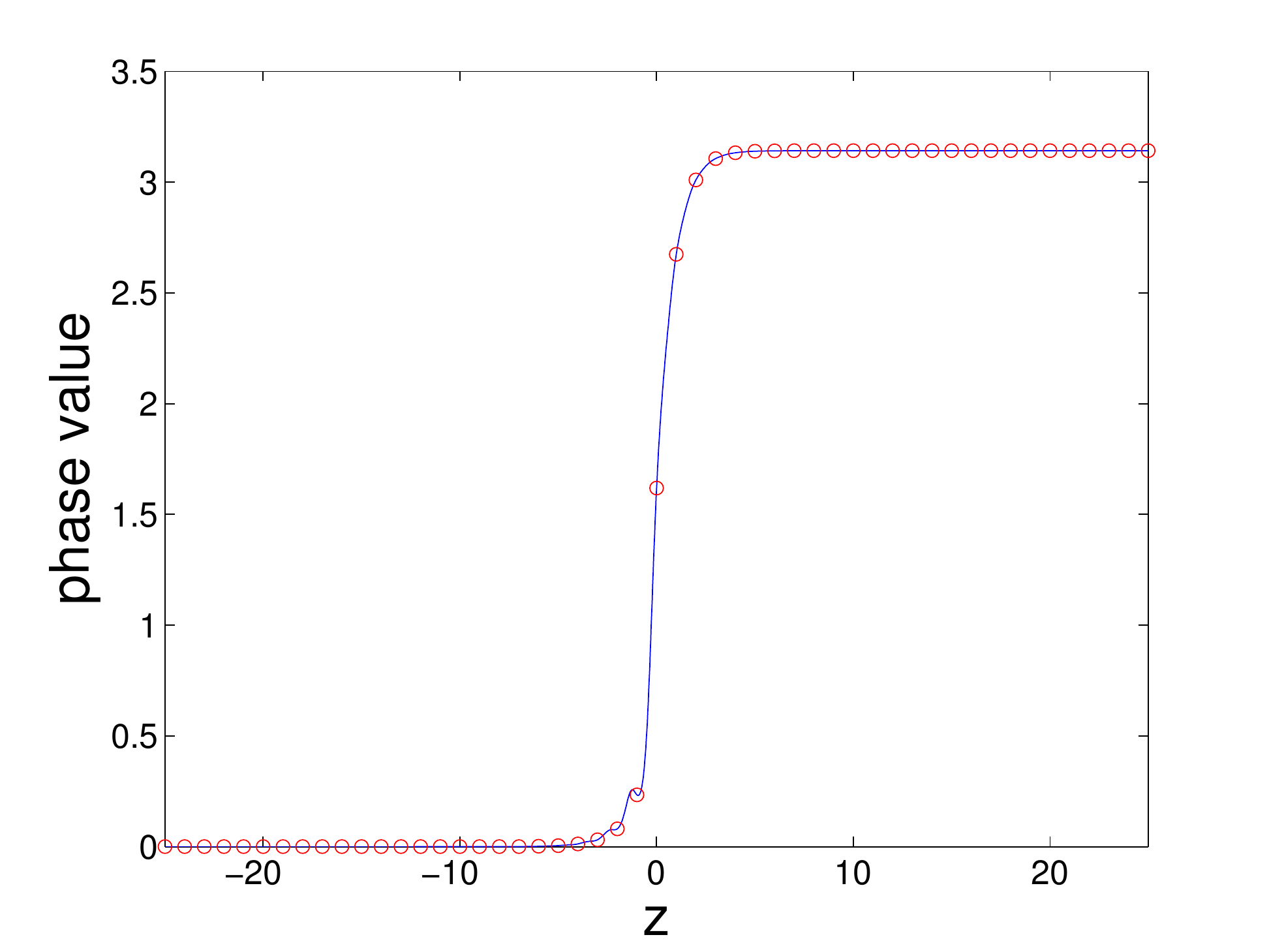}}}
{\subfloat[]{\includegraphics[width=0.4\textwidth]{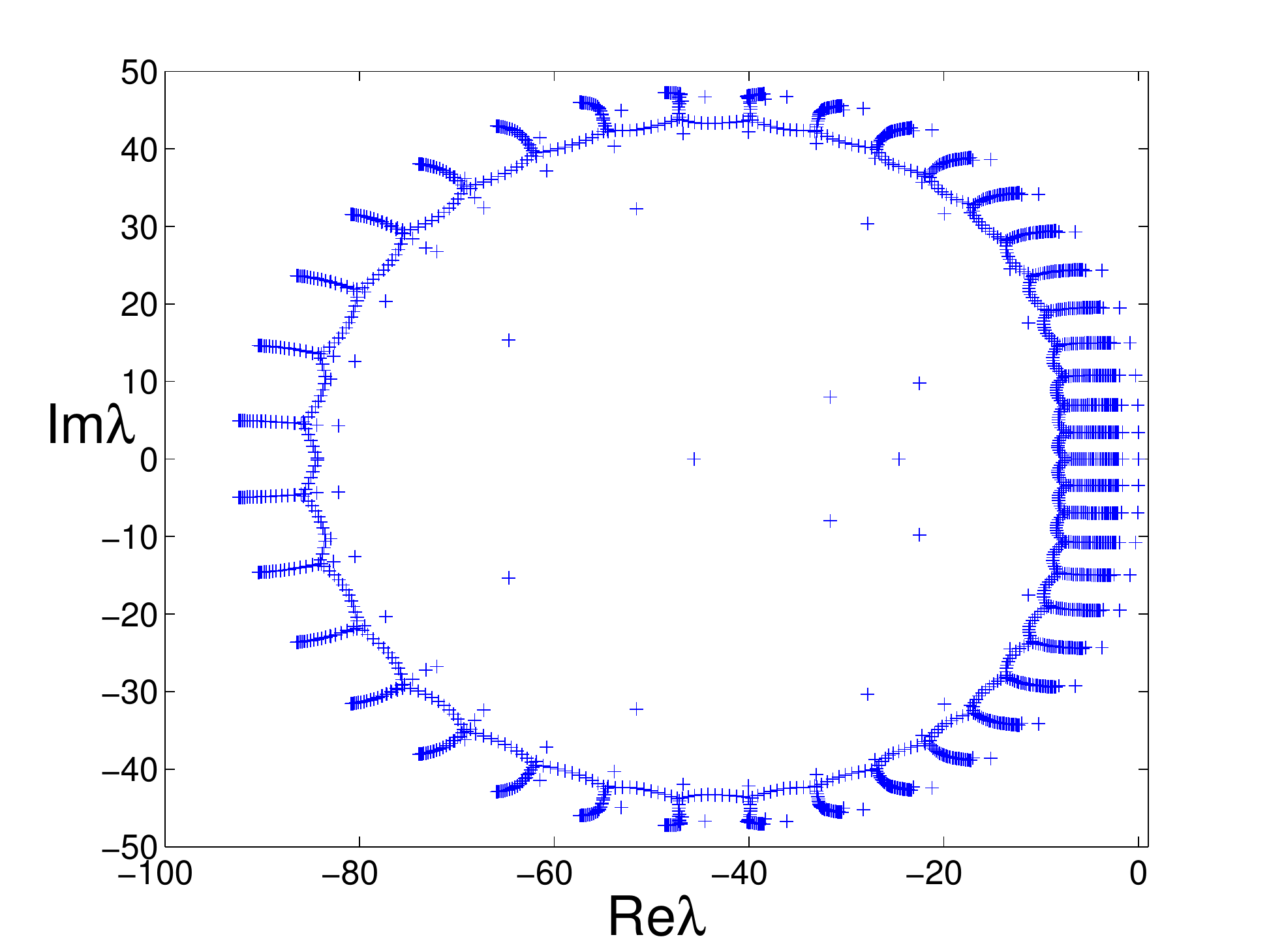}}}
{\subfloat[]{\includegraphics[width=0.4\textwidth]{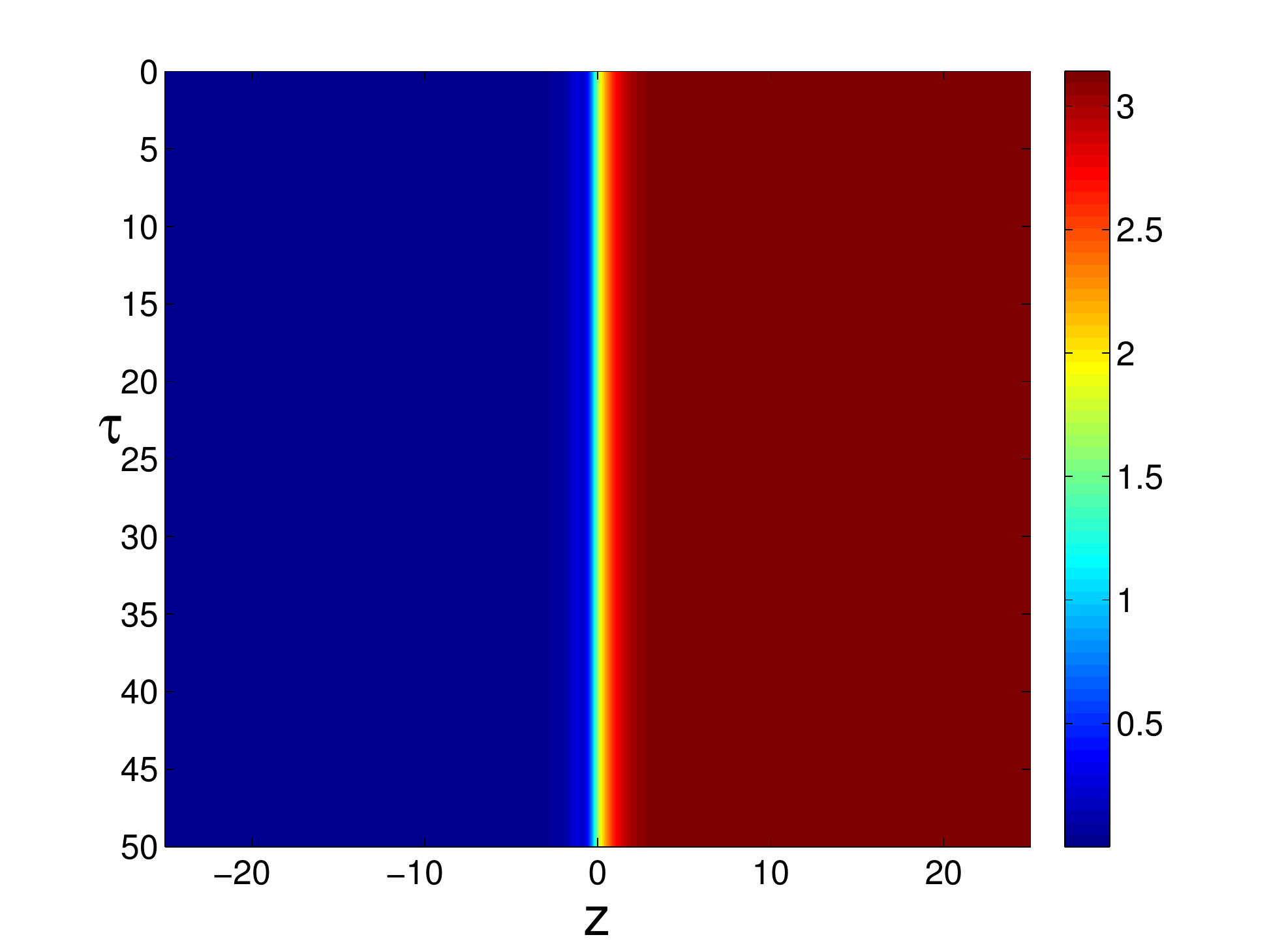}}}
{\subfloat[]{\includegraphics[width=0.4\textwidth]{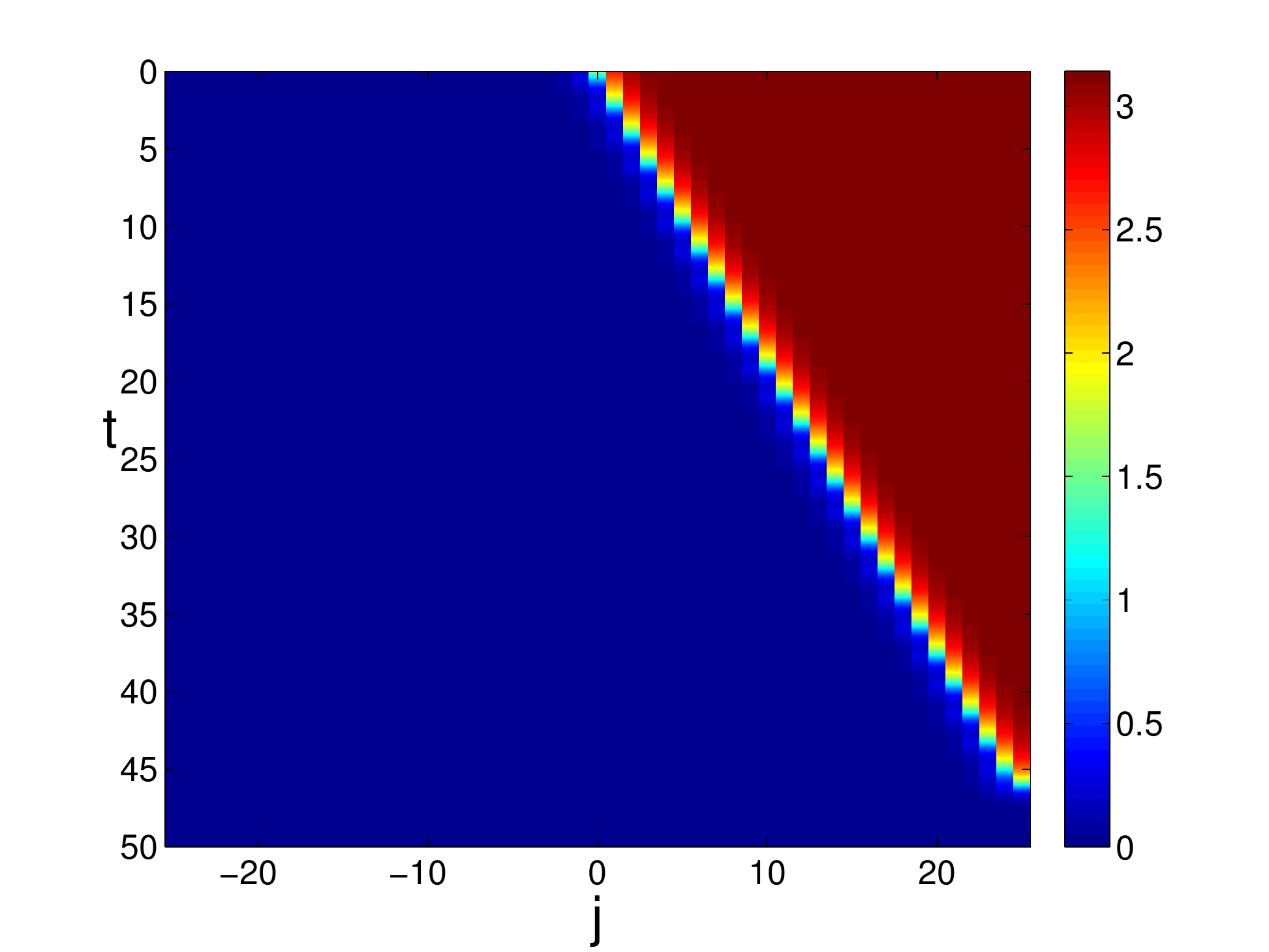}}}

\end{center}
\caption{Same plots as in Fig.~\ref{f200} but now for the stable
traveling wave solution with $k=1.5$ and $\mu=0.5$, yielding $c=0.5368$.}
\label{f202}
\end{figure}

\begin{figure}[!htb]
\begin{center}
{\subfloat[]{\includegraphics[width=0.4\textwidth]{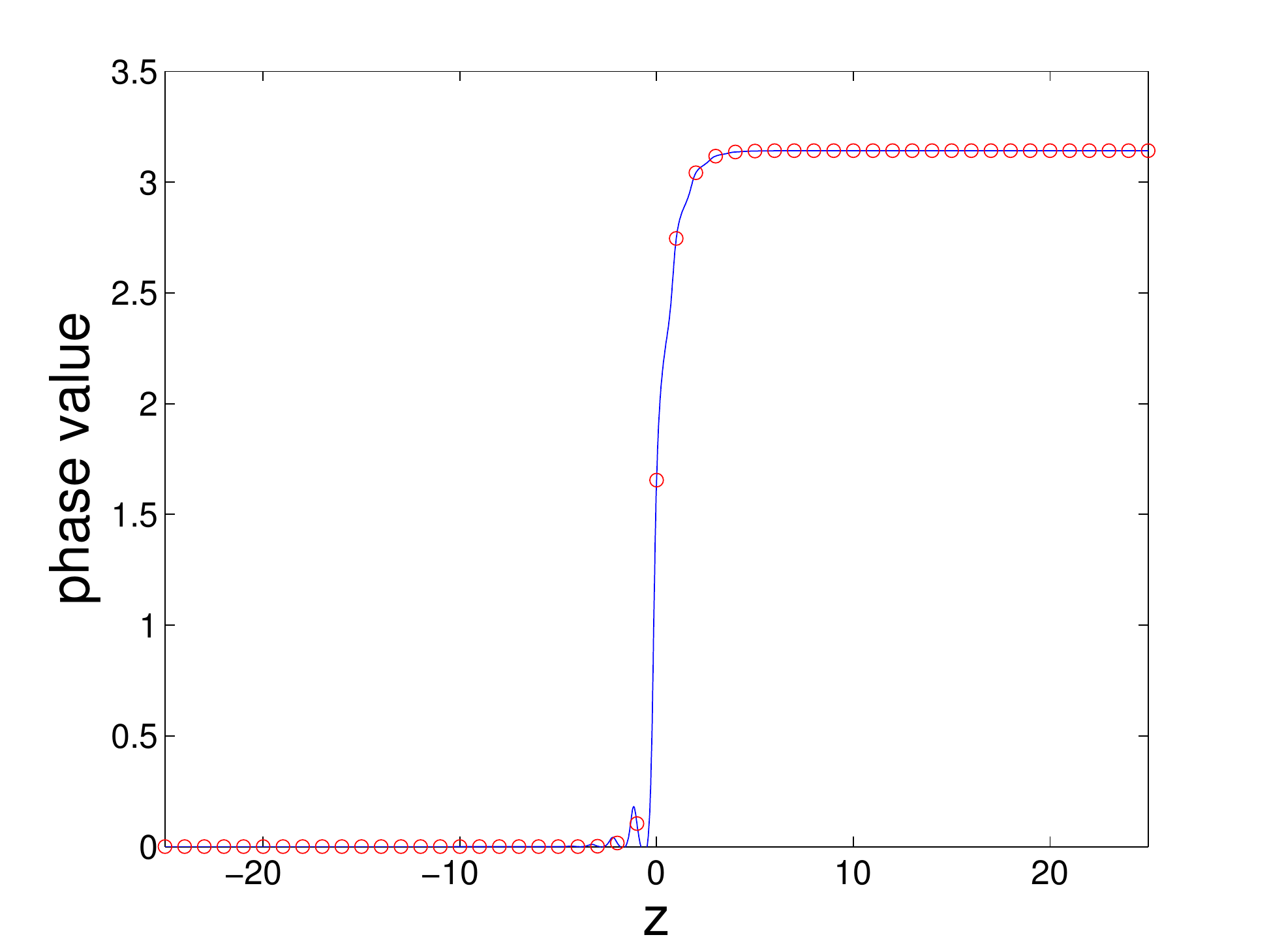}}}
{\subfloat[]{\includegraphics[width=0.4\textwidth]{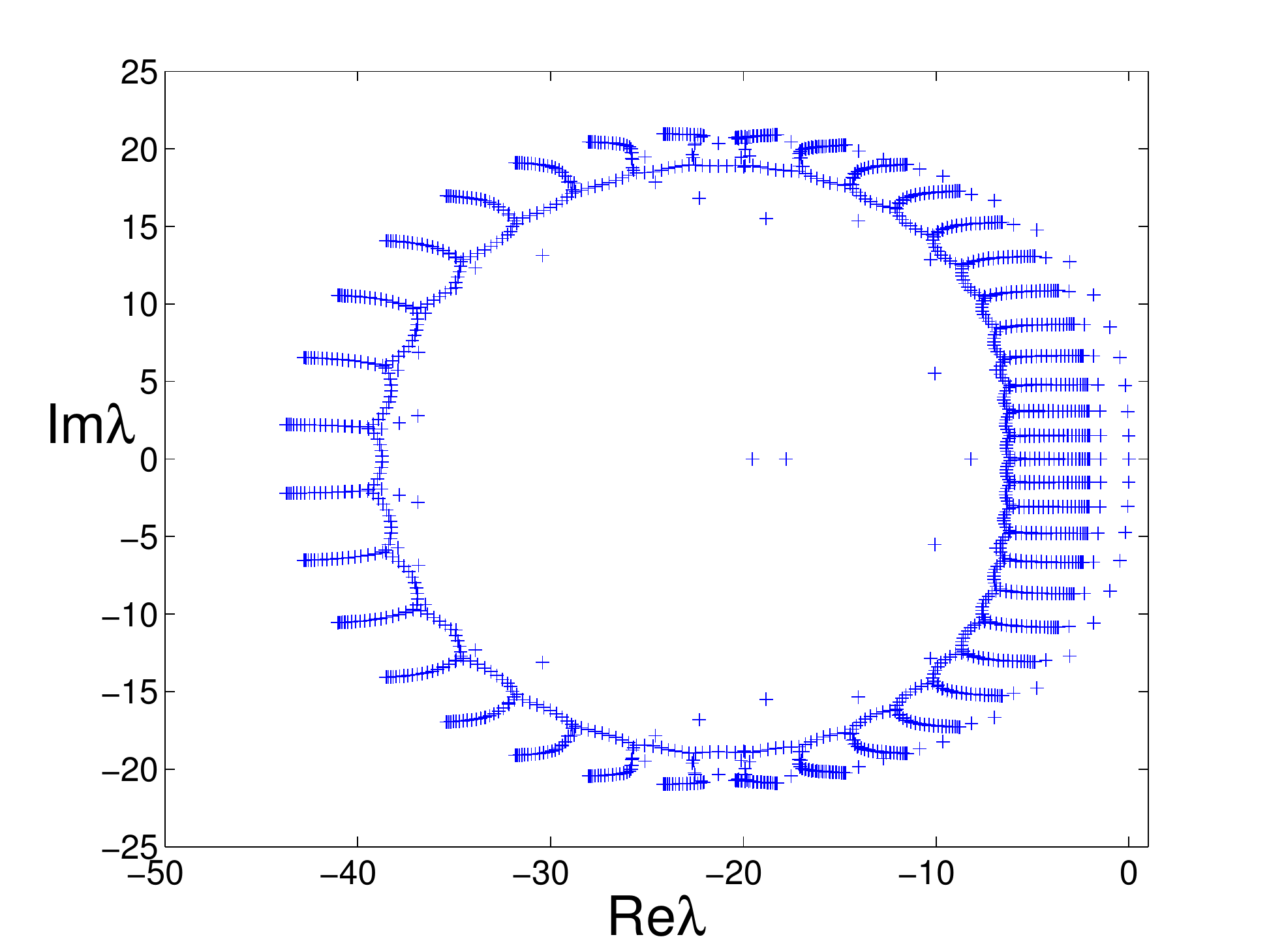}}}
{\subfloat[]{\includegraphics[width=0.4\textwidth]{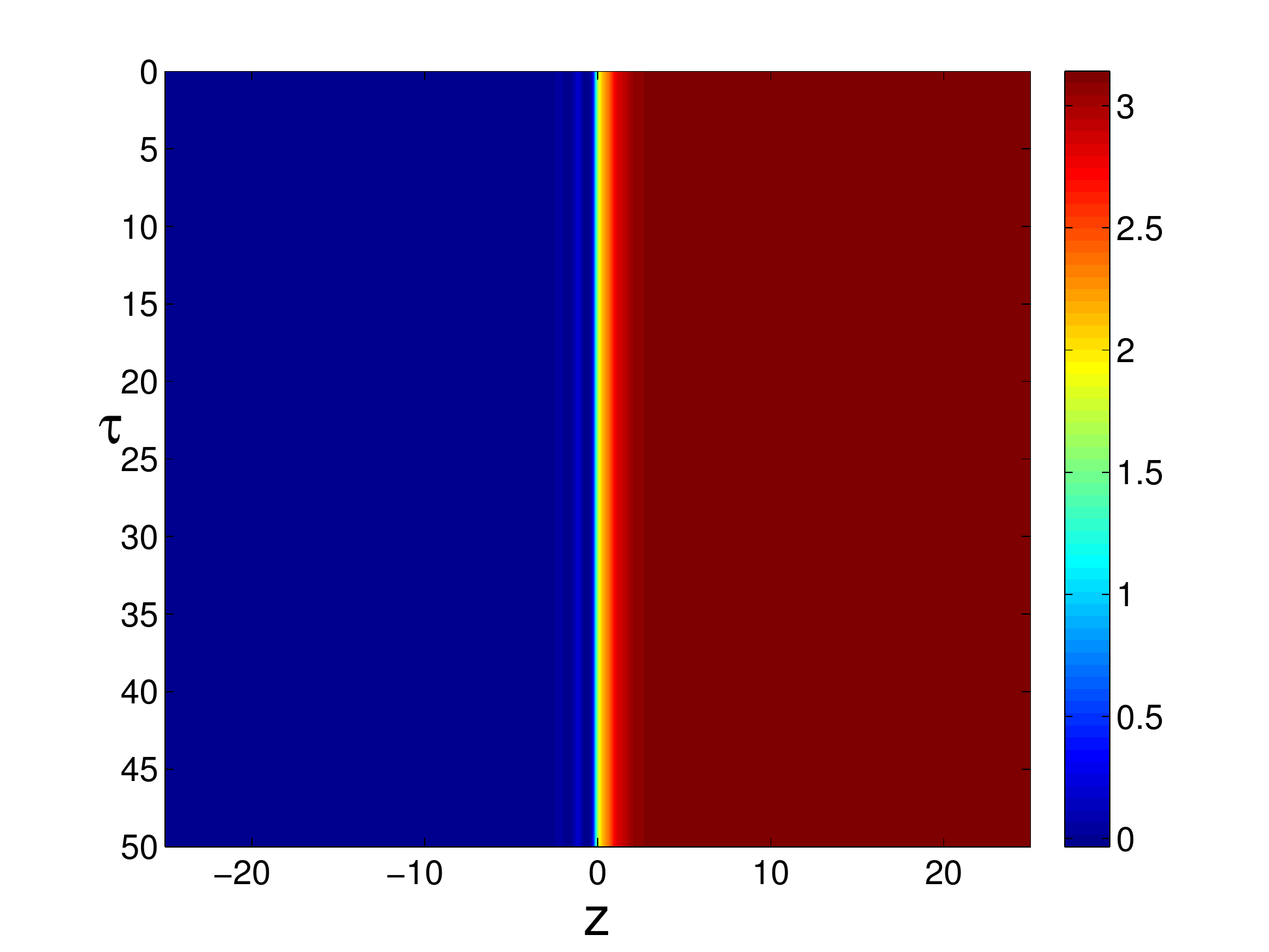}}}
{\subfloat[]{\includegraphics[width=0.4\textwidth]{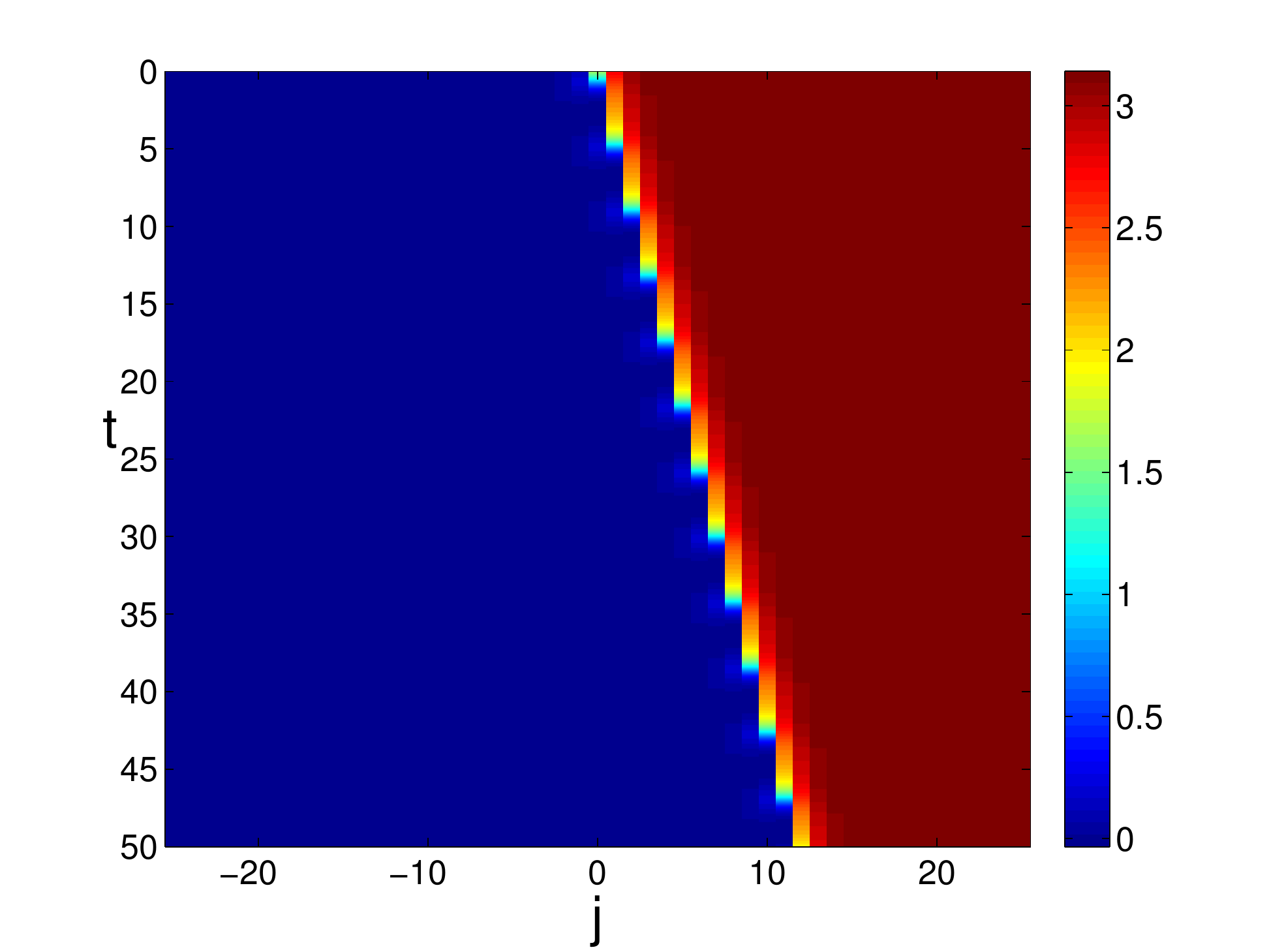}}}
\end{center}
\caption{Same plots as in Fig.~\ref{f200} but now for the stable
traveling wave solution with $k=1.1$ and $\mu=0.5$, yielding $c=0.2382$.}
\label{f204}
\end{figure}

Examples of traveling wave solutions are shown in Figs.~\ref{f200}a--\ref{f204}a corresponding to different
values of the parameters $k$ and $\mu$, characterizing the intersite coupling.
In each case, stability of the traveling wave solution $\phi(z)$
is analyzed by computing the spectrum of the Jacobian matrix obtained by linearizing Eq.~\eqref{tw1} about $\phi(z)$  with a forward difference approximation of the spatial derivative (see Appendix for further details); the resulting spectra are shown in Figs.~\ref{f200}b--\ref{f204}b.
We also check stability of the obtained solutions by solving both the ODE system \eqref{ee1} and the advance-delay PDE \eqref{tw1}
initialized at the traveling wave.
Figs.~\ref{f200}c--\ref{f204}c show space-time evolution of the solution of
the advance-delay PDE \eqref{tw1} with the initial condition $\Theta(z,0)=\phi(z)$. The solution is
obtained with the discretization grid used to obtain the traveling wave $\phi(z)$. As expected, the results show that $\phi(z)$
is a stationary solution of Eq.~\eqref{tw1}. Finally, in Figs.~\ref{f200}d--\ref{f204}d, we show the space-time evolution of the
solution of the ODE system \eqref{ee1}
with $N=2n+1=51$ oscillators initialized by the traveling wave solution $\phi(z)$ evaluated at the integer values of $z$, $\theta_j(0)=\phi(j)$, $j=-25,\dots,25$.
The time evolution of the PDE \eqref{tw1} and the ODE system \eqref{ee1} is accomplished using the classical fourth order Runge-Kutta method.

In Fig.~\ref{f200}a, the traveling wave solution $\phi(z)$ (solid line) is shown at $k=2.25$ and $\mu=0.5$, which yields $c=0.8123$.
Fig.~\ref{f200}b indicates that the eigenvalues
of the linearization Jacobian of the PDE \eqref{tw1}, evaluated at this solution,
have only negative real parts, implying the linear (spectral)
stability of the traveling wave solution as a stationary solution of Eq.~\eqref{tw1}.
To further examine this prediction, the traveling wave solution is subsequently
used as the initial data in Eq.~\eqref{tw1}, $\Theta(z,0)=\phi(z)$, and the PDE is solved numerically, with the space-time contour of the evolution
of $\Theta(z,\tau)$ shown in Fig.~\ref{f200}c confirming the robust evolution of this steady state. Finally, Fig.~\ref{f200}d shows the evolution
of the solution of the discrete ODE system \eqref{ee1} initialized by the traveling wave solution, $\theta_j(0)=\phi(j)$.
As anticipated by the traveling wave ansatz, the ODE solution is found to traverse the lattice at the velocity $c=0.8124$, which
agrees with the
traveling wave velocity very well \footnote{The speed of the front is numerically calculated as follows. We choose an integer grid point $k$ and
place the right-traveling front to its left, near an integer grid point $m$ with $m<k$ such that $|\theta_m-\frac{\pi}{2}| < |\theta_j-\frac{\pi}{2}|$ for all $j$. Then
 $\theta$ is close to $\frac{\pi}{2}$ at the integer grid point $m$ than at any other integer grid point. The front is allowed to
evolve using the standard fourth order Runge-Kutta method. We count the number of time steps $p$
such that $|\theta_k-\frac{\pi}{2}| < |\theta_j-\frac{\pi}{2}|$ for all $j$. The total time where $\theta= \frac{\pi}{2}$ lies
in the interval $[k-\frac{1}{2},k+\frac{1}{2}]$ is denoted as $p\Delta t$ where $\Delta t$ is the Runge-Kutta time step,  and the approximation to the
speed is then $\frac{1}{p \Delta t}$. The error is observed to be $O(\Delta t)$. If $p=0$, meaning that the front never entered the
interval $[k-\frac{1}{2},k+\frac{1}{2}]$, then the speed is zero, assuming that the method was allowed to run long enough.}.
%

Similar results are shown in Figs.~\ref{f202} and~\ref{f204} for
the cases of $k=1.5$ and  $k=1.1$ at $\mu=0.5$, with traveling wave velocities $c=0.5368$ and $c=0.2382$, respectively, obtained by solving Eq.~\eqref{tw11}. These are in excellent agreement with the velocities $c=0.5367$ and $c=0.2377$, respectively, obtained from the solution of the ODE system \eqref{ee1} initialized by the traveling wave (d panels).
Observe that as $k$ decreases at fixed $\mu$, the velocity $c$ of the traveling wave decreases, and
the eigenvalues shown in Figs.~\ref{f200}b, \ref{f202}b and \ref{f204}b  move closer to the imaginary axis as well.  Nonetheless, the obtained eigenvalues in all three cases
remain in the left half-plane $\text{Re} \lambda<0$, suggesting that the relevant
wave is stable as a stationary solution of the co-traveling wave PDE \eqref{tw1}. We
anticipate that this spectral stability of the solution for the PDE implies the dynamical stability
of the traveling wave as the solution of the ODE system \eqref{ee1}, as confirmed by our simulations. However, the general
demonstration of such a connection is, to the best of our knowledge, an intriguing open problem in analysis.
%

\subsection{Existence and stability of traveling waves for the chain}

\begin{figure}[tbp]
\begin{center}
{\subfloat[]{\includegraphics[width=0.45\textwidth]{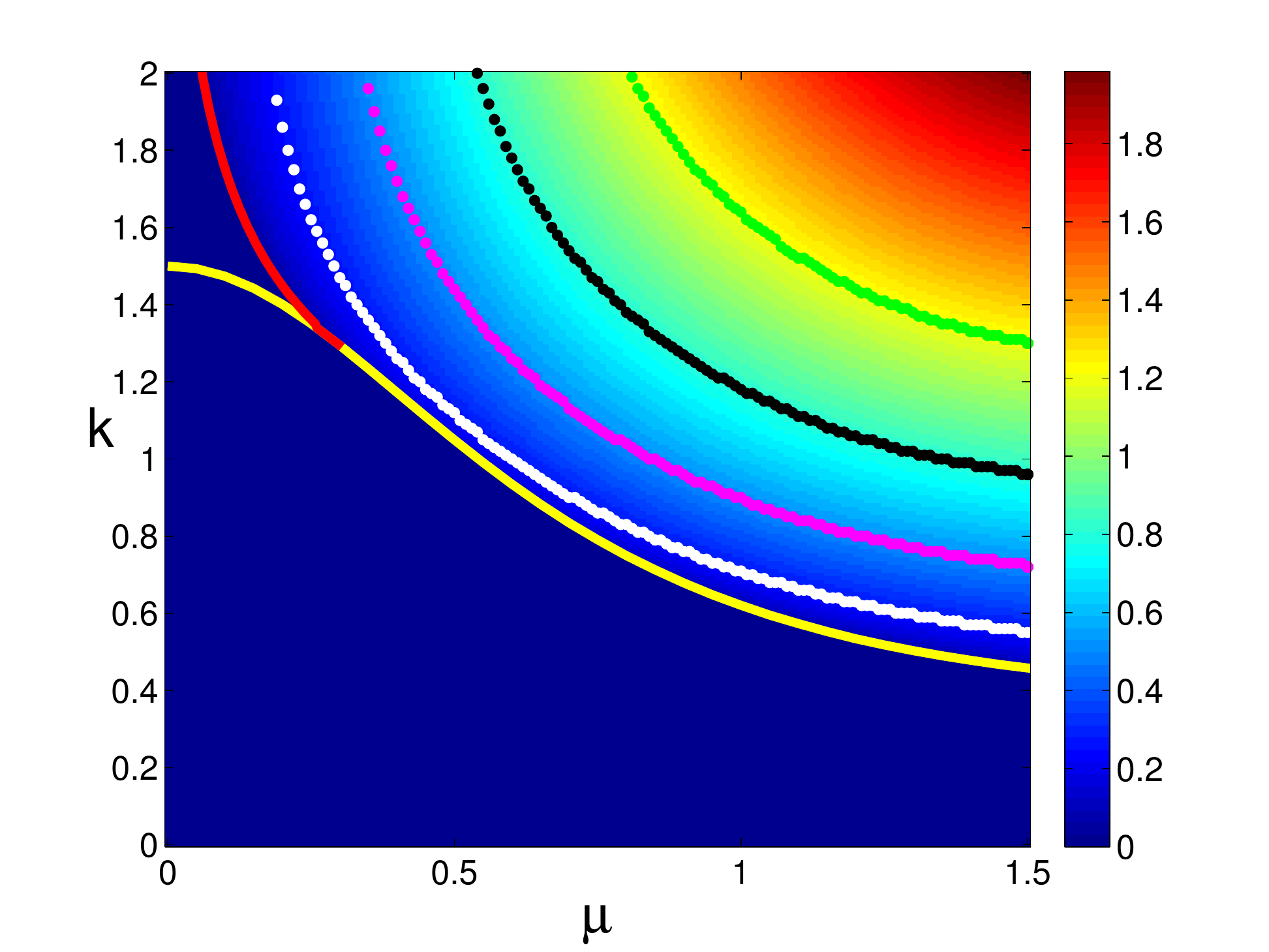}}} \quad
{\subfloat[]{\includegraphics[width=0.45\textwidth]{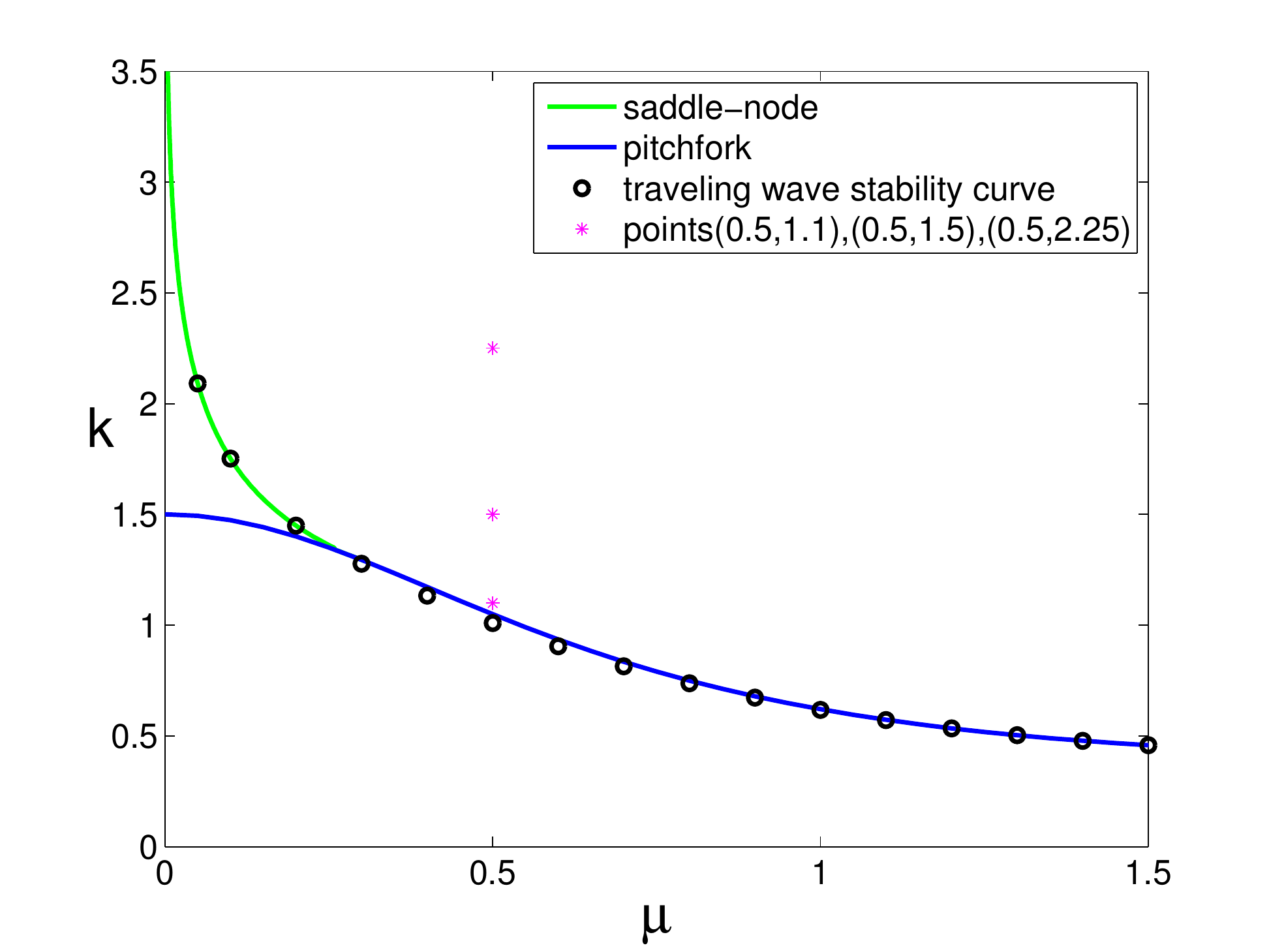}}}
%
\end{center}
\caption{(a) Contour plot of the velocity $c$ of the traveling waves in the $(\mu,k)$ plane obtained by solving Eq.~\eqref{tw11} and isocontours of $c=0.25$ (white), $c=0.5$ (magenta), $c=0.8$ (black) and $c=1.2$ (green), shown together with the pitchfork (yellow) and saddle-node (red) bifurcation curves obtained  for the equilibrium states of the 51-oscillator chain. (b) The curve (open circles) where the traveling waves have (almost) zero velocity nearly coincides with the saddle-node (green) and pitchfork (blue) bifurcation curves; parameter values used in Figures \ref{f200},\ref{f202},\ref{f204} are also shown (magenta dots).}
\label{f218}
\end{figure}

We now explore more systematically the existence and stability of traveling waves in the $(\mu,k)$ parameter plane.
The existence results for sufficiently small $\mu$ ($0 \leq \mu \leq 1.5$) are summarized in Fig.~\ref{f218}a,
which shows the contour plot of velocity of the traveling wave computed at different $\mu$ and $k$ using Eq.~\eqref{tw11} and
complements the steady state analysis of \cite{parks} shown in Fig.~\ref{f00}. For comparison,
pitchfork (yellow) and saddle-node (red) bifurcation curves for a 51-oscillator chain
(and shown by solid and dashed curves, respectively, in Fig.~\ref{f00}) are also included. To obtain the pitchfork curve,  stationary solutions
of Eq.~\eqref{ee1} are obtained
using Newton's method, starting with a value $k=0.4$ for given $\mu$ and using continuation in $k$ until the Jacobian of the relevant system
becomes singular. To obtain the saddle-node curve, a stable stationary solution
is identified in region B of Fig.~\ref{f00}. This is obtained by
integrating Eq.~\eqref{ee1} using an unstable solution initial condition
over a long time horizon until a stable stationary solution is obtained
(as an attractor of the dynamics).
This solution is used in Newton's method as a good initial guess in
order to obtain the stable
waveform within region B. Numerical continuation is then used
with increasing $\mu$ until the Jacobian becomes singular. This
monoparametric process is repated for different values of $k$ within region B.
The
curves obtained are almost indistinguishable from
the ones in Fig.~\ref{f00}.
While we perform our computations in a finite chain, having examined chains
of different sizes, we expect our principal conclusions to
persist (essentially without modification)
for the case of the infinite chain.
On the flip side, we can identify traveling waves
only in region C,
while below it their speed degenerates to $0$, leading to
standing waves (split-state equilibria),
in agreement with the findings of~\cite{parks} for the finite chain case
(and our discussion of standing wave states above).

We now explore this comparison in more quantitative detail
in Fig.~\ref{f218}b.
Here the open circles mark the curve above which numerical simulations of ODE system \eqref{ee1} yield stable  traveling waves.
To obtain these points, we fixed $\mu$ and solved \eqref{tw11} for the traveling wave solution $\phi(z)$ and its velocity at $(\mu,k)$ starting with $k$ just above the bifurcation curves and then progressively decreasing its value. To verify the velocity of the traveling wave at given $(\mu,k)$, we then used $\theta_j(0)=\phi(j)$ as initial conditions in the simulations of \eqref{ee1} and computed the velocity of the propagating front. The open circles were obtained by finding the values of $k$ where the traveling wave solution has zero velocity up to the numerical error in both methods.
The comparison strongly suggests that the disappearance of standing
wave solutions of~\cite{parks} gives rise to the traveling wave solutions
analyzed here.

\begin{figure}[!htb]
\begin{center}
{\subfloat[$\mu=0.5$]{\includegraphics[width=0.4\textwidth]{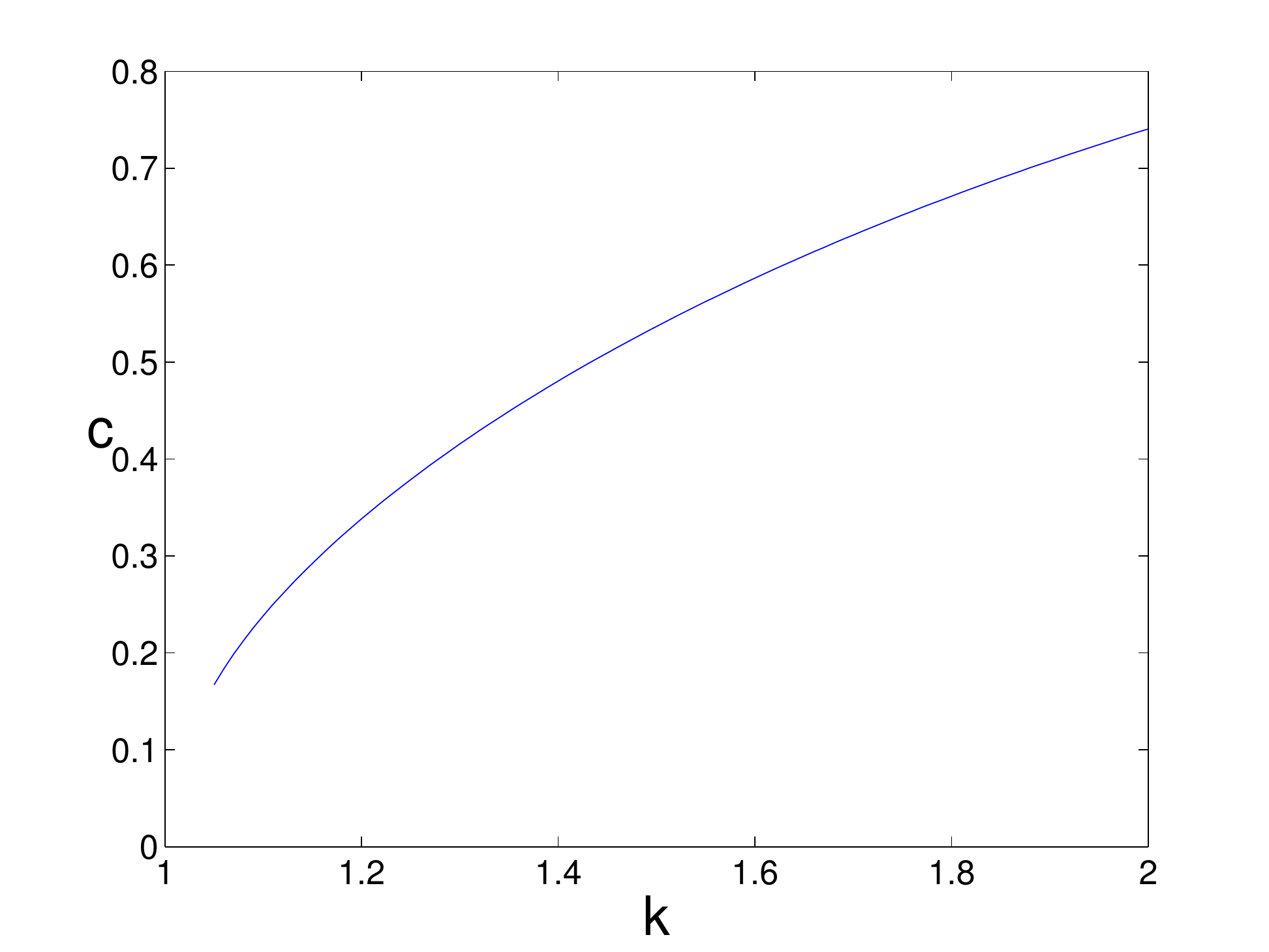}}}
{\subfloat[$\mu=1$]{\includegraphics[width=0.4\textwidth]{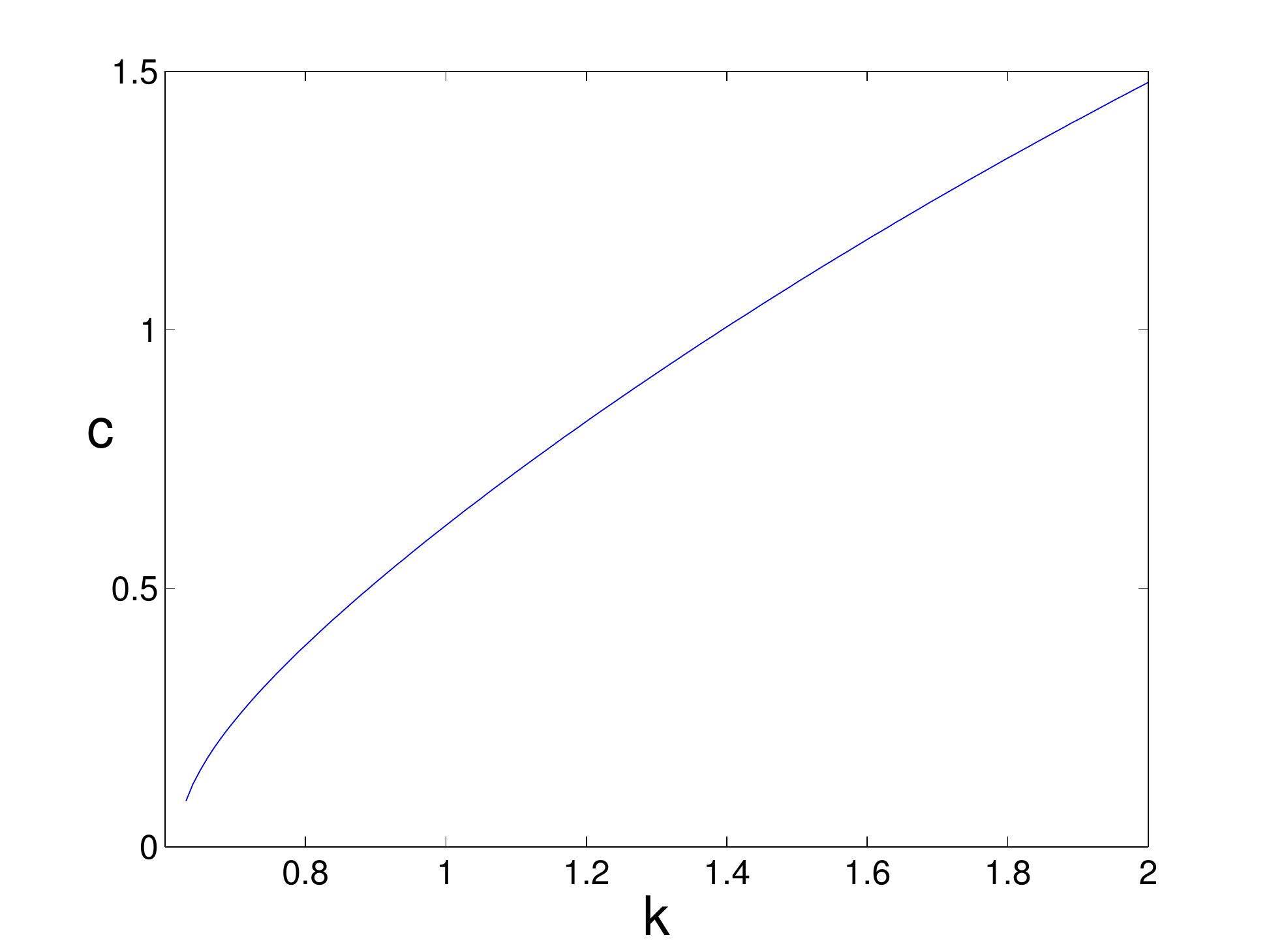}}}
{\subfloat[$k=1.01$]{\includegraphics[width=0.4\textwidth]{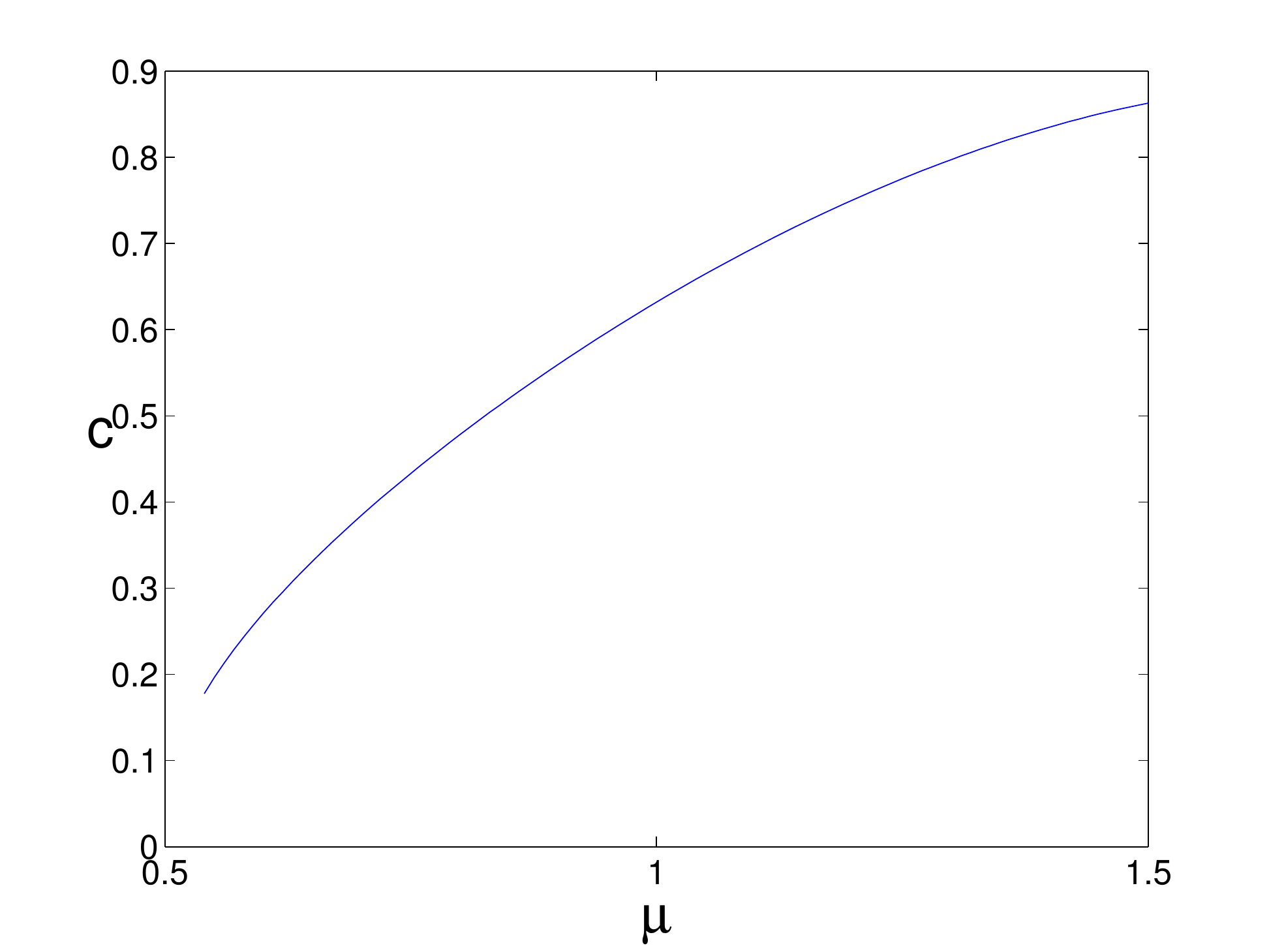}}}
{\subfloat[$k=1.46$]{\includegraphics[width=0.4\textwidth]{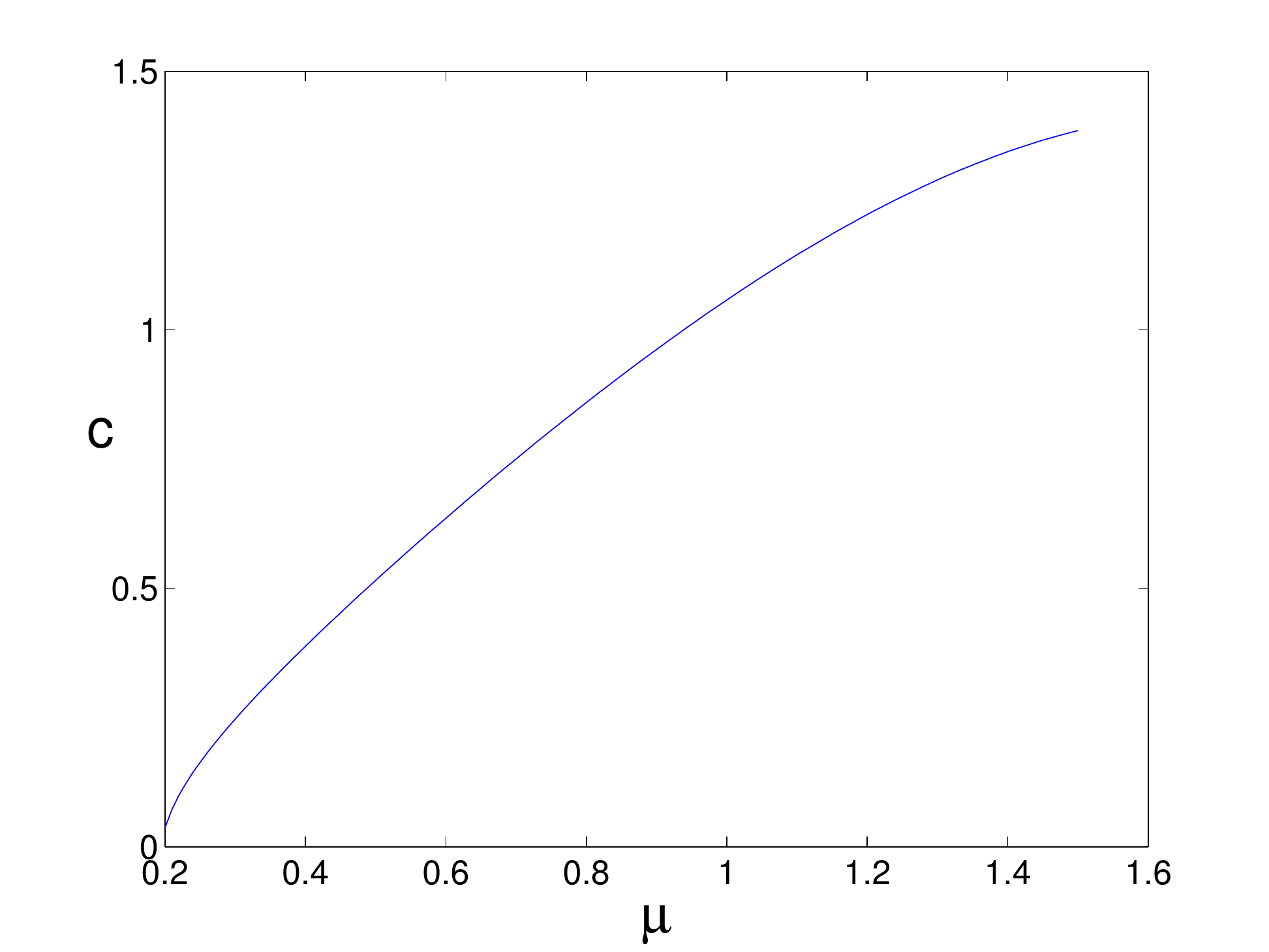}}}
\end{center}
\caption{Velocity $c$ as a function of $k$ at fixed $\mu$ (panels a and b) and as a function of $\mu$ at fixed $k$ (panels c and d).}
\label{speedvsk}
\end{figure}

The isocontours
associated with different velocities, shown in Fig.~\ref{f218}a, illustrate how the velocity $c$ of traveling waves, as obtained from the PDE, depends
on the two parameters $\mu$ and $k$.
To complement these results with a monoparametric
visualization, we also show some plots of $c(k,\mu)$ at fixed $\mu$ in Fig.~\ref{speedvsk}a,b and of $c$ as a function of $\mu$ at fixed $k$ in Fig.~\ref{speedvsk}c,d.
These curves show that the velocity increases with $k$ and $\mu$. As the velocity decreases, these curves approach the region containing
the standing waves.


%

We now discuss an important aspect of the obtained spectral pictures (e.g. the ones shown in
Figs.~\ref{f200}b, \ref{f202}b, \ref{f204}b) that concerns stability of the background state with
$\Theta_0=0$ or $\Theta_0=\pi$. Our solutions clearly have support on both of
these states, so the spectrum of these homogeneous states should
be mirrored within that of our (standing or traveling) waves.
Substituting $\Theta(z,\tau)=\Theta_0+\epsilon v(z,\tau)$ in Eq.~\eqref{tw1}, where $\epsilon$ is a small parameter, and linearizing about $\Theta_0$, one obtains
the following linear advance-delay partial differential equation with constant coefficients for $v(z,\tau)$:
\begin{equation}
v_{\tau} - cv_z= k \cos(\mu) (v(z+1,\tau)-2v(z, \tau)+v(z-1,\tau)) -2v(z,\tau).
\label{tw22}
\end{equation}
Seeking solutions of \eqref{tw22} in the form $v=e^{\lambda \tau} e^{i p z}$ for real $p$, we have
\begin{equation}
\lambda = -2(2k\cos(\mu)\sin^2(p/2)+1)+ i c p .
\label{tw3}
\end{equation}
The equilibrium state $\Theta(z,\tau) \equiv \Theta_0$ is unstable when $\text{Re}\lambda>0$, i.e. when
\[
k\cos(\mu) < -\frac{1}{2\sin^2(p/2)}
\]
for some real $p$. Since $k>0$, this can happen for example at large enough $k$ and
$\frac{\pi}{2} < \mu < \frac{3 \pi}{2}$. More precisely, the
background solution is always unstable for
\begin{equation}
k > \frac{1}{2} |\sec(\mu)|, \quad \cos(\mu)<0.
\label{instab}
\end{equation}
Additionally, since \eqref{tw3} implies that $\text{Re}\lambda=-2(2k\cos(\mu)\sin^2(p/2)+1)$ and $\text{Im}\lambda=c p$, these eigenvalues determine
the following locus of points in the spectral plane:
\begin{equation}
\text{Re}\lambda=-2\left\{2k\cos(\mu)\sin^2\left(\dfrac{\text{Im}\lambda}{2 c}\right)+1\right\},
\label{locus}
\end{equation}
which should be traceable in the linearization spectra of the traveling waves considered
here.

\begin{figure}[tbp]
\begin{center}
{\subfloat[]{\includegraphics[width=0.4\textwidth]{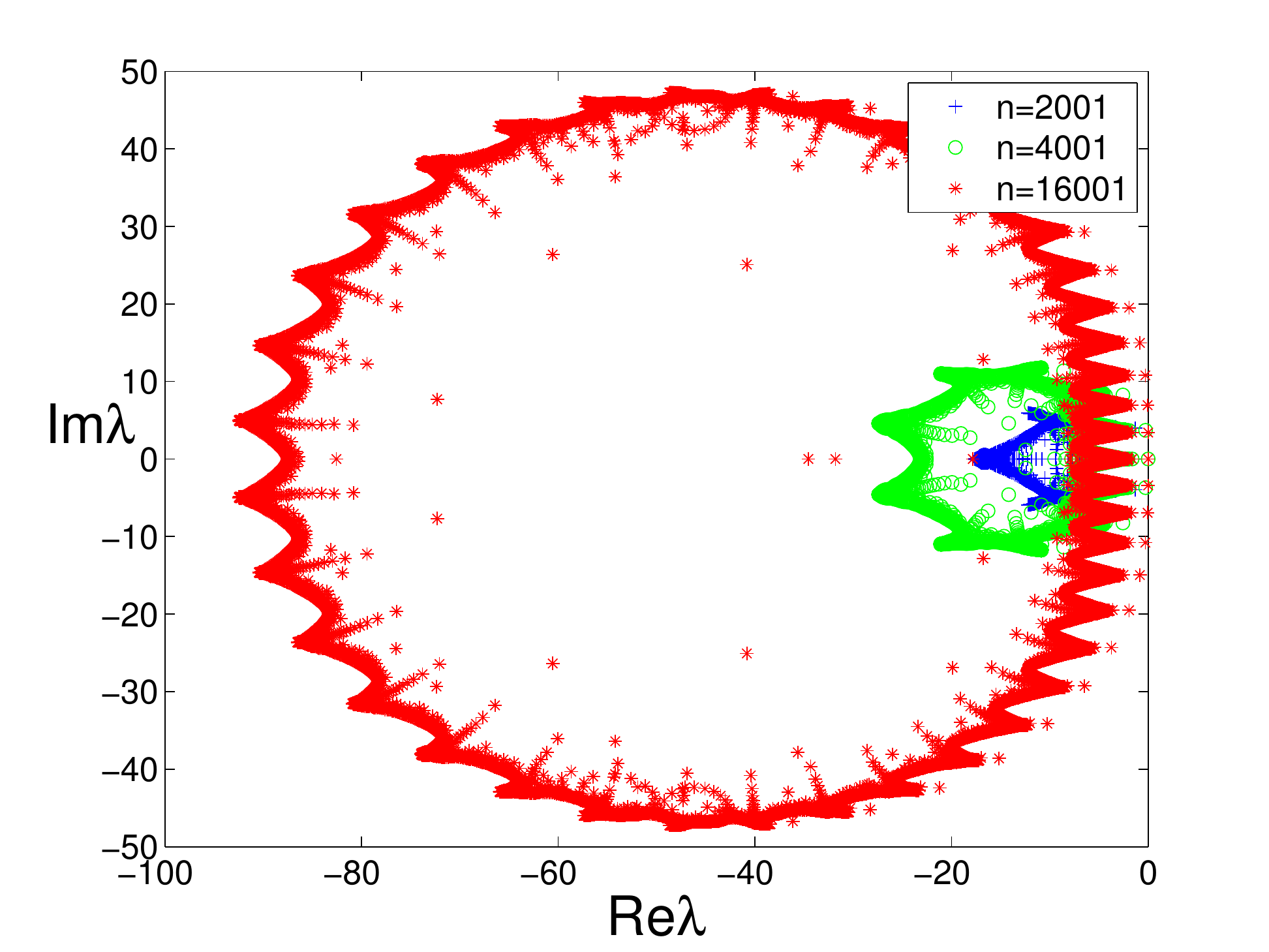}}}
{\subfloat[]{\includegraphics[width=0.4\textwidth]{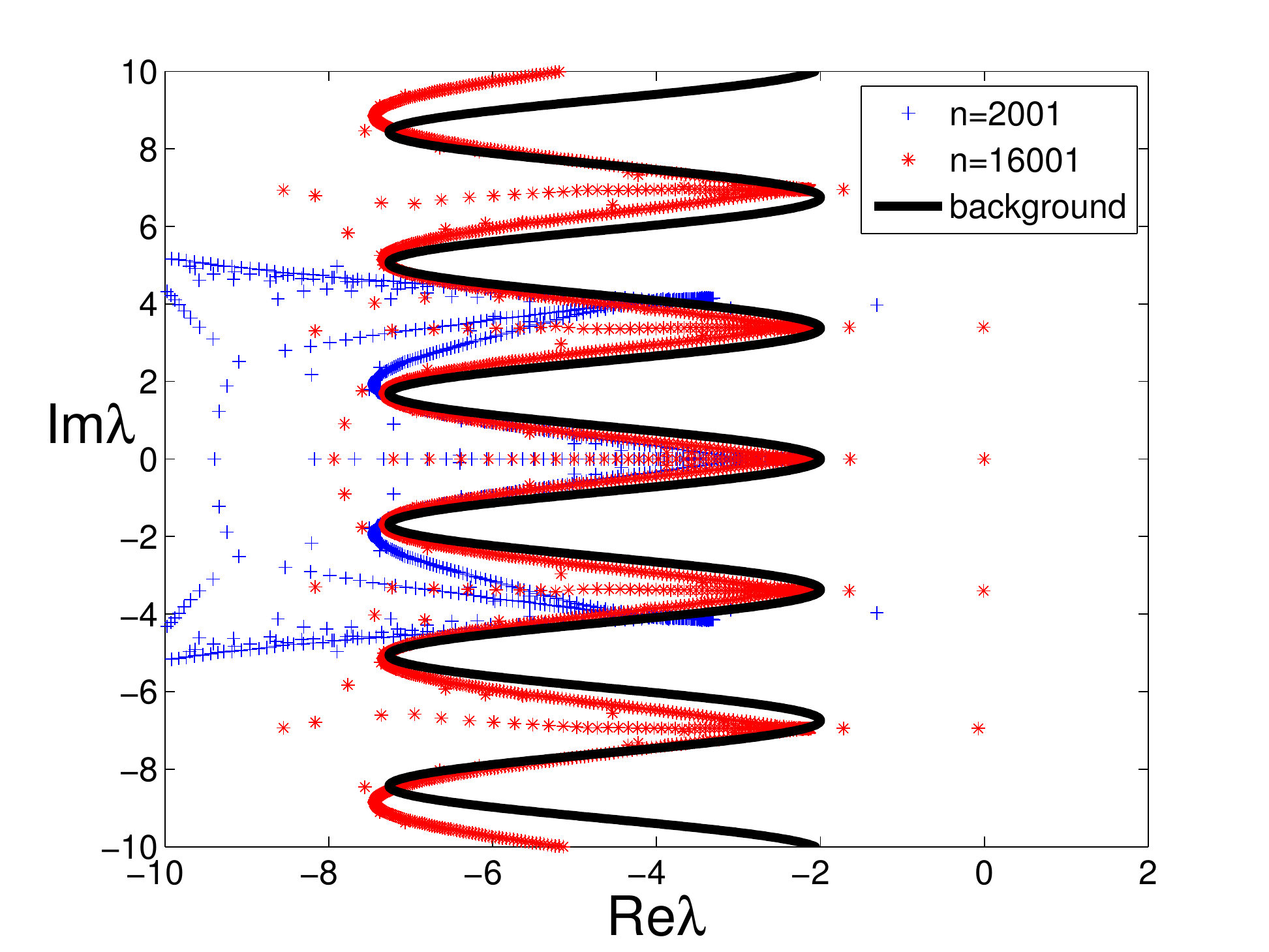}}}

\end{center}
\caption{(a) The eigenvalues of the Jacobian for the traveling wave solution at $\mu=0.5$, $k=1.5$ discretized on the interval $[-200,200]$ with
$n=2001$, $4001$ and $16001$ nodes. (b) The enlarged version of the $n=2001$ and $n=16001$ cases from (a), along with the continuous background spectrum curve given by Eq.~\eqref{locus}, near the imaginary axis.}
\label{f318}
\end{figure}

For $0 \le \mu \le \frac{\pi}{2}$ and $\frac{3\pi}{2} \leq \mu \leq 2\pi$, based on equation (\ref{locus}), the background is always stable, so any instability must come from the
front itself. For example, Fig.~\ref{f318}a shows  the eigenvalues of the
Jacobian matrix associated with the traveling wave solution at $\mu=0.5$ and $k=1.5$.
The eigenvalues are given for $n=2001$, $4001$ and $16001$ nodes in our discretization of the co-traveling wave problem \eqref{tw1} on the interval $[-200,200]$, with the step size $h=400/(n-1)$. The eigenvalues of the linearized equation about the background solutions $\Theta_0=0$ and $\Theta_0=\pi$ are given
by Eq.~\eqref{tw3}. For these parameter values, we
theoretically have $-7.27 \leq \text{Re}\lambda \leq -2$. In Fig.~\ref{f318} one can see a band of eigenvalues folding
towards the imaginary axis as $n$ increases.  Zooming in on the region near the imaginary axis in Fig.~\ref{f318}b,
we see that the eigenvalues are converging to the background spectrum as the continuum limit is approached, as predicted.
In the Appendix, we explain further how the forward difference scheme we use
results in the particular structure of the continuous spectrum of the
problem (inducing the parabolic shape
observed in Fig.~\ref{f318}a) and how a centered difference scheme would affect the corresponding spectral and stability picture.
Similar results are found for the cases of $k=2.25$ and $k=1.1$ (not shown here).

\begin{figure}[!htb]
\begin{center}
{\includegraphics[width=0.45\textwidth]{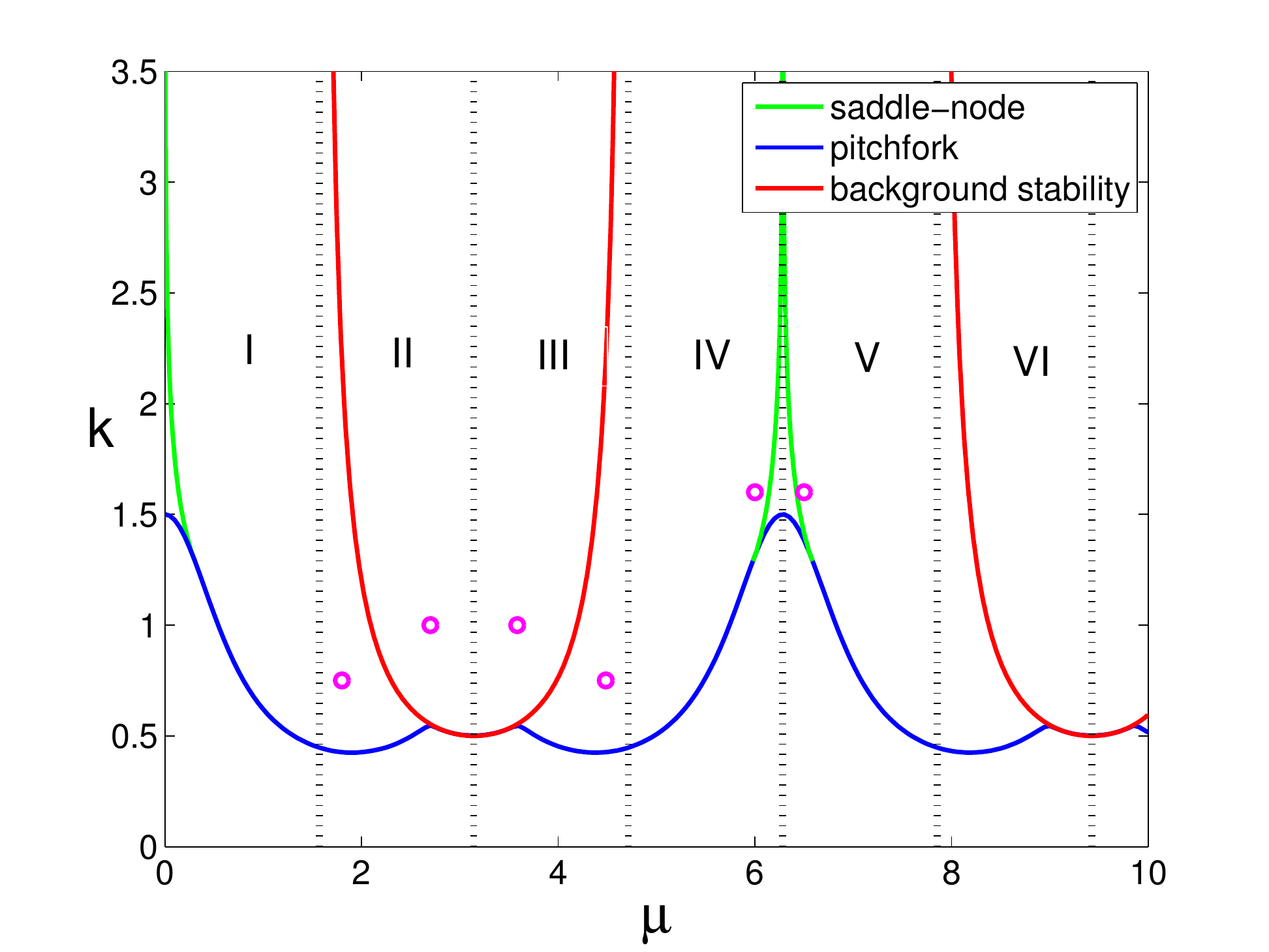}}
\end{center}
\caption{Extended pitchfork (blue) and saddle-node (green) bifurcation curves and the background instability curve (red).  Magenta circles
denote the parameter values where the
stability of the traveling waves was probed via
direct numerical computations.}
\label{fig4}
\end{figure}

In the above discussion,
we extended the analysis of the standing waves in~\cite{parks} to traveling wave
solutions. However, somewhat in line with the considerations in \cite{parks}, this study until now was
limited to $0 \leq \mu \leq 1.5$. We now extend our analysis to the entire upper half of the $(\mu,k)$ plane.
So far, we have obtained stability curves (pitchfork and saddle-node) separating
traveling and standing waves.
Extending these notions to a broader $\mu$-interval in the $(\mu,k)$ plane,
we obtain the bifurcation curves for $0 \le \mu \le 10$ as shown in Fig.~\ref{fig4}.
Along with extended pitchfork and saddle-node curves we show the curves above which the background is unstable, according to the inequality \eqref{instab} obtained above for the linearized problem. The curve separating standing and traveling waves coincides with the saddle-node bifurcation curves where they exist and follows the pitchfork bifurcation curve otherwise.

\begin{figure}[!htb]
\begin{center}
{\subfloat[]{\includegraphics[width=0.45\textwidth]{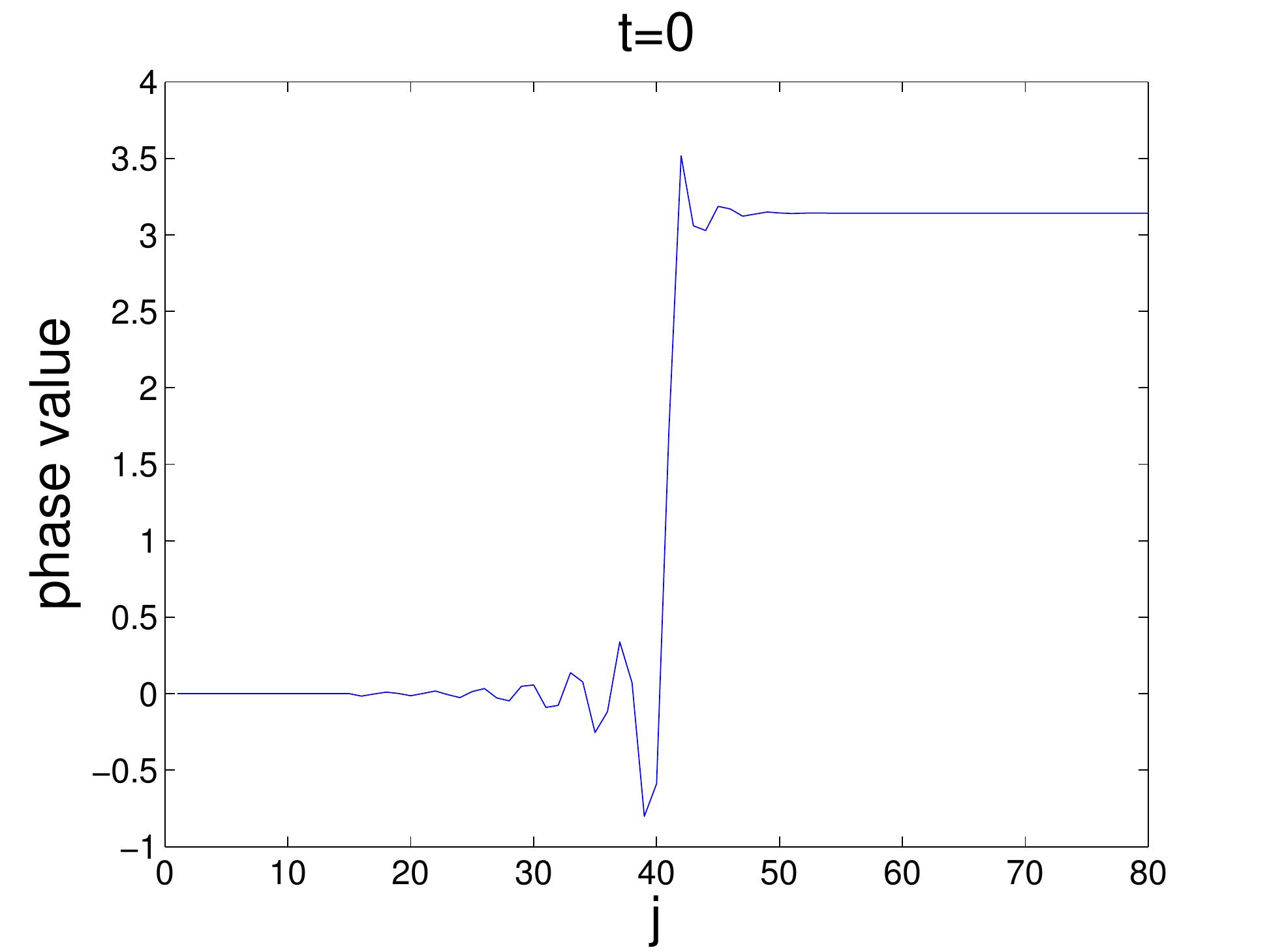}}}
{\subfloat[]{\includegraphics[width=0.45\textwidth]{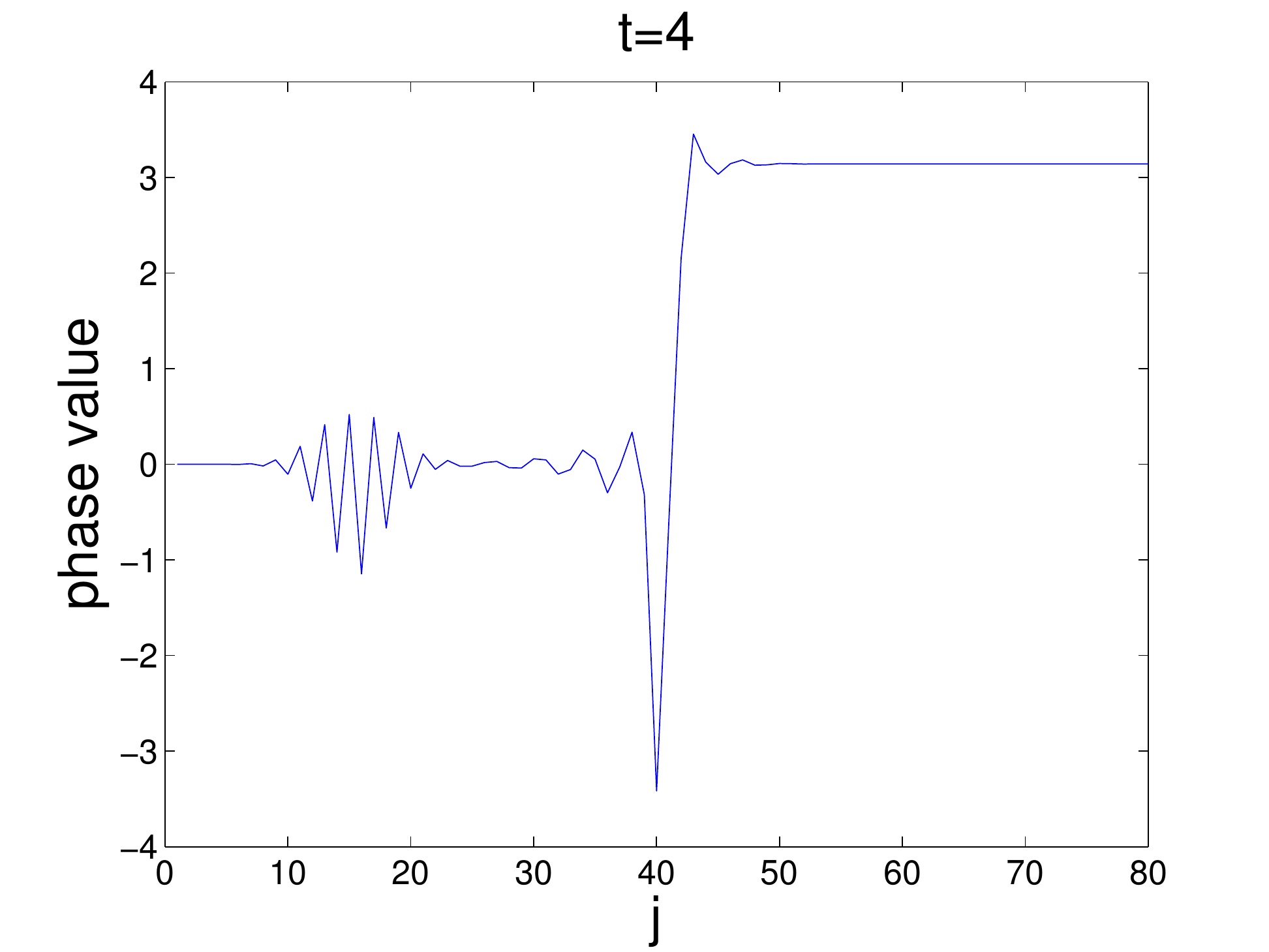}}}
{\subfloat[]{\includegraphics[width=0.45\textwidth]{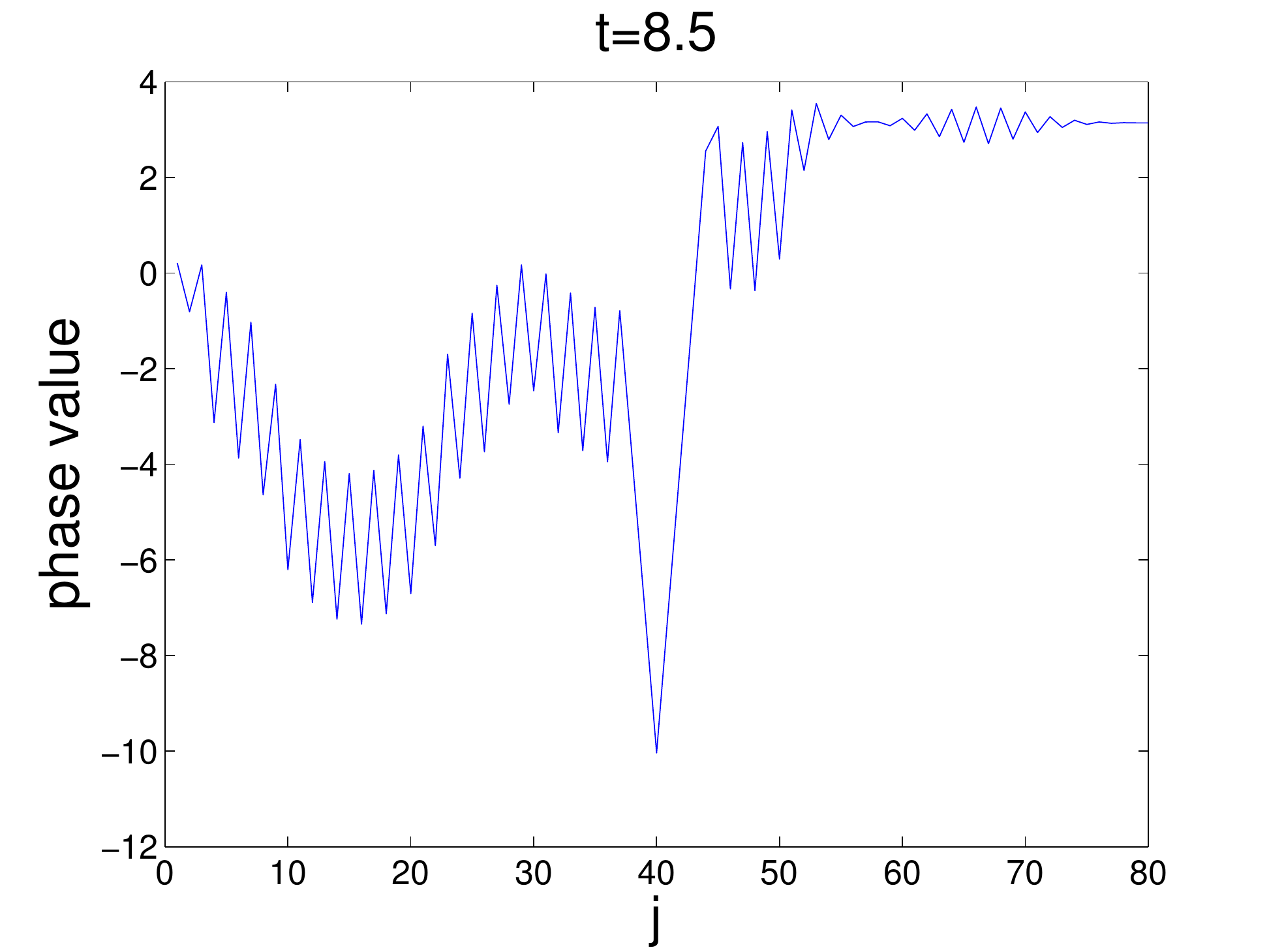}}}
{\subfloat[]{\includegraphics[width=0.45\textwidth]{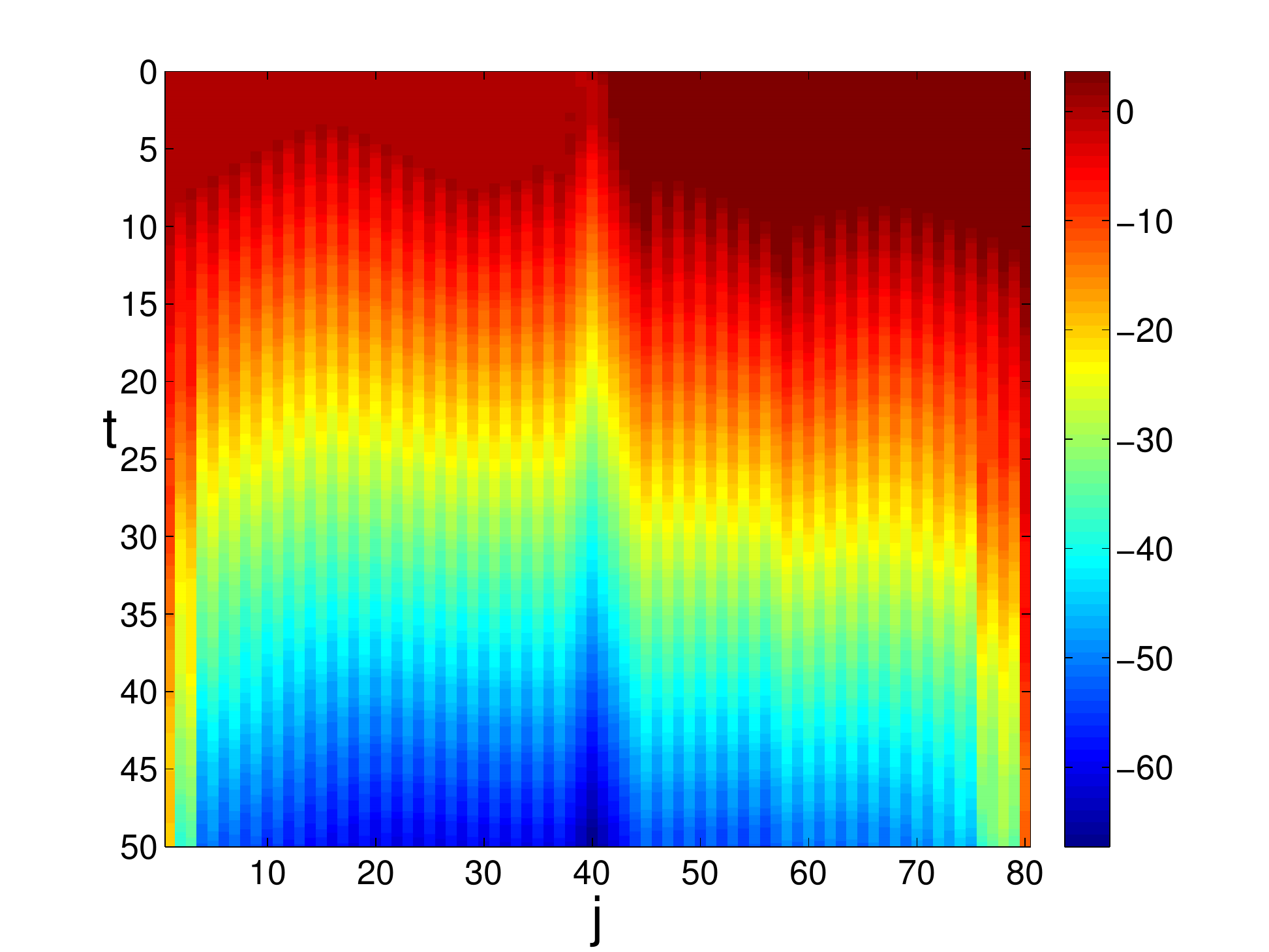}}}
\end{center}
\caption{Unstable traveling wave at $(\mu,k)=(2.7, 1)$ with $c=0.2233$: (a) Initial state ($t=0$) obtained from the solution of Eq.~\eqref{tw11}. (b) Solution of Eq.~\eqref{ee1} at $t=4$ where the left side of the solution, with $\theta_j \approx 0$, manifests destabilization. (c) Solution at $t=8.5$ where the right side, with $\theta_j \approx \pi$,
also shows destabilization.
(d) Contour plot of space-time evolution until time $t=50$.  Here the color code represents $\theta_j$ values on the real line, rather than in $[0,2\pi)$ mod $2\pi$.}
\label{fig11a}
\end{figure}

\begin{figure}[!htb]
\begin{center}
{\subfloat[]{\includegraphics[width=0.45\textwidth]{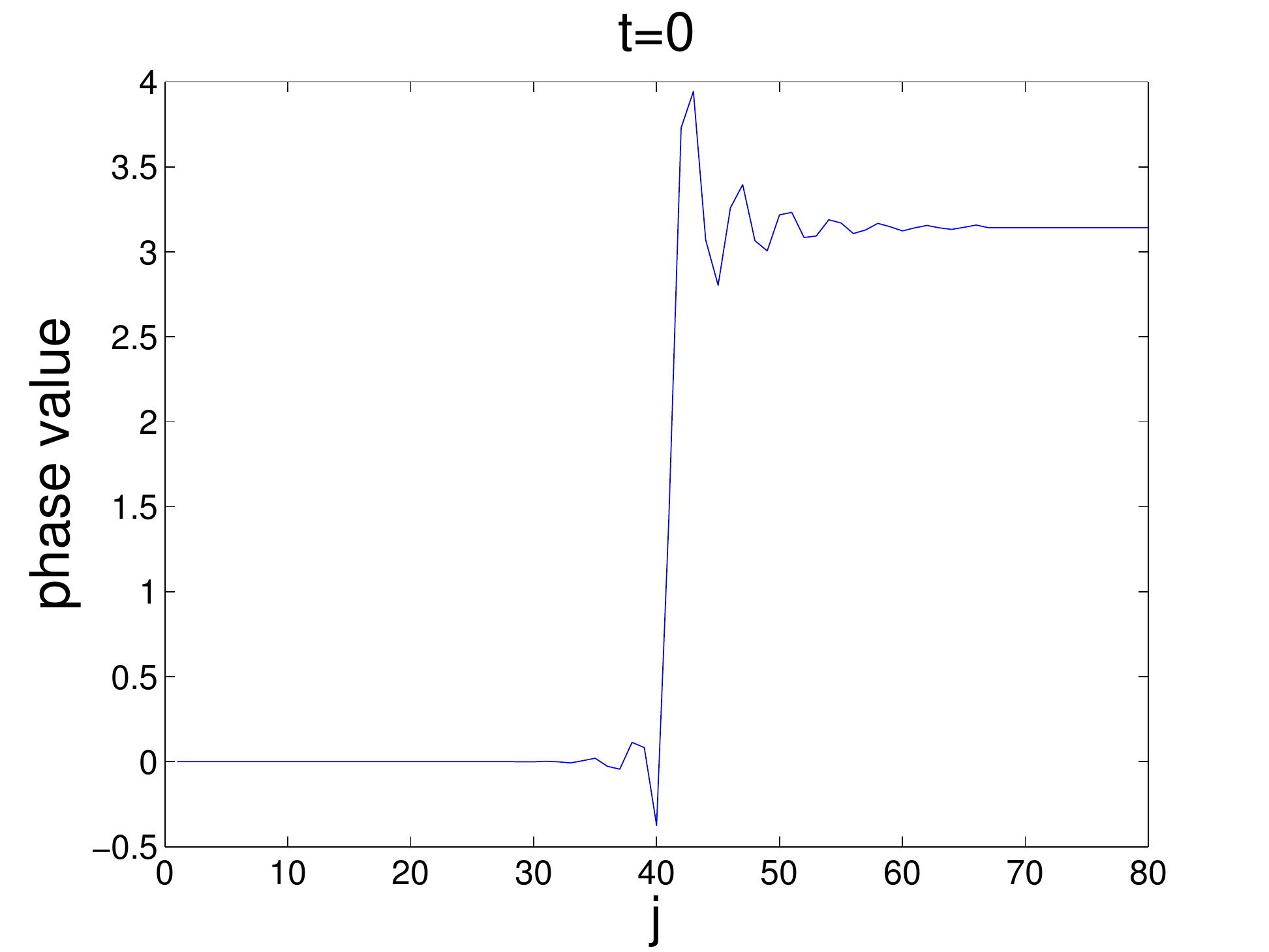}}}
{\subfloat[]{\includegraphics[width=0.45\textwidth]{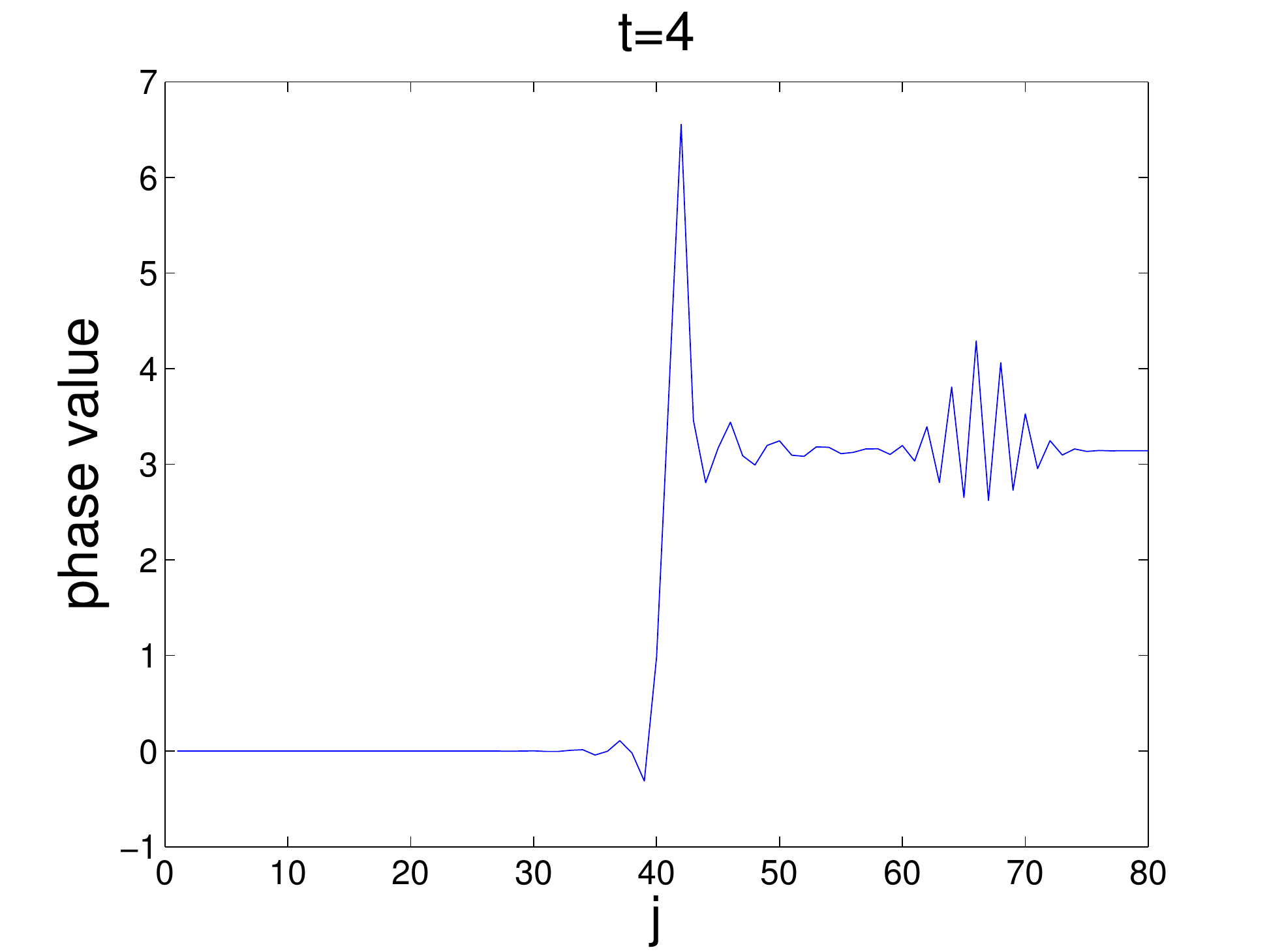}}}
{\subfloat[]{\includegraphics[width=0.45\textwidth]{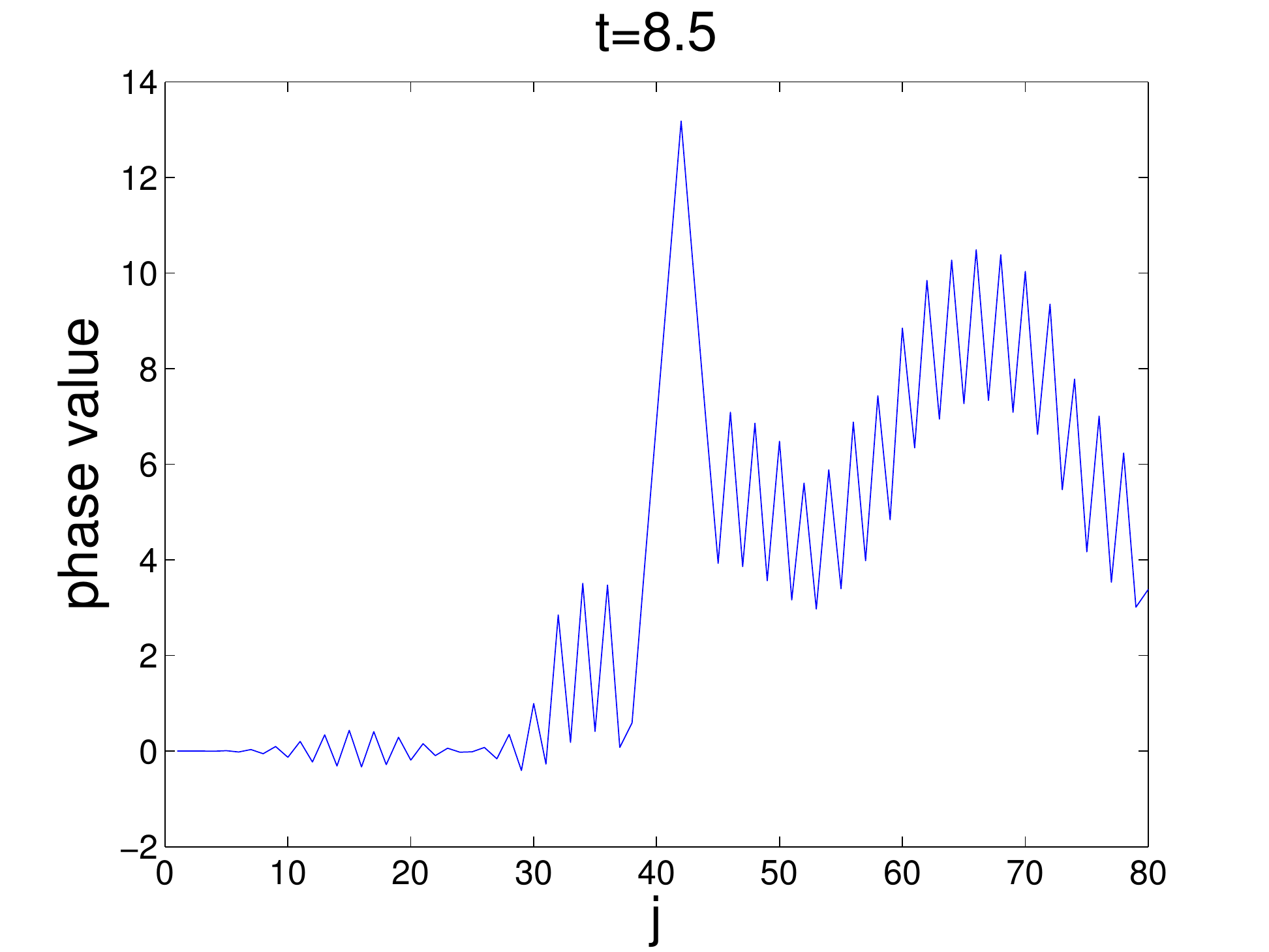}}}
{\subfloat[]{\includegraphics[width=0.45\textwidth]{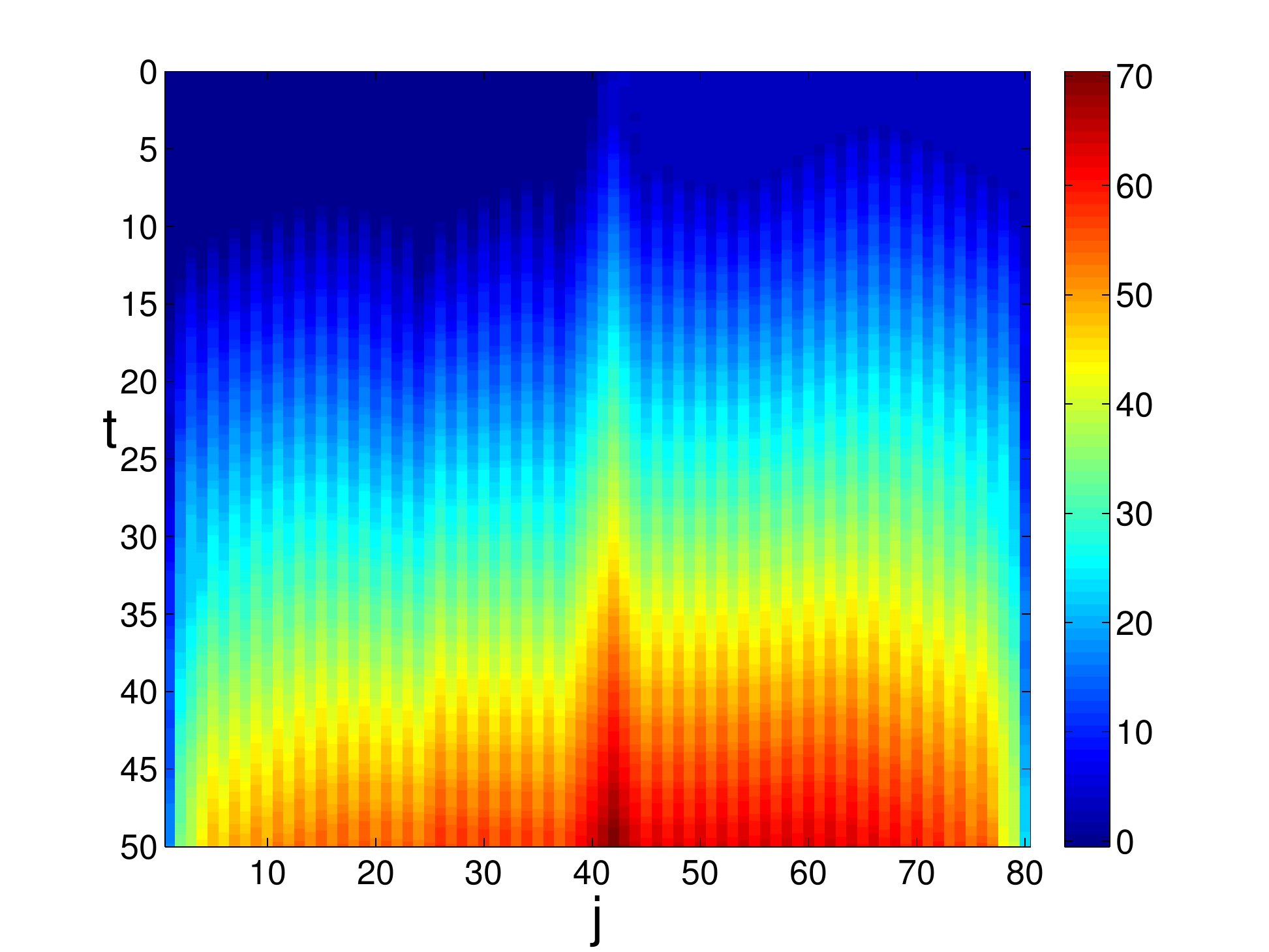}}}
\end{center}
\caption{Unstable traveling wave at $(\mu,k)=(2\pi-2.7, 1)$ with $c=-0.2233$: (a) Initial state ($t=0$) obtained from the solution of Eq.~\eqref{tw11}. (b) Solution of Eq.~\eqref{ee1} at $t=4$ where the right side of the solution, with $\theta_j \approx \pi$ imanifests destabilization. (c) Solution at $t=8.5$ where the left side, with $\theta_j \approx 0$, suffers a similar effect. (d) Contour plot of space-time evolution until time $t=50$.  Again, the color code represents $\theta_j$ values on the real line, rather than in $[0,2\pi)$ mod $2\pi$.}
\label{fig12a}
\end{figure}

To explore the stability of traveling wave solutions where they exist,
we divide the parameter space into several regions corresponding to different intervals of $\mu$. These are
the $\mu$ regions $I$ ($ [0, \frac{\pi}{2}]$), $II$ ($[\frac{\pi}{2},\pi]$),
$III$ ($[\pi, \frac{3\pi}{2}]$), $IV$ ($[\frac{3\pi}{2}, 2\pi]$), $V$ ($[2\pi, \frac{5\pi}{2}]$) and $VI$ ($[\frac{5\pi}{2}, 3\pi]$).
In the regions from $II$ to $V$, we choose points $(\mu, k)$ and solve the traveling wave Eq.~\eqref{tw11} for $\phi(z)$ and velocity $c$.
Using the forward difference
or in some cases the centered difference approximation
yields convergence to the relevant solutions. As before, we then use
the resulting traveling wave as initial data to solve the ODE system \eqref{ee1} in order to check its stability and visualize the resulting dynamics. To obtain the initial condition, we evaluate the traveling wave solution at $51$ integer points, and then place it on a grid of $81$ points by padding it with $0$s on the left and $\pi$s on the right:
\[
\theta_j(0)=\begin{cases}
            0, & j=-40,\dots,-26\\
            \phi(j), & j=-25,\dots,25\\
            \pi, & j=26,\dots,40.
            \end{cases}
\]
The padding introduces a small perturbation of the initial curve, which, when the solution is unstable, initiates the
instability. The points $(\mu,k)= (1.8,0.75)$, $(2.7,1)$, $(2\pi-2.7,1)$, $(2 \pi-1.8,.75,)$, $(6, 1.6)$ and $(6.5, 1.6)$
(magenta circles in Fig.~\ref{fig4}) are chosen as representative points
from the regions $II$, $III$, $IV$ and $V$.
 The points $(\mu,k)=(2.7, 1)$ and $(2\pi-2.7, 1)$ lie in the region where the background state was shown to be unstable, above the red curve in Fig.~\ref{fig4}. Perturbed traveling waves for these points are shown in Fig.~\ref{fig11a} and Fig.~\ref{fig12a} respectively.

When $(\mu,k)=(2.7,1) \in II$, the traveling
wave is predicted to move to the right
with speed $c=0.2233$.  In panel (a) of Fig.~\ref{fig11a}, the initial condition is shown. In panel (b), the solution is shown at time $t=4$, when the 
perturbation of the unstable solution has caused the relevant dynamical instability to be manifested in the $\theta \approx 0$ part of the solution.
In panel (c), the solution is shown at time ($t=8.5$), when the 
instability of the background is manifested at both
ends of the domain. Finally, panel (d) gives the space-time contour plot
up to time $t=50$, clearly illustrating the destabilization of the background
and where it is initiated.

In Fig.~\ref{fig12a}, when $(\mu,k)=(2\pi-2.7,1) \in III$,
the front is predicted to move
with speed $c=-0.2233$.   The results are similar to the ones discussed above and observed in Fig.~\ref{fig11a}.  Observe that the traveling wave solutions $\phi_1(z)$ at $(\mu,k)=(2.7, 1)$, shown in Fig.~\ref{fig11a}a, and $\phi_2(z)$ in Fig.~\ref{fig12a}a at $(\mu,k)=(2\pi-2.7, 1)$ exhibit the symmetry $\phi_2(z)=\pi-\phi_1(-z)$, and their velocities are equal in absolute value and opposite in sign.
This symmetry is discussed below. Importantly, in both cases the intervals where the padding is added, and hence slight perturbations  are introduced,
are exactly where the instability of the
background initially
manifests itself (in panels (b) and (c)) numerically.
%

\begin{figure}[!htb]
\begin{center}
{\subfloat[]{\includegraphics[width=0.45\textwidth]{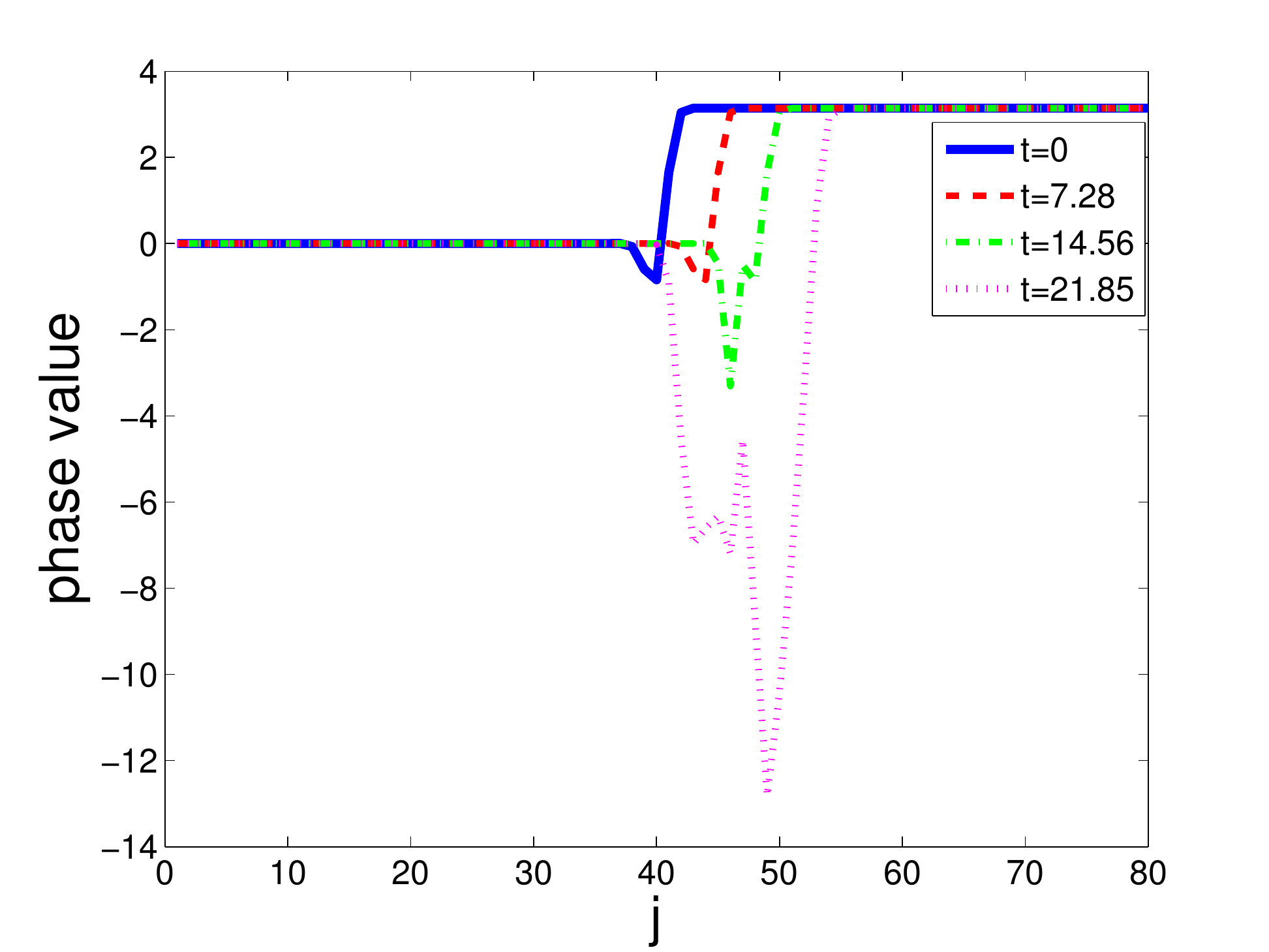}}}
{\subfloat[]{\includegraphics[width=0.45\textwidth]{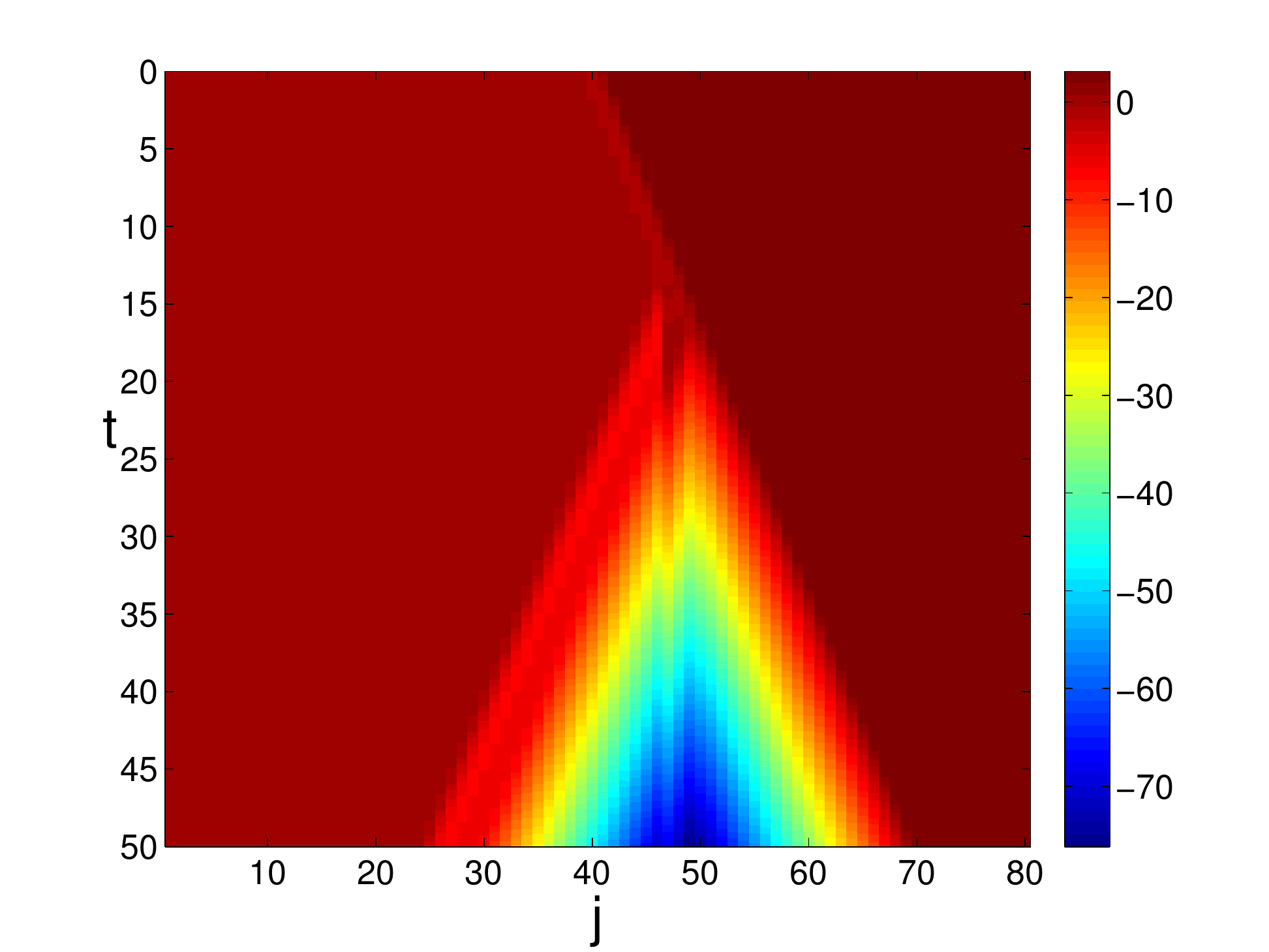}}}\end{center}
\caption{Unstable traveling wave at $(\mu,k)=(1.8, 0.75)$ with $c=0.5493$: (a) Snapshots of the solution
of Eq.~\eqref{ee1} initialized at the traveling wave at times $t=0$ (solid blue curve), $7.28$ (approximate onset of instability, dashed red), $14.56$ (dash-dotted green) and $21.85$ (dotted magenta).
(b) The contour plot of space-time evolution of
the solution (with $\theta \in \mathbb{R}$) until time $t=50$. The traveling wave becomes unstable after an initial transient propagation period and splits into two fronts propagating in the opposite directions with the same speed as the initial wave.}
\label{fig11}
\end{figure}
\begin{figure}[!htb]
\begin{center}
{\subfloat[]{\includegraphics[width=0.45\textwidth]{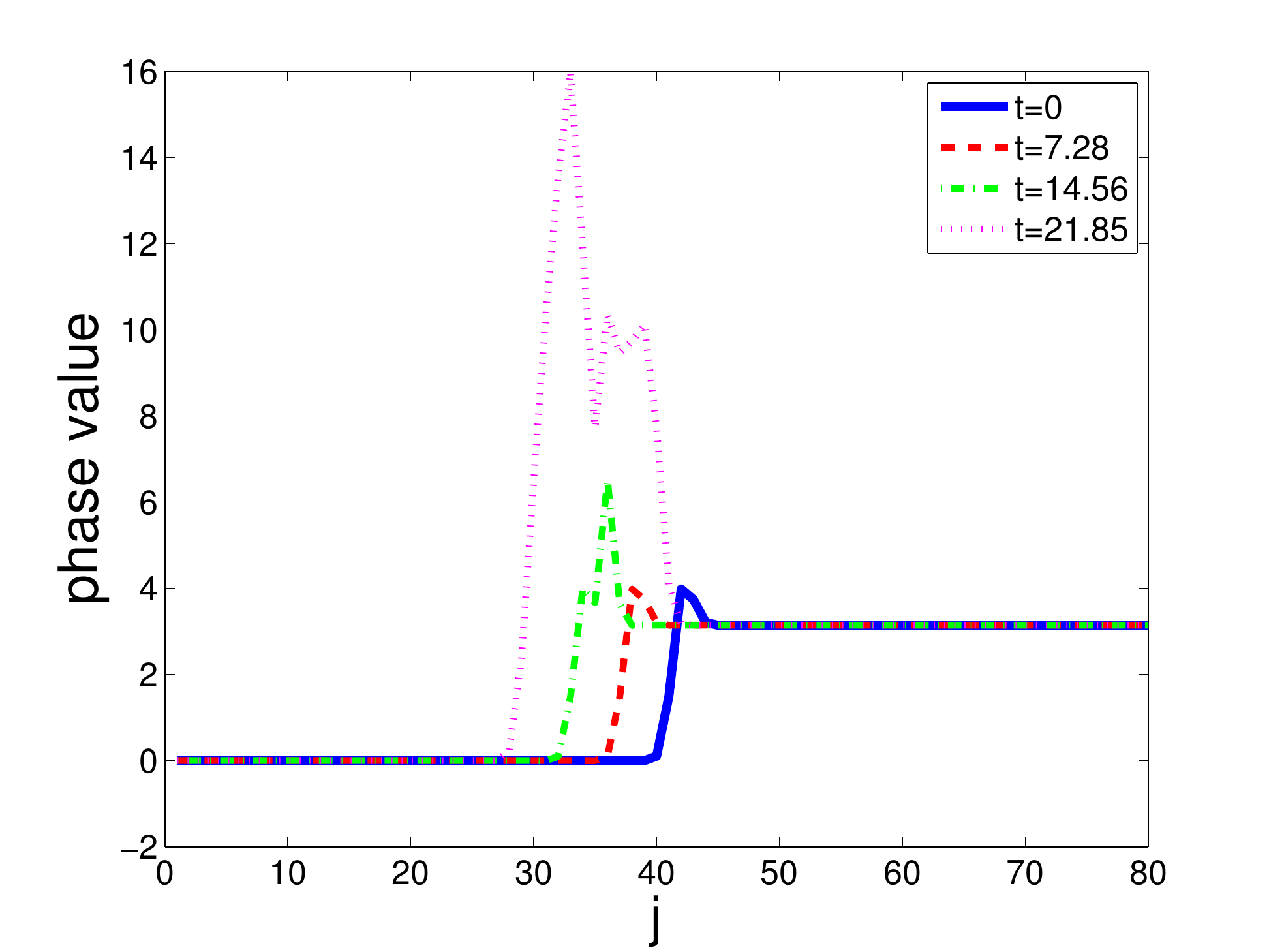}}}
{\subfloat[]{\includegraphics[width=0.45\textwidth]{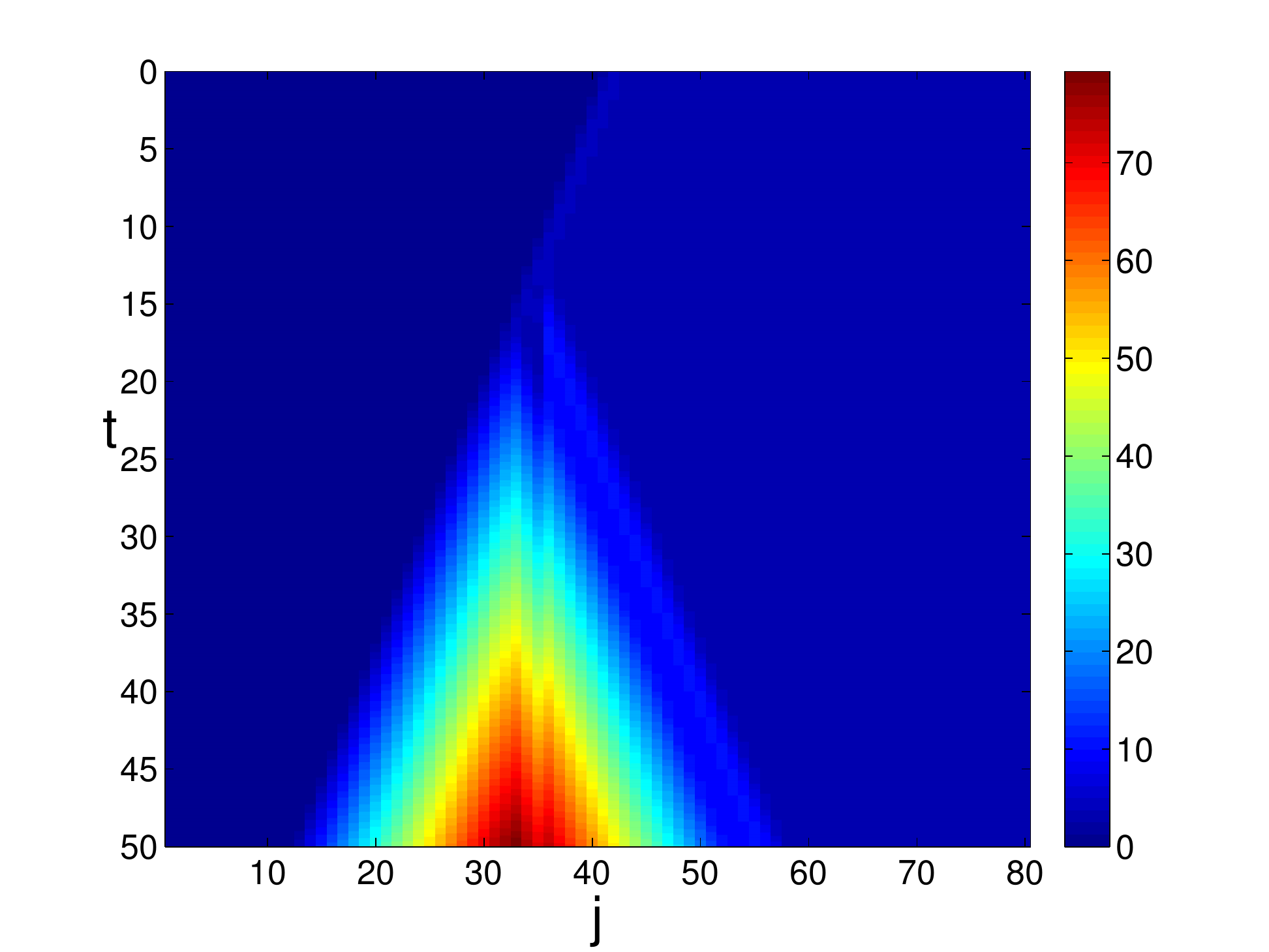}}}
\end{center}
\caption{Unstable traveling wave at $(\mu,k)=(2\pi-1.8, 0.75)$ with $c=-0.5493$: (a) Snapshots of the solution
of Eq.~\eqref{ee1} initialized at the traveling wave at times $t=0$ (solid blue curve), $7.28$ (dashed red), $14.56$ (approximate onset of instability, dash-dotted green) and $21.85$ (dotted magenta).
(b) The contour plot of space-time evolution of
the solution (with $\theta \in \mathbb{R}$) until time $t=50$. The traveling wave becomes unstable after an initial propagation transient period and splits into two fronts propagating in opposite directions with the same speed as the initial wave.}
\label{fig12}
\end{figure}

Fig.~\ref{fig11}a shows
time snapshots of the solution of the ODE system \eqref{ee1} initialized by the traveling wave (solid blue curve) obtained from Eq.~\eqref{tw11}
at  $(\mu, k)$= $(1.8, 0.75)$, which lies in the region $II$ below the background instability curve.
The space-time contour plot of the solution is shown in Fig.~\ref{fig11}b.
The traveling wave is initially traveling to the
right at the positive predicted velocity $c=0.5493$. However, a frontal instability of the traveling wave
leads to the formation of two fronts near $t=7.28$ that
can be seen at $t=14.56$ (dashed-dotted green curve) and $t=21.85$ (dotted magenta curve) in Fig.~\ref{fig11}a,
as well as in the contour plot shown in Fig.~\ref{fig11}b. Interestingly, the two fronts propagate in the opposite directions with the same speed as the initial unstable traveling wave.
%
A similar instability is observed at $(\mu, k)=(2 \pi-1.8, 0.75)$ in region $III$ in Fig.~\ref{fig12} .
However, in this case the predicted velocity of the traveling wave is found to be negative, $c=-0.5493$. As explained below, this pair of solutions also exhibits the symmetry mentioned above for the solutions shown in Fig~\ref{fig11a}a and \ref{fig12a}a.
In contrast to those solutions, which featured background instability, no instability of the background
is observed in these two examples, and only the frontal instability arises.
At $k=0.75$, the frontal instability exists at least for
$1.65 < \mu < q$, where $q$ satisfies $0.75=0.5 | \sec(q) |$; that is,
until $\mu$ reaches the background instability region.
No instability is observed for traveling waves at $\mu \le 1.65$ and $k=0.75$ propagated until time $t=500$.
The results of our stability investigations are summarized in Table I. We emphasize that curves in the $(\mu,k)$-plane separating stable and unstable traveling waves sampled above \emph{do not} coincide with the vertical lines $\mu=\pi/2$ and $\mu=3\pi/2$ bounding regions II and III where unstable waves were found. In particular, our numerical results indicate that the curve separating stable traveling waves at small $\mu$ from the unstable ones in region II is located in region II slightly to the right of the vertical line $\mu=\pi/2$. By the symmetry described below we anticipate a similar stability boundary in region III slightly to the left of the line $\mu=3\pi/2$.
%

\begin{figure}[!htb]
\begin{center}
{\subfloat[]{\includegraphics[width=0.45\textwidth]{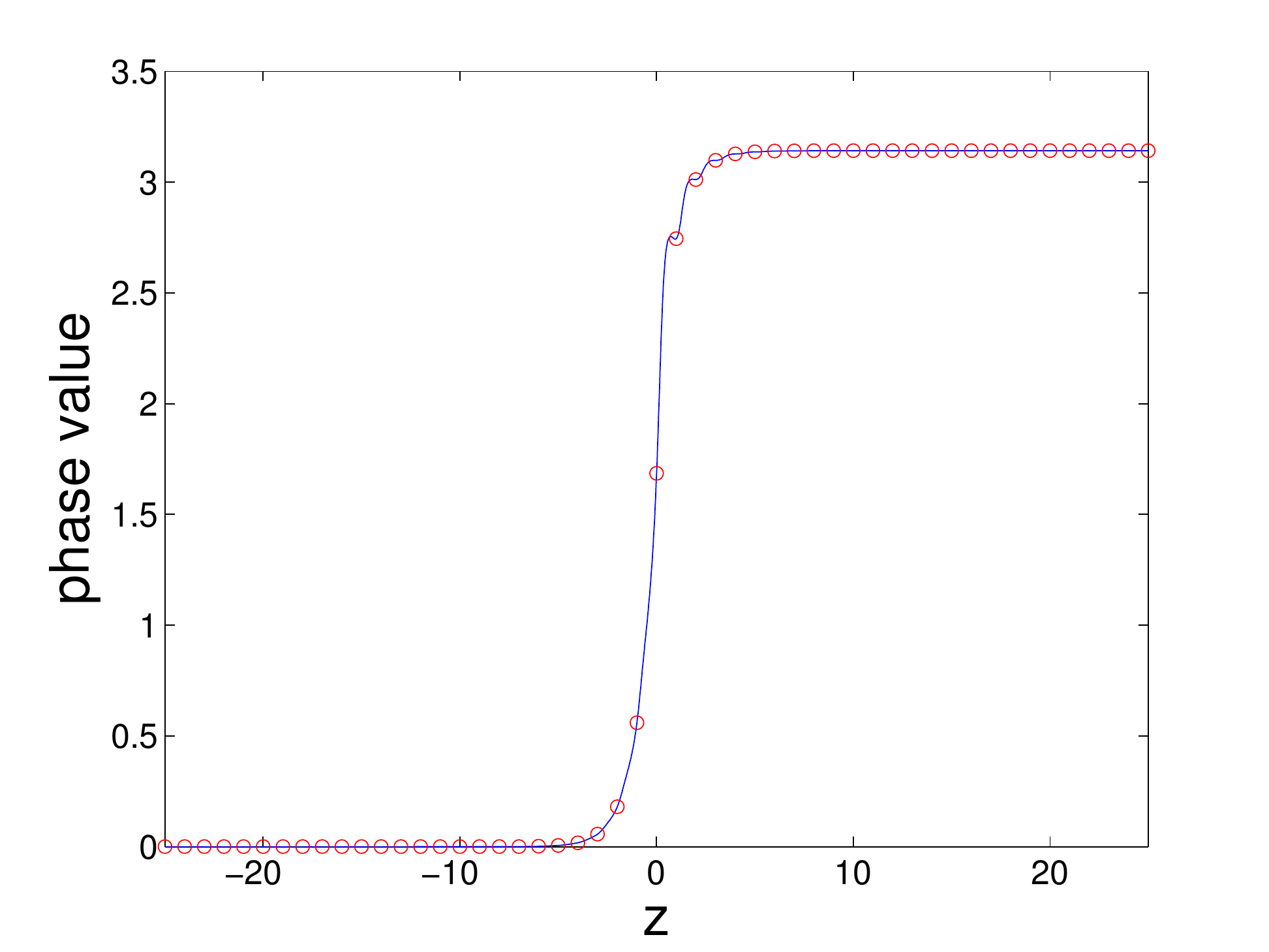}}}
{\subfloat[]{\includegraphics[width=0.45\textwidth]{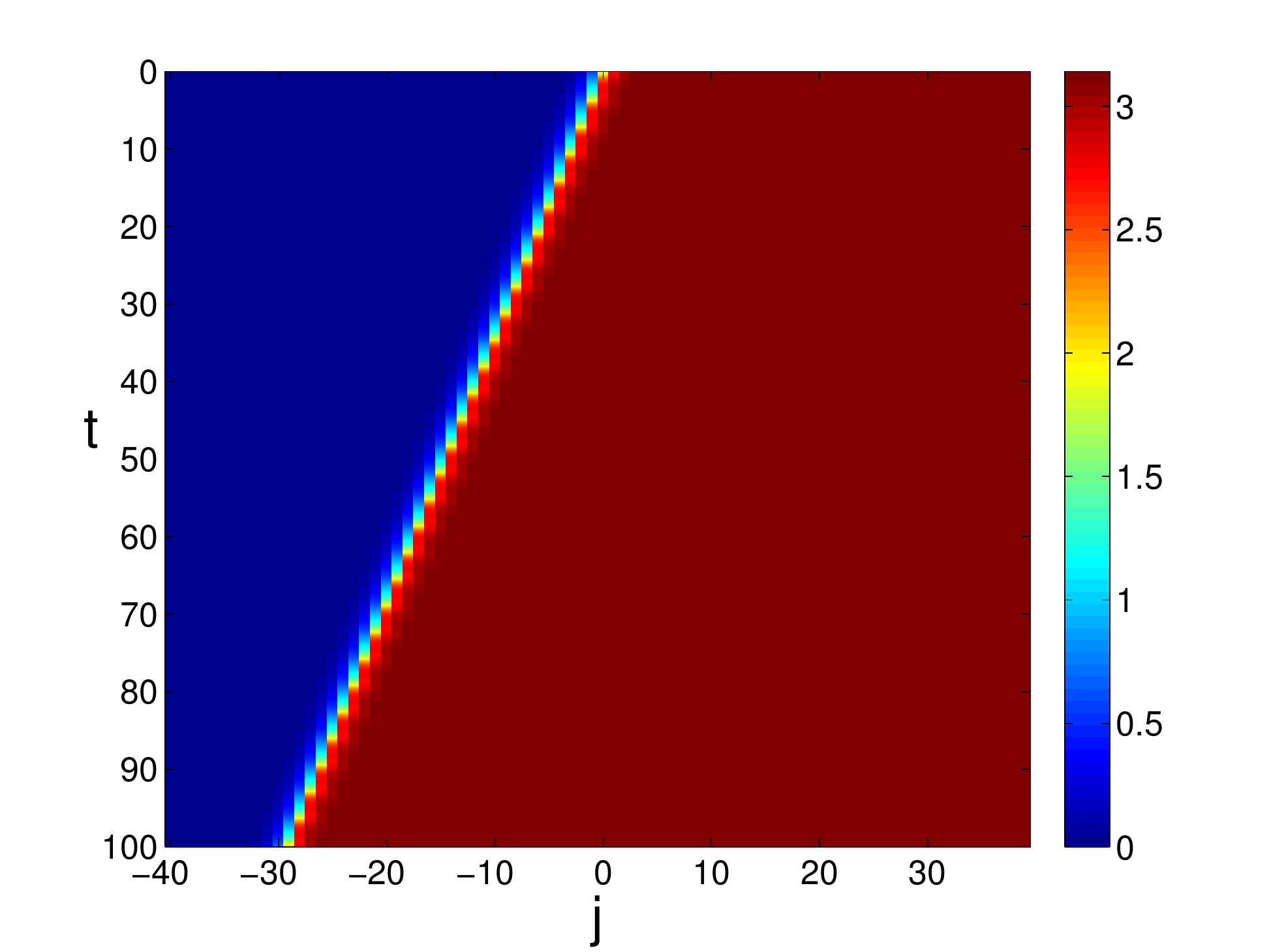}}}
\end{center}
\caption{Stable traveling wave at $(\mu,k)=(6, 1.6)$ with $c=-0.2919$:
(a) The traveling wave $\phi(z)$ (solid line) and the initial condition $\theta_j(0)=\phi(j)$ (circles) for the simulation of the ODE system \eqref{ee1}.
(b) The contour plot of the space-time evolution of the solution of ODE system \eqref{ee1} until time $t=100$. In accordance with the predicted
negative velocity, the traveling wave propagates to the left. }
\label{fig9}
\end{figure}

\begin{figure}[!htb]
\begin{center}
{\subfloat[]{\includegraphics[width=0.45\textwidth]{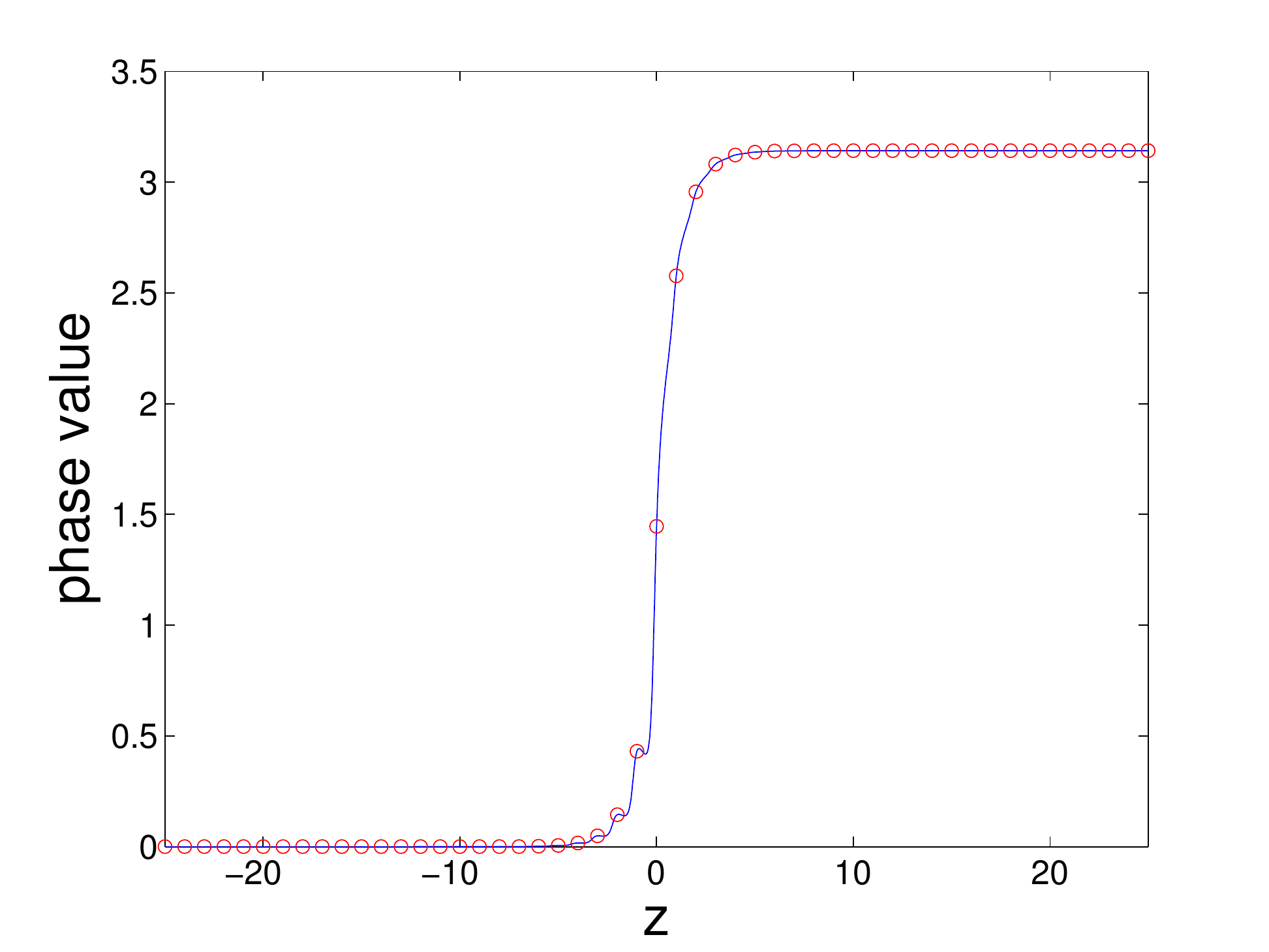}}}
{\subfloat[]{\includegraphics[width=0.45\textwidth]{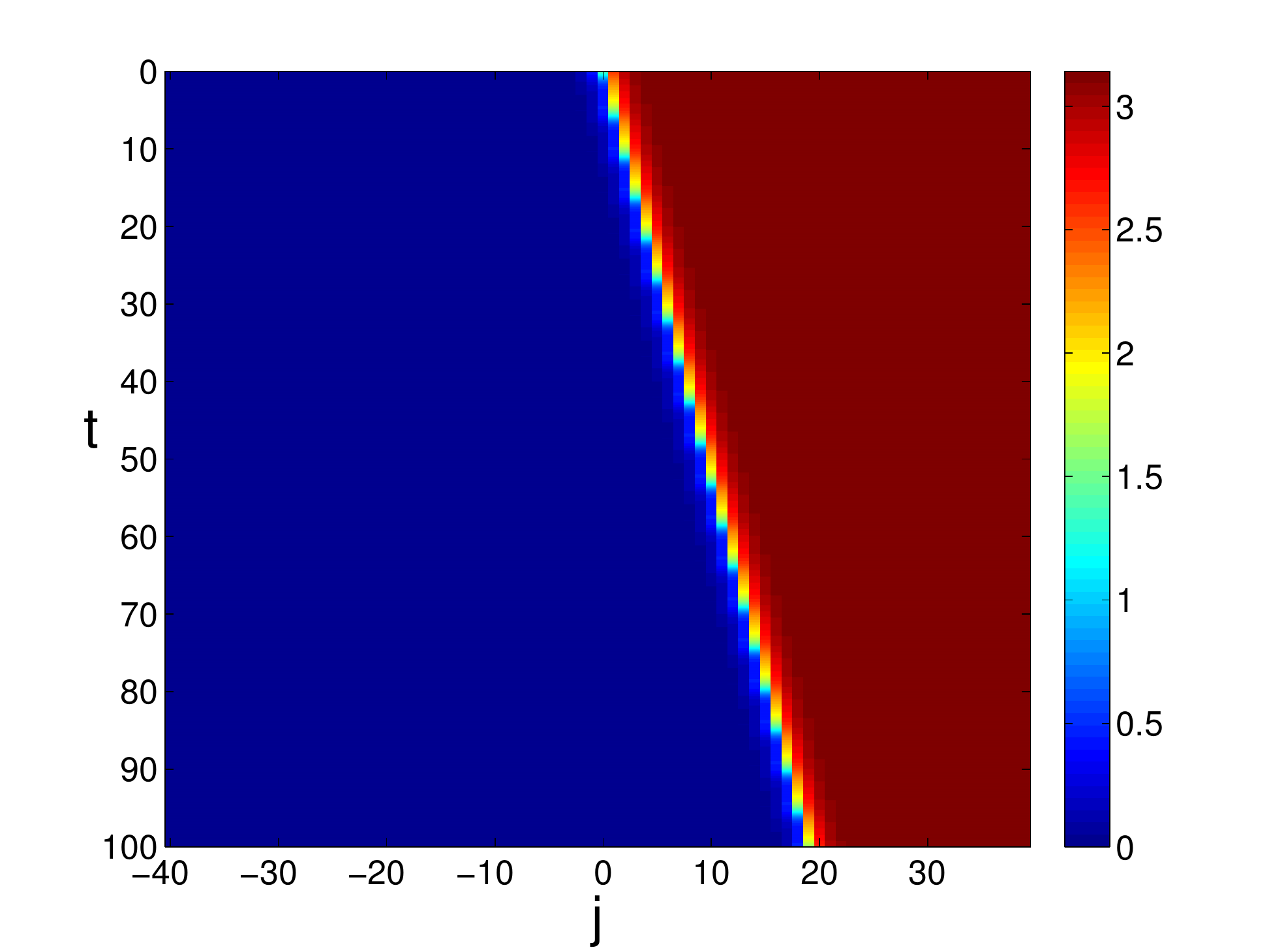}}}

\end{center}
\caption{Stable traveling wave at $(\mu,k)=(6.5, 1.6)$ with $c=0.1894$:
(a) The traveling wave $\phi(z)$ (solid line) and the initial condition $\theta_j(0)=\phi(j)$ (circles) for the simulation of the ODE system \eqref{ee1}.
(b) The contour plot of the space-time evolution of ODE system \eqref{ee1} until time $t=100$. In accordance with the predicted
positive velocity, the traveling wave now propagates to the right.}
\label{fig8}
\end{figure}

Recall that the background is stable in the regions $IV$ and $V$. To explore the traveling wave stability in these regions, we
solve Eq.~\eqref{tw11} at points $(6,1.6)$ and $(6.5,1.6)$ in regions $IV$ and $V$, respectively.
The fronts and their space-time evolution are shown in Fig.~\ref{fig9} for $(\mu,k)=(6,1.6)$ and Fig.~\ref{fig8} for $(\mu,k)=(6.5,1.6)$.
At the point $(6,1.6)$ in region $IV$, the wave is stable and travels to the left, in accordance with the negative velocity $c=-0.2919$
predicted by the traveling wave equation; see Fig.~\ref{fig9}.
At the point $(6.5,1.6)$ in region $V$, the wave is also stable and travels to the right with the positive velocity $c=0.1894$,
as shown in Fig.~\ref{fig8}.
%

%
\begin{table}[htdp]
\caption{Stability and velocities of traveling waves} 
\centering 
\begin{tabular}{c c c c } 
\hline\hline 
point&region&stability&velocity\\ [0.5ex] 
\hline
(2.7,1)&$II$&unstable (background)&0.2233\\
($2\pi-2.7$,1)&$III$&unstable (background)&-0.2233\\
(1.8,0.75)&$II$&unstable (frontal)&0.5493\\
($2\pi-1.8$, 0.75)&$III$&unstable (frontal)&-0.5493\\
(6,1.6)&$IV$&stable&-0.2919\\
(6.5,1.6)&$V$&stable&0.1894\\
\hline
\end{tabular}
\label{t1}
\end{table}%

To extend these results to the whole upper half of the parameter plane, we observe that the right hand side of the traveling wave Eq.~\eqref{tw11} is
$2 \pi$-periodic in $\mu$, implying that its solutions are also $2 \pi$-periodic in $\mu$.
For example, the traveling wave solutions in region $I$ are isomorphic to the ones in
region $V$ ($II$ with $VI$ etc.). Note also that the traveling
(and standing) wave solutions are symmetric about $\mu=\pi$.
To show this, we define new variables
$\bar{c}=-c$, $\bar{\mu}=2\pi-\mu$, $\bar{z}=-{z}$, $\bar{\phi}(\bar{z})=\pi-\phi(z)$. This yields $\phi'(z)=\bar{\phi}'(\bar{z})$,
$\phi(z+1)=\pi-\bar{\phi}(\bar{z}-1)$ and $\phi(z-1)=\pi-\bar{\phi}(\bar{z}+1)$. Substituting this in Eq.~\eqref{tw11} for our choice of $H(\theta)$ and $f(\theta)$ we obtain,
after simplification,
\[
-\bar{c}\bar{\phi}'(\bar{z})=
k(\sin(\bar{\phi}(\bar{z}+1)-\bar{\phi}(\bar{z})+\bar{\mu})-\sin(\bar{\mu})+
\sin(\bar{\phi}(\bar{z}-1)-\bar{\phi}(\bar{z})+\bar{\mu})-\sin(\bar{\mu}))-\sin(2\bar{\phi}(\bar{z})),
\]
which in fact coincides with Eq.~\eqref{tw11}.
Hence, $\bar{\phi}(\bar{z})$ is a
solution to the traveling wave equation at $\bar{\mu}=2\pi-\mu$ with velocity $\bar{c}=-c$.
In other words, for any $k>0$, given a traveling wave $\phi(z)$ with $\mu=\pi+b$ (for some $b$), there is a
traveling wave at $\mu=\pi-b$ obtained through the above transformations.
This traveling wave at $\mu=\pi-b$ has the same speed and
spectrum as the wave with $\mu=\pi+b$ but propagates in the direction opposite
to that of the original
traveling wave. This implies that the backward moving waves in regions $III$ and $IV$ are obtained
directly from the waves in regions $II$ and $I$, respectively, with the spectrum
(and hence stability properties)
preserved.  The observed symmetry obviously exists not only at $\pi$ but at $(2m-1)\pi$
for any integer $m$. Note that in accordance with these results, the extended curve shown in Fig.~\ref{fig4} repeats periodically once $\mu$ reaches $2\pi$ and is symmetric about $\mu=\pi$ and $\mu=3\pi$.  Since the waves under study are $2\pi$-periodic, this analysis can be extended to
the entire upper half of the $(\mu,k)$ plane.  In particular, it explains the symmetry observed above for the solution pairs at $(2.7,1)$, $(2\pi-2.7,1)$ and at $(1.8, 0.75)$, $(2\pi-1.8,0.75)$.

In summary, we have obtained an analytical expression for the boundary of a region in parameter space where traveling waves suffer a background instability.  Numerical simulations show that as parameters are varied within from the zero-speed boundary of the traveling wave region, traveling waves can be stable or can undergo either  a background instability (within the region predicted analytically) or a frontal instability (between the stable region and the background instability region).  Due to a symmetry in the model corresponding to the periodicity of $H$ and $f$, all results repeat periodically in $\mu$.

\subsection{Two-dimensional phase equation dynamics}
We now turn to  some explorations of the two-dimensional generalization
of our oscillator problem.
The extension of Eq.~\eqref{ee1} to two dimensions has the form
\begin{equation}
\begin{split}
&\dot{\theta}_{1,1} = k (H(\theta_{1,2}-\theta_{1,1})+H(\theta_{2,1}-\theta_{1,1}))+f(\theta_{1,1}) \\
&\dot{\theta}_{i,1} = k (H(\theta_{i,2}-\theta_{i,1})+H(\theta_{i+1,1}-\theta_{i,1})+H(\theta_{i-1,1}-\theta_{i,1}))+f(\theta_{i,1}), \quad i=2,\dots,2n-1,\\
&\dot{\theta}_{1,j} = k (H(\theta_{1,j+1}-\theta_{1,j})+H(\theta_{1,j-1}-\theta_{1,j})+H(\theta_{2,j}-\theta_{1,j}))+f(\theta_{1,j}), \quad j=2,\dots,2n-1, \\
&\dot{\theta}_{i,j}  = k (H(\theta_{i,j+1}-\theta_{i,j})+H(\theta_{i,j-1}-\theta_{i,j})+
H(\theta_{i+1,j}-\theta_{i,j})+H(\theta_{i-1,j}-\theta_{i,j}))+f(\theta_{i,j}), \\
& \qquad i,j=2,\dots,2n-1,\\
&\dot{\theta}_{i,2n} = k (H(\theta_{i,2n-1}-\theta_{i,2n})+H(\theta_{i+1,2n}-\theta_{i,2n})+H(\theta_{i-1,2n}-\theta_{i,2n}))+f(\theta_{i,2n}),\\
& \qquad i=2,\dots,2n-1,\\
&\dot{\theta}_{2n,j} = k (H(\theta_{2n,j+1}-\theta_{2n,j})+H(\theta_{2n,j-1}-\theta_{2n,j})+H(\theta_{2n-1,j}-\theta_{2n,j}))+f(\theta_{2n,j}),\\
& \qquad j=2,\dots,2n-1,\\
&\dot{\theta}_{2n,2n} = k (H(\theta_{2n,2n-1}-\theta_{2n,2n})+H(\theta_{2n-1,2n}-\theta_{2n,2n}))+f(\theta_{2n,2n}).
\end{split}
\label{td1}
\end{equation}

\begin{figure}[tbp]
\begin{center}
{\includegraphics[width=0.8\textwidth]{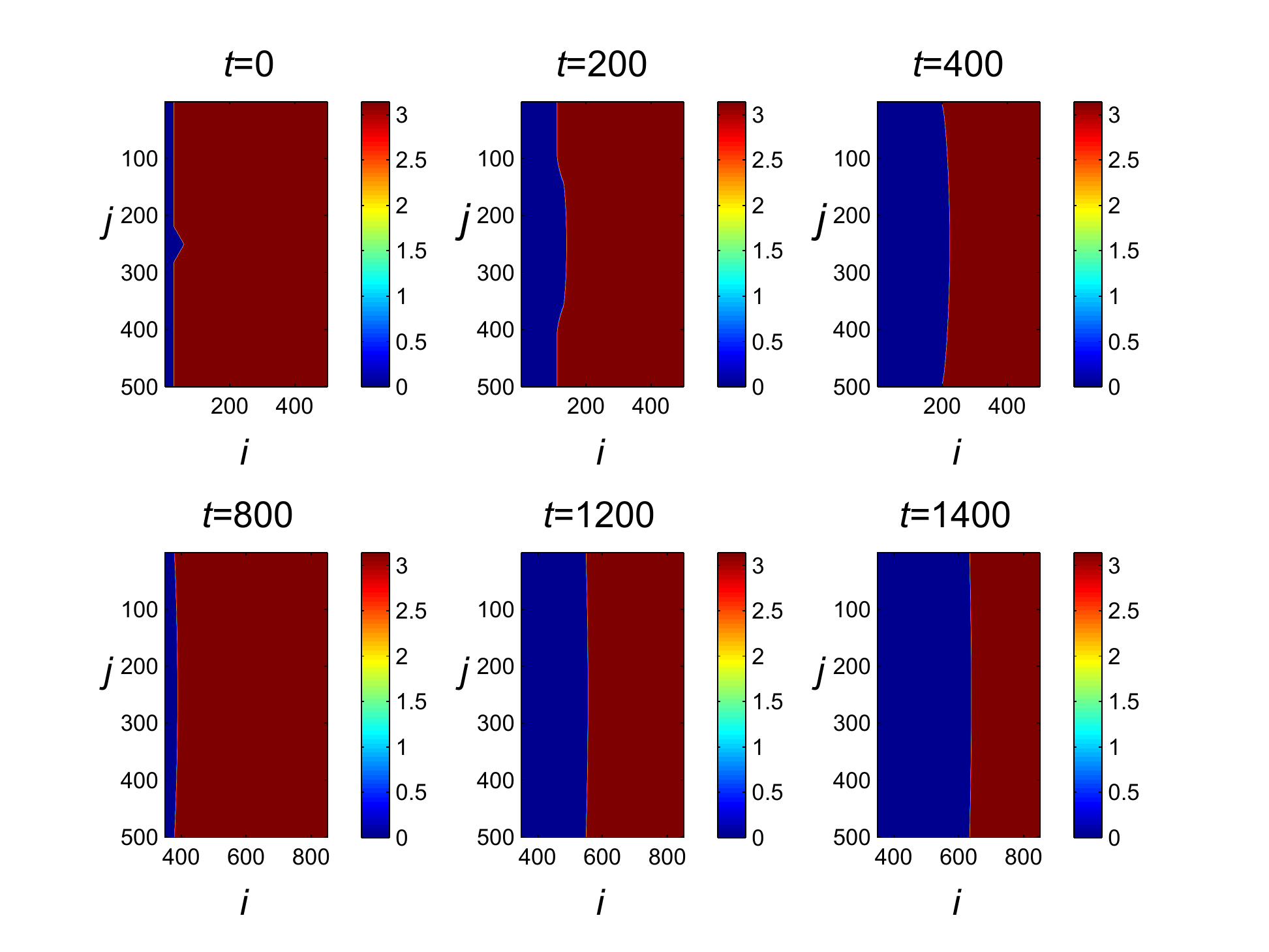}}     \quad
\end{center}
\caption{Snapshots of the evolution of a perturbed planar front introduced within a two-dimensional domain with $k=1.3$, $\mu=0.5$ at times $t=10$, $200$, $400$, $800$, $1200$ and $1400$.
The evolution shows the healing of the front and the decay of the associated perturbation that
restore the dynamically robust planar front traveling with velocity $c=0.4155$.}
\label{f4}
\end{figure}

We start by solving Eq.~\eqref{tw11} on the interval $[-25,25]$ at the parameter values
$k=1.3$ and $\mu=0.5$ to
obtain a one-dimensional traveling wave $\phi(z)$ with velocity $c=0.4155$. We then solve Eq.~\eqref{td1} for $\theta_{i,j}(t)$
using the classical fourth order Runge-Kutta method with the initial data
\[
\theta_{i,j}(0)=\begin{cases}
                 0, & i \leq 3\\
                 \phi(i-3), & 4 \leq i \leq 54\\
                 \pi, & i \geq 55,
\end{cases}
\]
thus studying the two-dimensional evolution of an initially planar front.
The planar front propagates in the horizontal direction at the velocity of the one-dimensional wave, suggesting that the solution $\theta_{i,j}(t)=\phi(i-ct)$,
which corresponds to a stable traveling wave in the one-dimensional problem, is also stable in the two-dimensional setting. To illustrate this stability,
we distort a segment of the planar front in the initial condition, as shown in the first panel ($t=0$) in Fig.~\ref{f4}.
The resulting evolution is shown in the remaining panels of Fig.~\ref{f4}, where for better visualization, the range of the $i$-axis is shifted to the left in the last three panels.
It can be seen that over time the system ``heals'' the perturbation
and gradually restores its quasi-one-dimensional planar front character,
while the solution eventually settles into traveling with
the velocity predicted by the one-dimensional results.

\begin{figure}[tbp]
\begin{center}
{\includegraphics[width=0.8\textwidth]{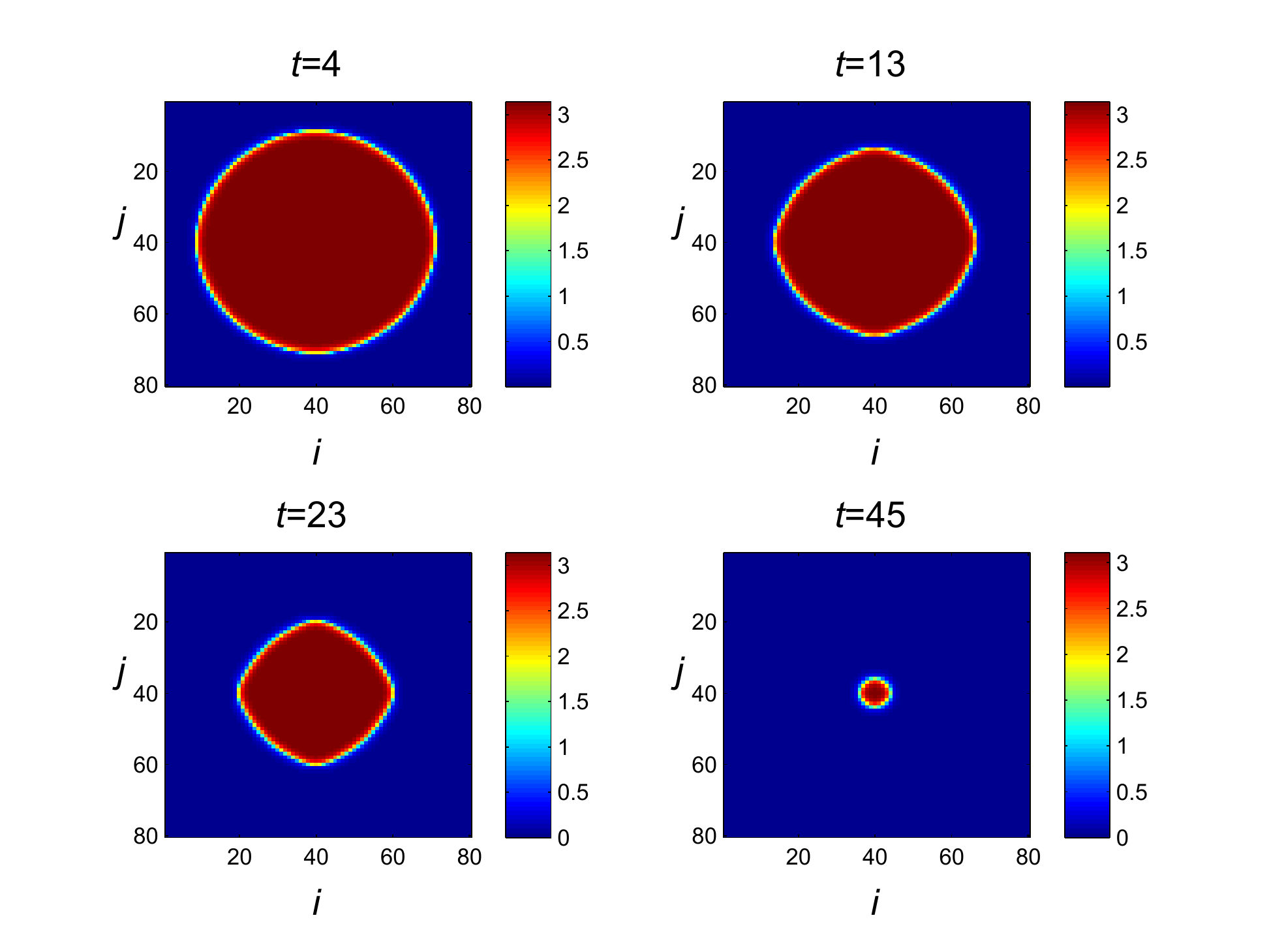}}
{\includegraphics[scale =0.4] {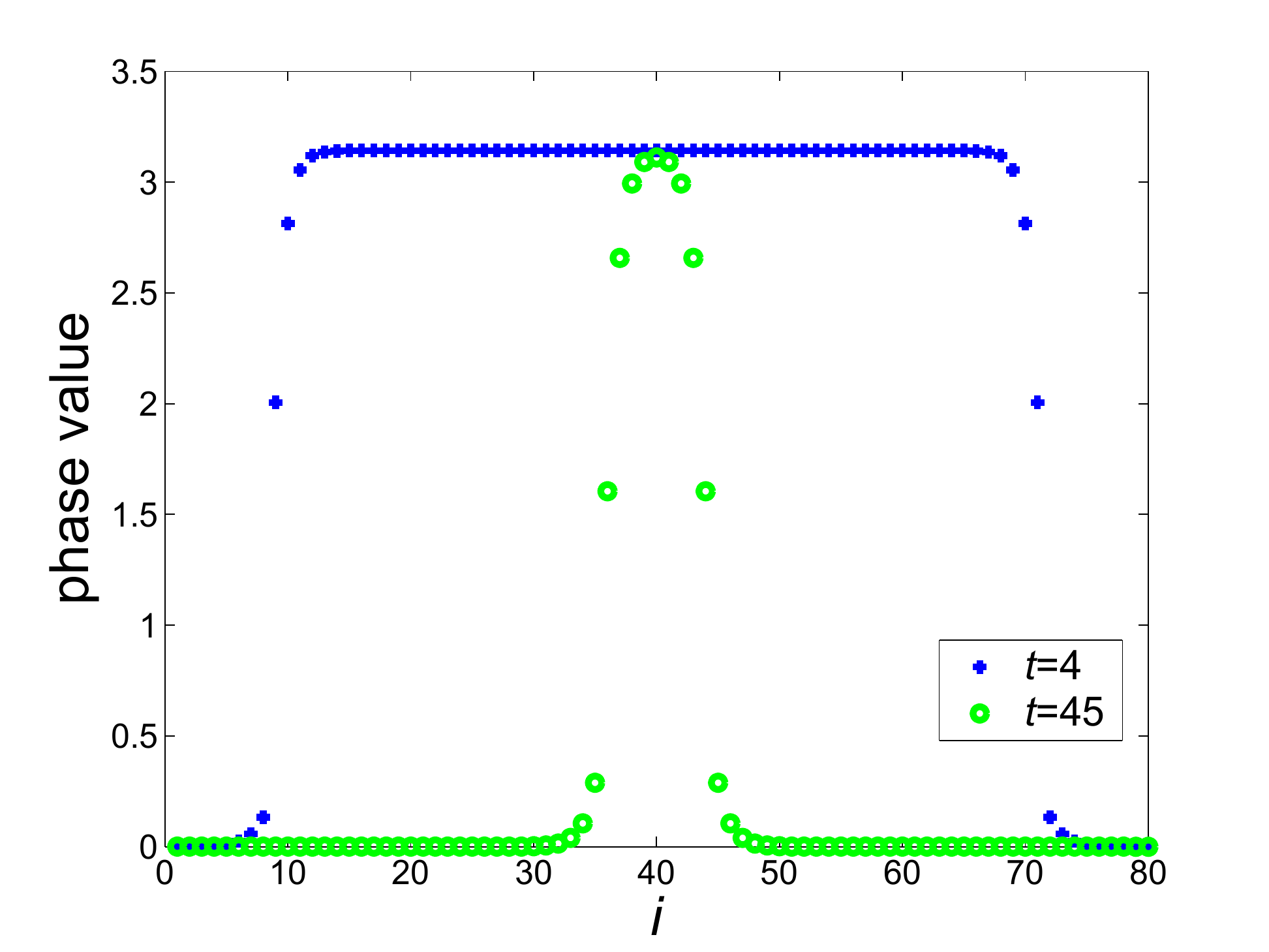}}
\end{center}
\caption{Snapshots of the evolution of a circular front with $k=1.3$, $\mu=0.5$ at times $t=4$, $13$, $23$ and $45$. The
front shrinks (and is eventually annihilated)
as time evolves. In the bottom panel, a horizontal section($j=40$)
of the front at $t=4$ is denoted by plus signs and a horizontal section of the front at $t=45$ (of smaller width) is
denoted by circles.}
\label{f400}
\end{figure}

Next we consider the evolution of a radial front.
In the first four panels of
Fig.~\ref{f400}, we show the evolution of an initially circular front for $k=1.3$ and $\mu=0.5$.
The initial condition is set to $\theta_{i,j}(0)=\pi$ for $(i,j)$ within the circle of radius $30$ centered at $(40,40)$, and $\theta_{i,j}(0)=0$ outside of the concentric circle of radius $36$.  The initial value of $\theta_{i,j}$ for $(i,j)$ between these circles is obtained by linear interpolation.
The front shrinks and is eventually annihilated (i.e., disappears).
The bottom panel of  Fig.~\ref{f400} displays horizontal slices of the solution at the initial and final time steps shown above.  Notice that in this case, apparently, the initial radial profile
of the front is gradually deformed to conform more suitably to
the square symmetry of the underlying lattice grid. Hence it appears
that linear (planar) fronts are fairly robust in this system, while
radial ones are clearly not as robust and eventually disappear.

\begin{figure}[!htb]
\begin{center}
{\includegraphics[width=0.8\textwidth]{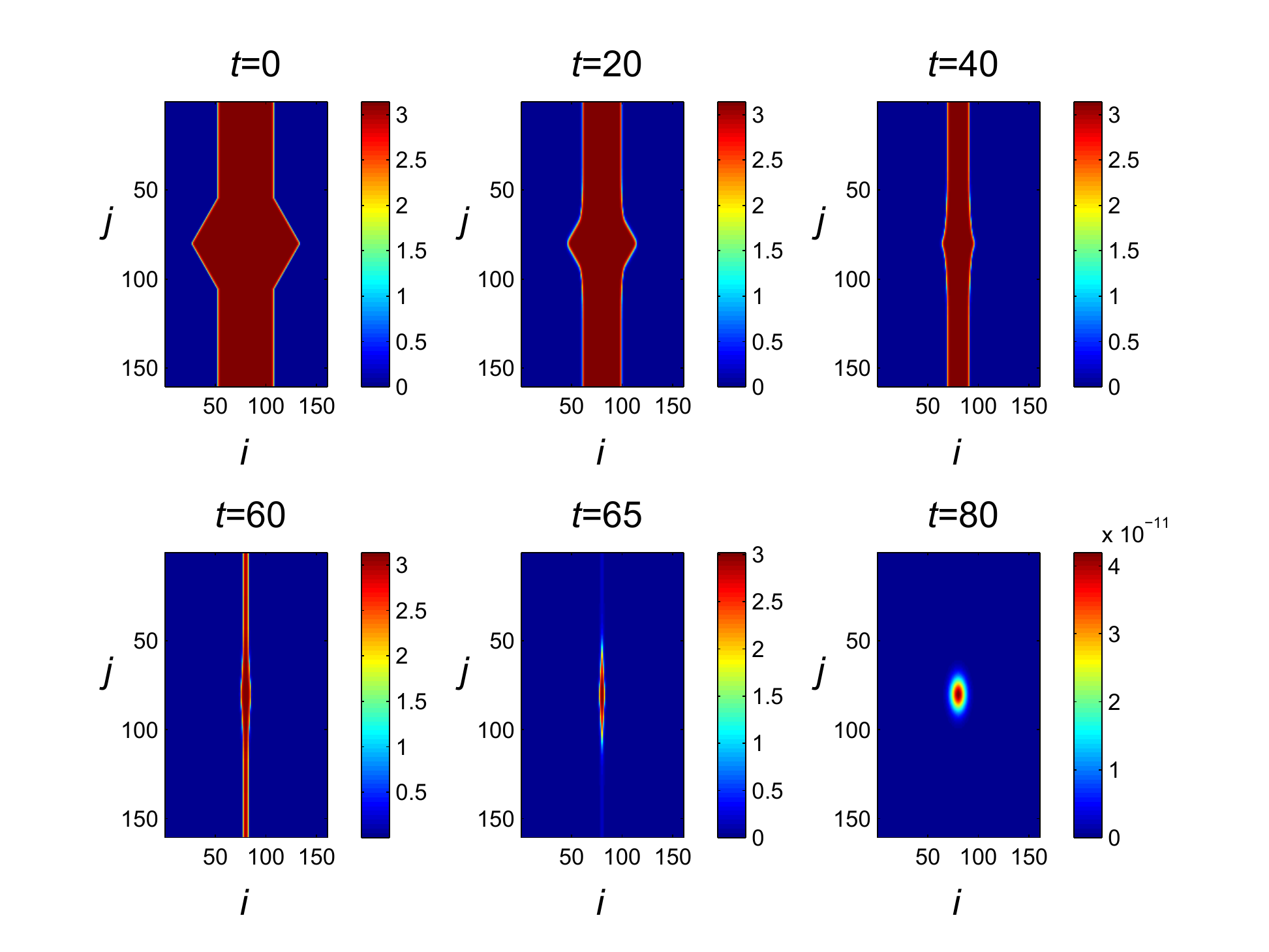}}
\end{center}
\caption{Snapshots of two  outwardly perturbed fronts at times $t=0$, $20$, $40$, $60$, $65$ and $80$ when  $k=1.3$, $\mu=0.5$.  The fronts meet near the domain edges, form a loop, and shrink.}
\label{2d_cir_3}
\end{figure}

\begin{figure}[!htb]
\begin{center}
{\includegraphics[width=0.8\textwidth]{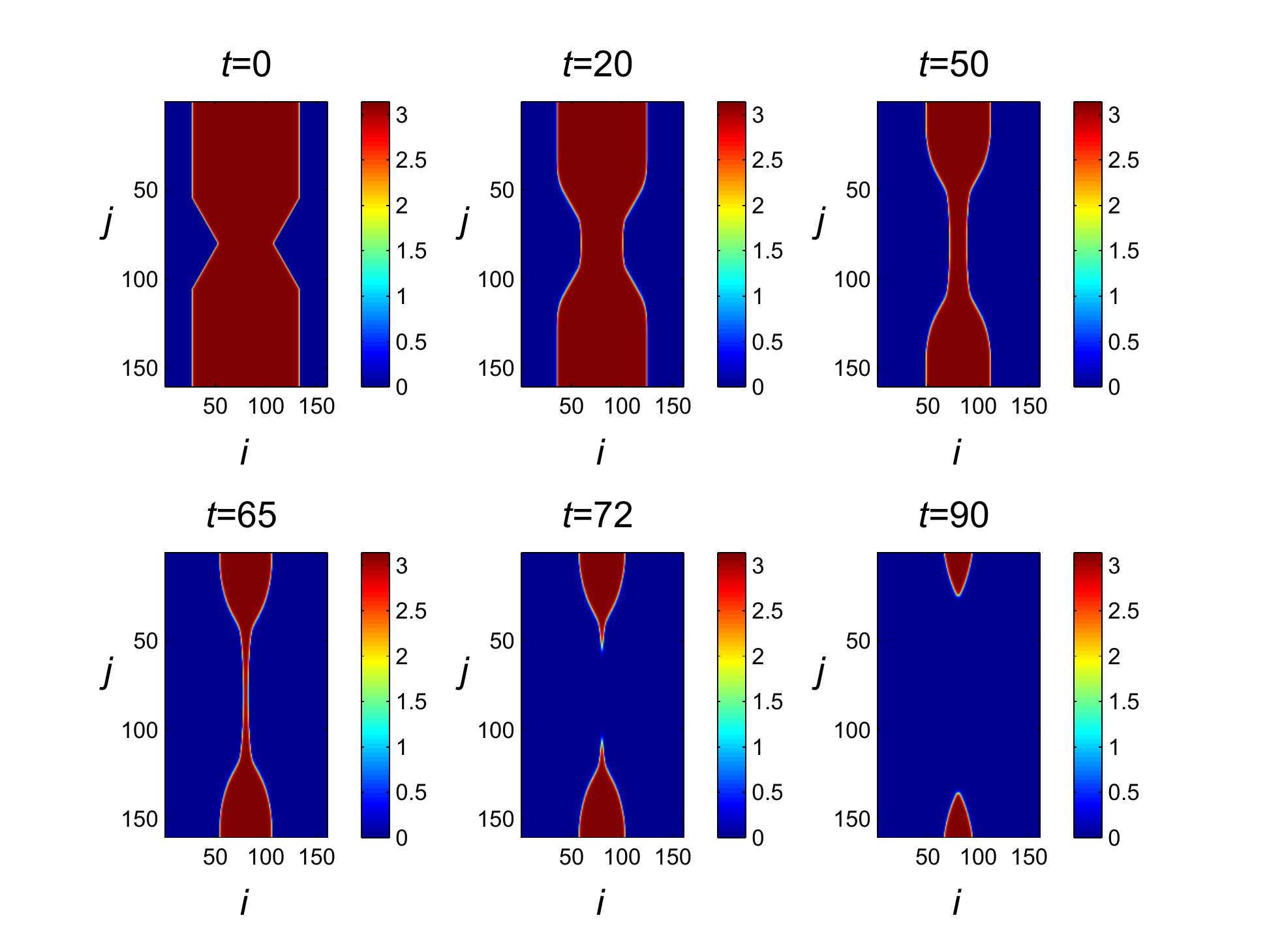}}
\end{center}
\caption{Snapshots of two inwardly perturbed fronts at times $t=0$, $20$, $50$, $65$, $72$ and $90$ when $k=1.3$, $\mu=0.5$.
The fronts meet near their centers,  separate, and evolve into two (upper and lower) parts.}
\label{2d_cir_4}
\end{figure}

\begin{figure}[!htb]
\begin{center}
{\includegraphics[width=0.8\textwidth]{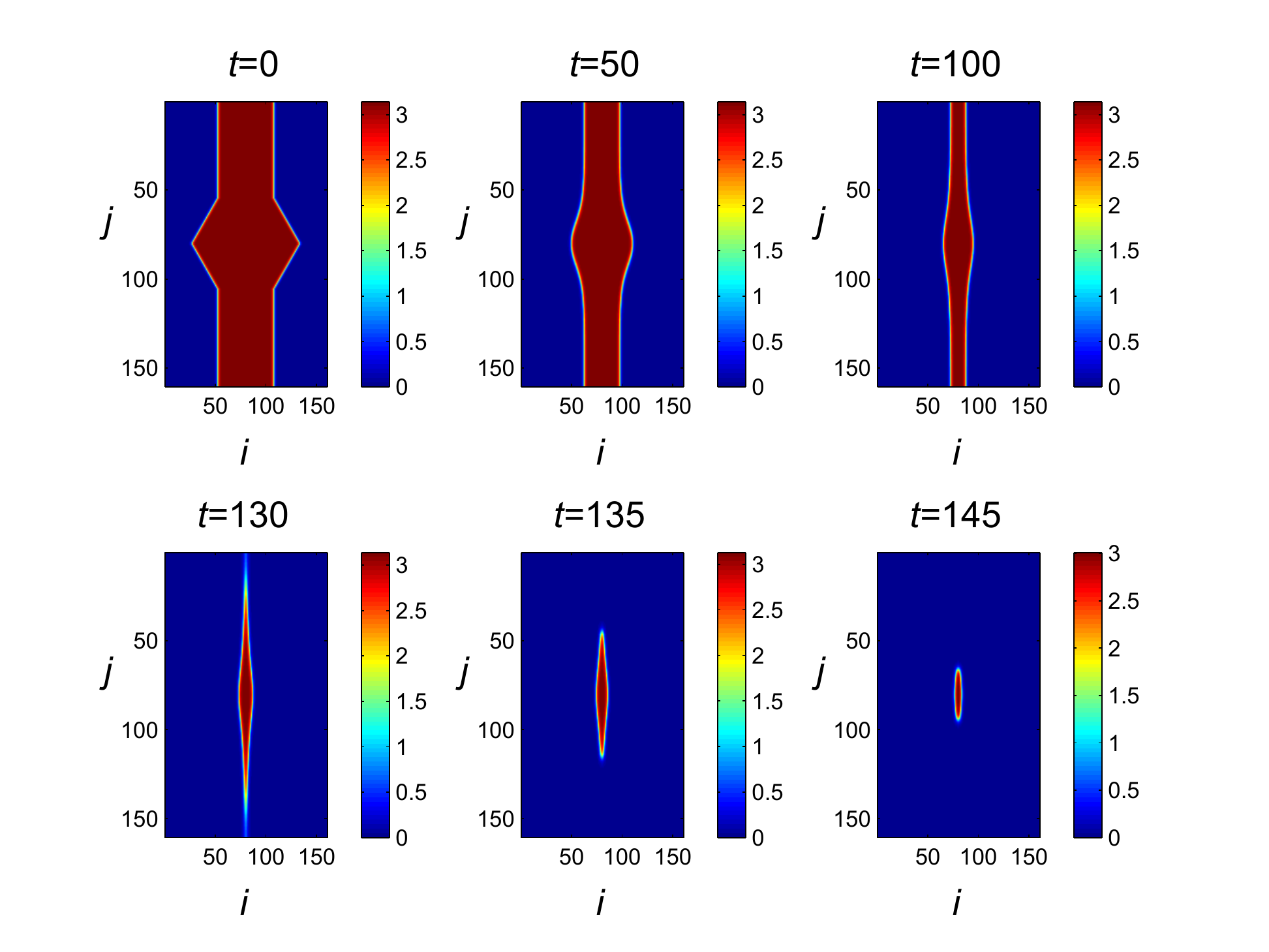}}
\end{center}
\caption{Snapshots of two  outwardly perturbed fronts at times $t=0$, $50$, $100$, $130$, $135$ and $145$ when $k=2$, $\mu=0.15$. As in Fig.~\ref{2d_cir_3}, the fronts meet near the domain edges, form a loop, and shrink.}
\label{2d_cir_15}
\end{figure}

\begin{figure}[!htb]
\begin{center}
{\includegraphics[width=0.8\textwidth]{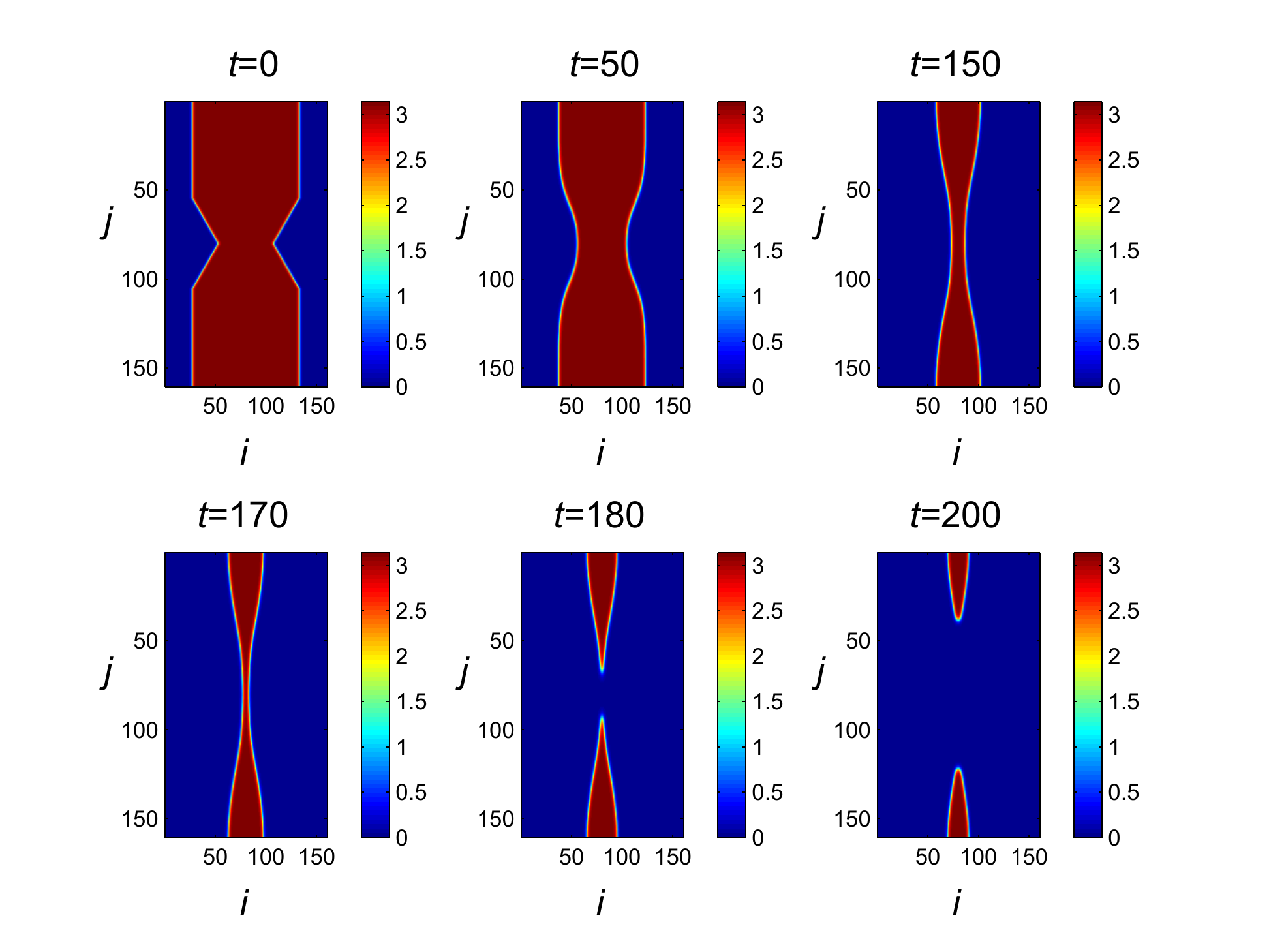}}
\end{center}
\caption{Snapshots of two  inwardly perturbed  fronts at times $t=0$, $50$, $150$, $170$, $180$ and $200$ when $k=2$, $\mu=0.15$. As in Fig.~\ref{2d_cir_4}, the fronts meet near their centers,  separate, and evolve into two (upper and lower) parts.}
\label{2d_cir_14}
\end{figure}

As noted in \cite{drover}, Carpenter \cite{carp} observed that experimentally induced phosphenes move according to the following rules in two dimensions:\\
\indent 1. Lines never cross through one another. Rather, they combine to form loops.  \\
\indent 2. A line never breaks apart unless it meets another line. \\
To test whether our two-dimensional model captures these features, we now consider simulations with two symmetric fronts that initially bulge either outward (Fig.~\ref{2d_cir_3} and Fig.~\ref{2d_cir_15}) or inward (Fig.~\ref{2d_cir_4} and Fig.~\ref{2d_cir_14}). These initial
conditions are shown in the first panel ($t=0$) of each figure.
The simulation results shown in Figs.~\ref{2d_cir_3}-\ref{2d_cir_14}, with $k=1.3$, $\mu=0.5$ in Fig.~\ref{2d_cir_3} and Fig.~\ref{2d_cir_4} and $k=2$, $\mu=0.15$ in Fig.~\ref{2d_cir_15} and Fig.~\ref{2d_cir_14}, are consistent with Carpenter's observations listed above. We can see that the outwardly bulging fronts eventually touch near the edges of the domain and form one loop. Meanwhile, the inwardly bulging fronts eventually  touch near their centers and the lines break into two parts. These features are very much in line
with the expectations of~\cite{drover} (compare with their Fig.~3).
However, it should also be mentioned that~\cite{drover} posit a
third and final feature, namely that: \\
\indent 3. Neighboring lines show a tendency to move in a similar manner. \\
Our simulations of the coupled oscillator model did {\it not}
reveal such a tendency. Whether the model can be improved to reflect this feature is a question that remains to be considered in future studies.

%
%
%

\section{Conclusions and Future Challenges}

In the present work we have revisited a generic nonlinear lattice model derived in \cite{parks} and
associated with the dynamics of  a forced network of coupled oscillators. A specific motivation for studying this system comes from its relevance to phosphenes, artificial perceptions of light arising in the visual system in which contours, possibly representing boundaries between sets of neurons in different activity states or phases, emerge and propagate.
We complemented the important initial steady state analysis of~\cite{parks} (see also~\cite{drover}) by exploring the possibility of traveling
waves in the system of ordinary differential equations. This led us to introduce the co-traveling frame (advance-delay) PDE and study
existence and stability properties of the traveling waves as both special periodic modulo shift solutions of the original systems of ODEs and stationary solutions of the co-traveling frame PDE.
Our results on traveling waves complement the work of \cite{parks} on the existence and stability of standing waves.
We found a curve above which the standing waves become unstable. This curve agrees with and extends the bifurcation curves obtained in \cite{parks} by analyzing the equilibrium states of the ODE system. We showed that the instability of standing waves above the curve leads to the emergence of stable traveling waves in some parameter regimes, while at other parameter values traveling waves exhibit either frontal or background instability. An analysis of the background steady states provided
information about the spectrum of these waves. From an applications perspective, a two-dimensional collection of oscillators, corresponding to cells in the retina or in a layer of visual cortex, is most relevant, and
for this reason we also considered some
prototypical 
examples of two-dimensional evolution. In particular, we demonstrated that
perturbed planar fronts can heal and resume their planar form, while radial fronts shrink. Our simulations of the system that initially has two fronts with bulging centers
are in qualitative agreement with Carpenter's experimental observations \cite{carp} (see also the discussion of~\cite{drover}).

The present work leads to numerous interesting questions for the further exploration of this and related systems.
In particular, obtaining an analytical handle on the spectrum of the front in the co-traveling wave PDE and connecting this
spectrum to the stability properties of the original ODEs would be
extremely valuable from a theoretical perspective, not only in the
context of the present setting but also for wide additional classes
of lattice dynamical problems bearing traveling waves, such as generalized Frenkel-Kontorova (see \cite{FK} and references therein),
Fermi-Pasta-Ulam (see e.g.~\cite{pegof,english,stekev}
and the discussion of Chapter 1 in~\cite{nester})
or nonlinear Schr{\"o}dinger systems (see e.g.~\cite{dep1,bar,champ,hadi}
and the discussion of
Chapter 21 in~\cite{dnls}).
On the numerical front, while we explored a few prototypical cases of two-dimensional evolution, a better understanding of the
stationary and traveling states in two dimensions clearly merits further investigation.
Another interesting direction is to formulate and study solutions and spectra of a two-component problem that
may support traveling pulses.  

On the applications side, once the basic properties of this model system are  more fully understood, it may serve as a form of computational test bed for exploring and making predictions about visual phenomena evoked by electrical stimulation.  In this context, it may be interesting to explore the influence of changes in the amplitude or qualitative form of the forcing function on wave propagation \cite{tehovnik2013} and to investigate waves induced by presenting traveling wave stimuli (e.g., \cite{folias, jalics}), representing objects passing through the visual field, in addition to periodic forcing.
Indeed, these adjustments may help resolve the model's current failure to capture the observation that neighboring fronts tend to propagate in a similar manner.
At a more fundamental level, the present class of models appears to
be a (modified) overdamped variant of the widely studied, so-called sine
lattices (see e.g.~\cite{takeno} and, for a discussion of some
of the relevant applications, \cite{feizhang}), hence it would
be particularly interesting to explore hybrid variants of these
models having as special case limits the overdamped and the undamped
cases previously explored.
Some of these issues are currently under investigation and
will be reported upon in  future publications.

\begin{acknowledgments}

P.G.K.~gratefully acknowledges the support of
the US-AFOSR under grant FA9550-12-1-0332,
and the ERC under FP7, Marie
Curie Actions, People, International Research Staff
Exchange Scheme (IRSES-605096).
P.G.K.'s work at Los Alamos is supported in part by the U.S. Department
of Energy.
JR was partially supported by the NSF award DMS-1312508.

\end{acknowledgments}

\appendix
\section*{Appendix: some details on numerical methods}
To solve the traveling wave Eq.~\eqref{tw11}, we used a second order forward difference scheme
and sometimes a second order centered difference scheme to
approximate the term $\phi'(z)$.  For a grid point $j$, these approximations are of the form
$(-3\theta_{j}+4\theta_{j+1}-\theta_{j+2})/(2 \Delta x)$
and $(\theta_{j+1}-\theta_{j-1})/(2 \Delta x)$, respectively, where $\Delta x$ is the grid spacing. In some cases,
the forward difference approximation used in solving Eq.~\eqref{tw11} did not converge but the centered difference approximation
did. Whichever approximation was used to solve Eq.~\eqref{tw11} for the traveling wave $\phi(z)$, it was checked that the
solution of the ODE system (\ref{ee1}) with the initial condition $\theta_i(0)=\phi(i)$ produced results that were consistent with the obtained traveling wave solution.

\begin{figure}
\begin{center}
{\subfloat[]{\includegraphics[width=0.4\textwidth]{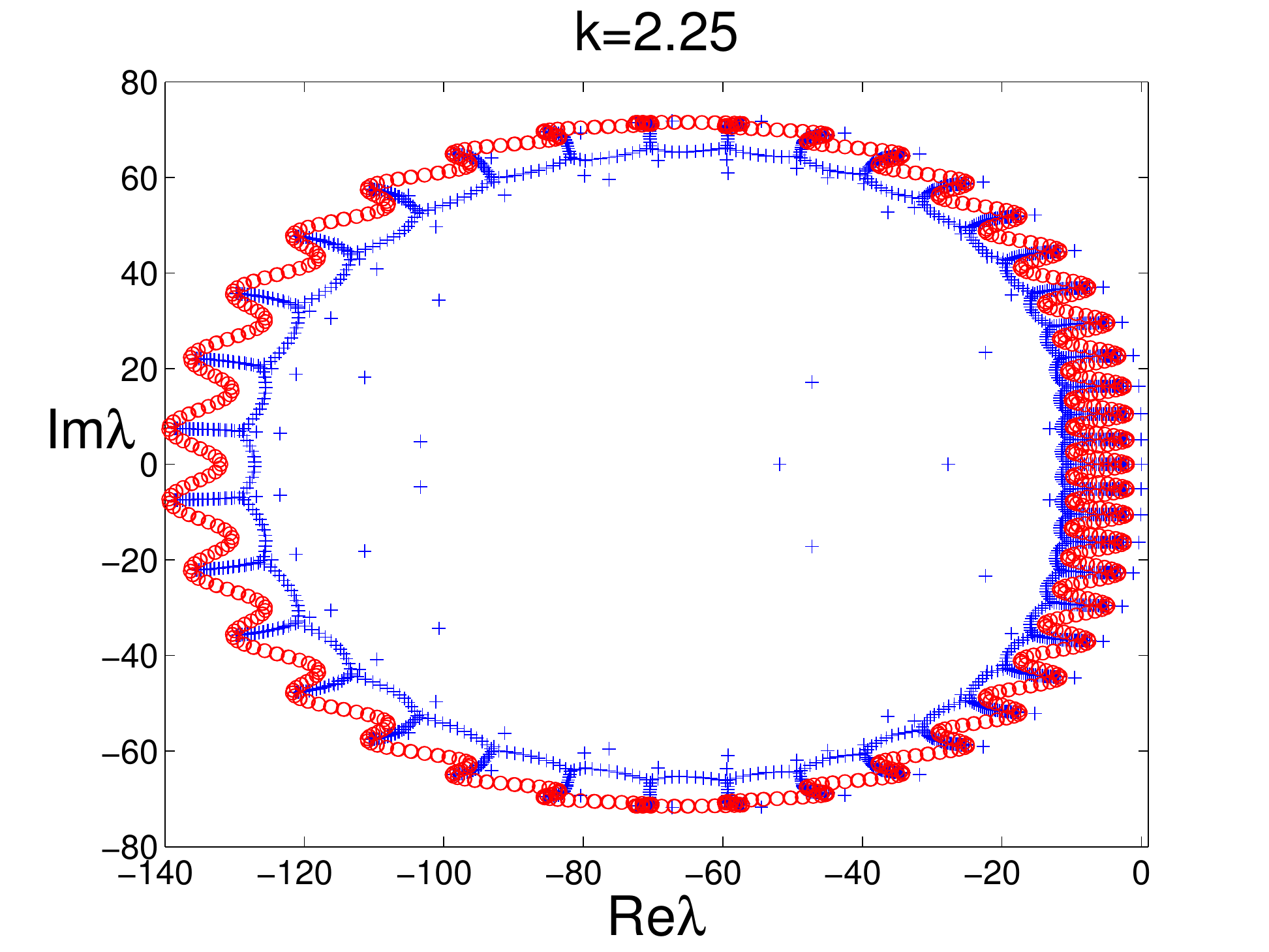}}}
{\subfloat[]{\includegraphics[width=0.4\textwidth]{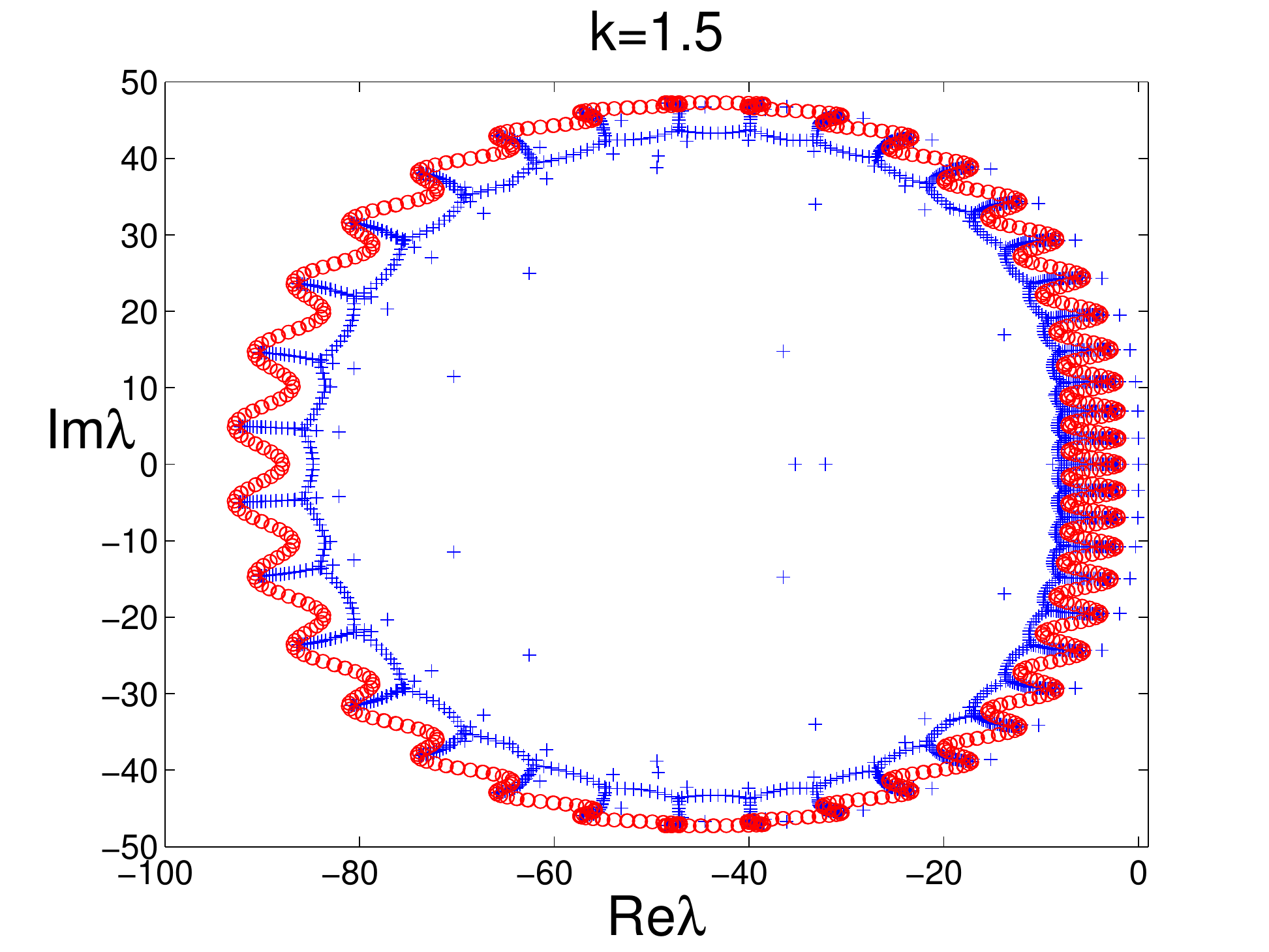}}}
{\subfloat[]{\includegraphics[width=0.4\textwidth]{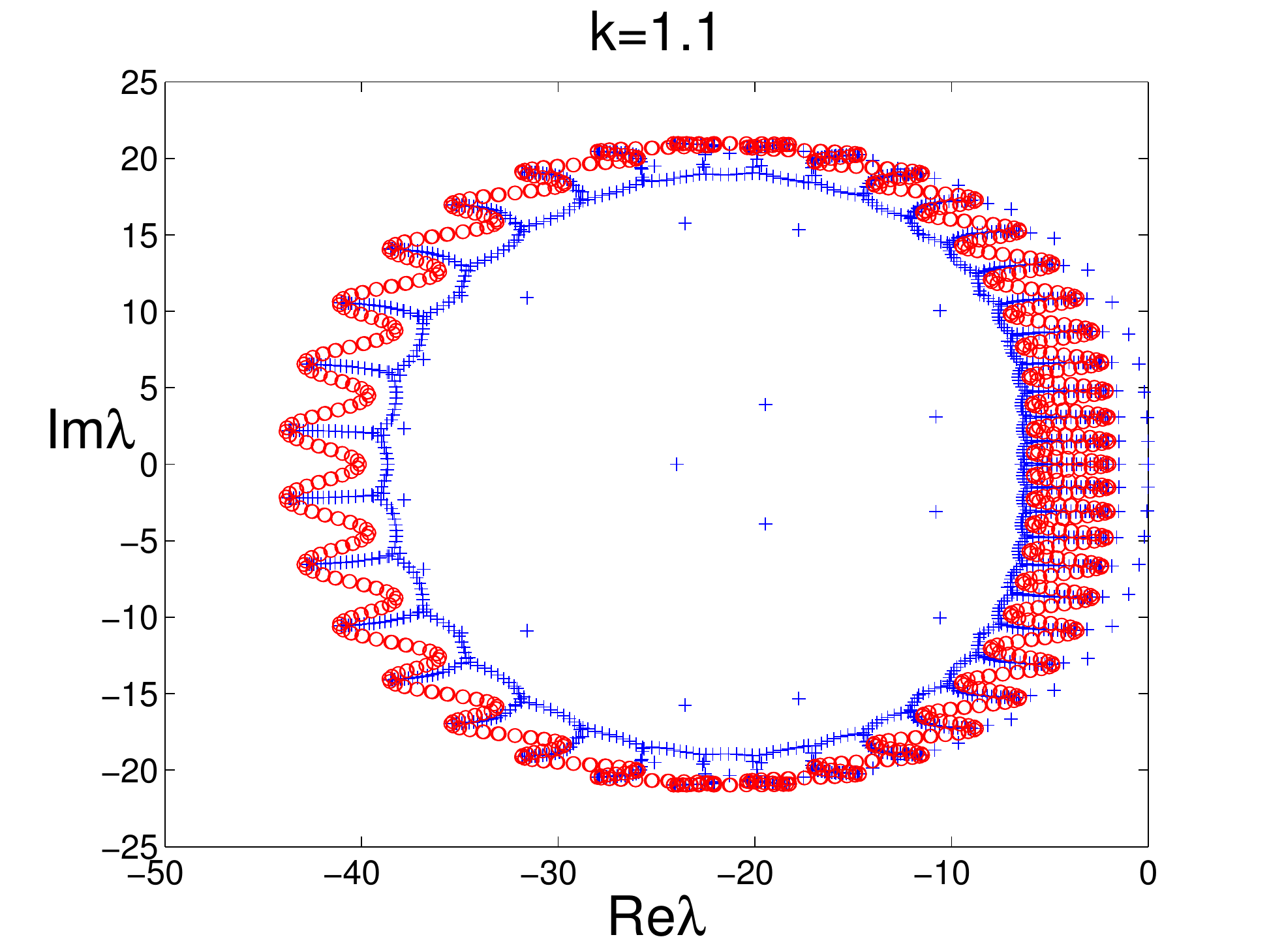}}}

\end{center}
\caption{Plot of the eigenvalues of the Jacobian associated with the linearization of Eq.~\eqref{tw1} about the traveling wave solution (blue pluses),
solved on $[-25, 25]$ using $2001$ nodes and a forward difference scheme,
and the eigenvalues for the background equilibrium state (red circles) given by Eq.~\eqref{tw24}. Here
$\mu=0.5$ and $k=2.25$, $1.5$ and $1.1$.}
\label{back}
\end{figure}

In the following we analyze the stability of the background state in the discretized Eq.~\eqref{tw11} with the forward and centered difference
approximations described above. First, the forward difference approximation is analyzed. In Fig.~\ref{f318}, we see that the eigenvalues of the Jacobian for Eq.~\eqref{tw11} approximate the eigenvalues of the continuum background as
the number of points is increased, although the full structure of the forward difference spectral locus is more complex.
The eigenvalues for the background can also be obtained in the context of the forward difference as follows. In this case, Eq.~\eqref{tw22}
obtained by linearizing Eq.~\eqref{tw1} about the background equilibrium state $\Theta_0=0$ or $\Theta_0=\pi$ is replaced by
\begin{equation}
\begin{split}
V_{\tau} &- c\left[ \frac{-V(J+2,\tau)+4V(J+1,\tau)-3V(J,\tau)}{2 \Delta x}\right]=\\
&k \cos(\mu) [(V(J+q,\tau)-2V(J, \tau)+V(J-q,\tau))] -2V(J,\tau),
\end{split}
\label{tw23}
\end{equation}
where $V(J,\tau)$ is the approximation of $v(z,\tau)$ at a grid point $z_J=J\Delta x$ for integer $J$ and $q$ is an integer such that $q \Delta x=1$.
Seeking solutions in the form $V_J(\tau)=e^{\lambda \tau} e^{iJp \Delta x}$, where $p$ is the wave number,
and solving for $\lambda$, we find that
\begin{equation}
\lambda=c \left[ \frac{-\cos(2p\Delta x)+4\cos(p\Delta x)-3}{2\Delta x}\right]+2k\cos{\mu}(\cos(p)-1)-2+
i c\left[ \frac{-\sin(2p\Delta x)+4\sin(p\Delta x)}{2\Delta x} \right].
\label{tw24}
\end{equation}
The real and imaginary parts of these eigenvalues parametrized by $p$ are shown by red circles in Fig.~\ref{back} at $k=2.25$, $1.5$ and $1.1$ and $\mu=0.5$.
For comparison, the eigenvalues of the Jacobian associated with the traveling wave solution are shown by blue pluses (recall also Fig.~\ref{f318}, where these
eigenvalues are shown for the case $k=1.5$ and $\mu=0.5$ for different numbers of nodes in the discretization.)
To solve (\ref{tw11}), $2001$ nodes are used in $[-25,25]$. The plots show that the two sets of eigenvalues are close to each other.
%

Expanding Eq.~\eqref{tw24} in Taylor series at small $\Delta x$, we obtain
\[
\lambda=-\frac{c p^4 (\Delta x)^3}{4}+2\{k\cos(\mu)(\cos(p)-1)-1\}+i(cp+c \frac{p^3}{3} (\Delta x)^2 +O((\Delta x)^4)),
\label{tw25}
\]
which yields Eq.~\eqref{tw3} in the limit $\Delta x \rightarrow 0$. The principal part
of the error in the real and imaginary parts is $-c\frac{p^4}{4} (\Delta x)^3$ and $c\frac{p^3}{3} (\Delta x)^2$ respectively.
As the wave number $p$ increases, this error pushes the real part of $\lambda$ to $-\infty$ and the imaginary
part to $+\infty$ and $-\infty$, again in line with the observations
of Fig.~\ref{f318}.

It is interesting to explore the spectral properties of the
background of the traveling wave using a centered difference approximation instead of
the forward difference approximation. Eq.~\eqref{tw23} is then replaced by
\[
V_{\tau} - c\frac{V(J+1,\tau)-V(J-1,\tau)}{2 \Delta x}= k \cos(\mu) (V(J+q,\tau)-2V(J, \tau)+V(J-q,\tau)) -2V(J,\tau),
\]
and the eigenvalues $\lambda$ are given by
\begin{equation}
\lambda=2k\cos(\mu)(\cos(p)-1)-2+i\dfrac{c \sin(p \Delta x)}{\Delta x}.
\label{tw27}
\end{equation}
In this case there is no error in the real part of $\lambda$.

\begin{figure}
\begin{center}
{\subfloat[]{\includegraphics[width=0.4\textwidth]{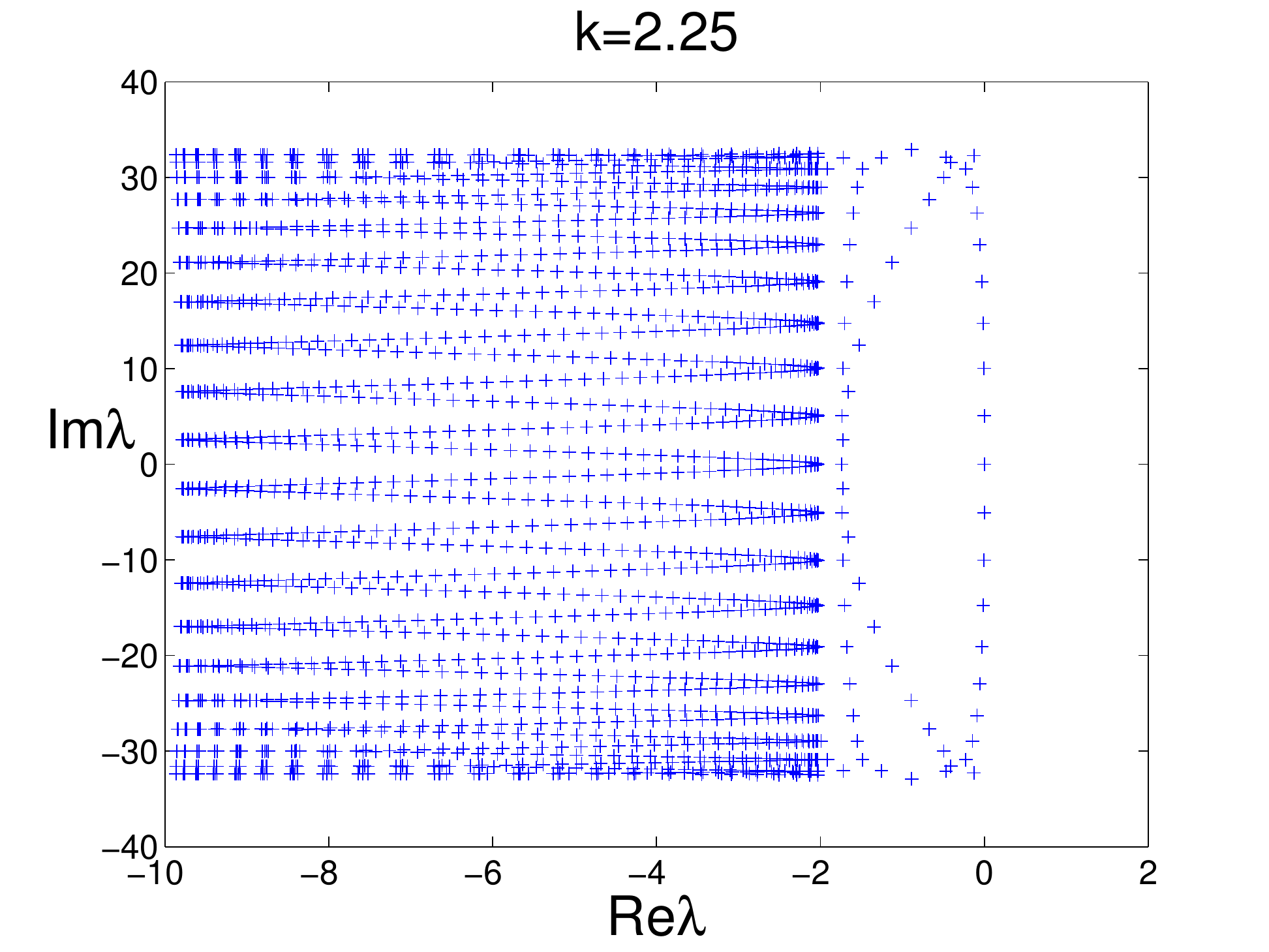}}}
{\subfloat[]{\includegraphics[width=0.4\textwidth]{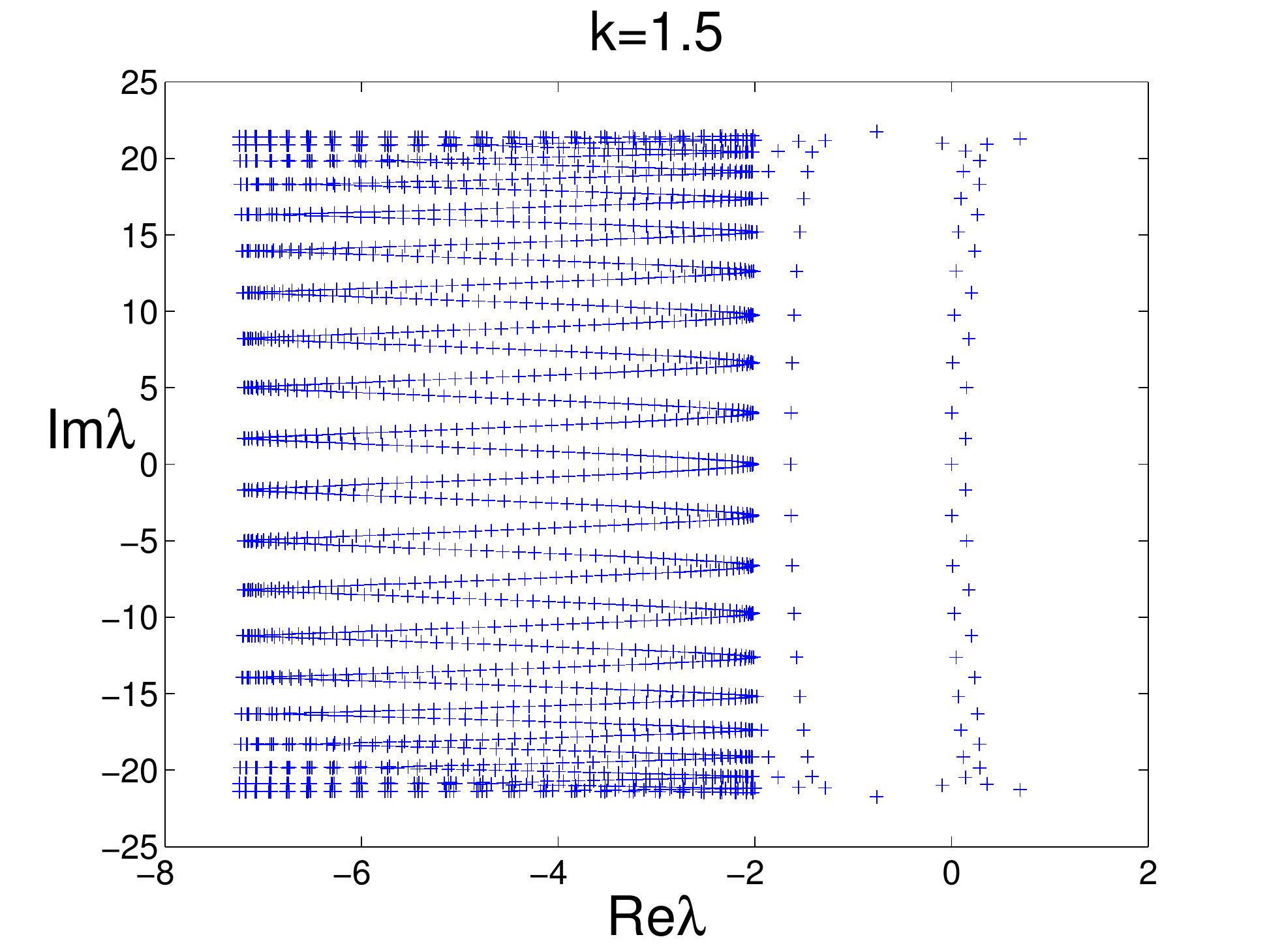}}}
{\subfloat[]{\includegraphics[width=0.4\textwidth]{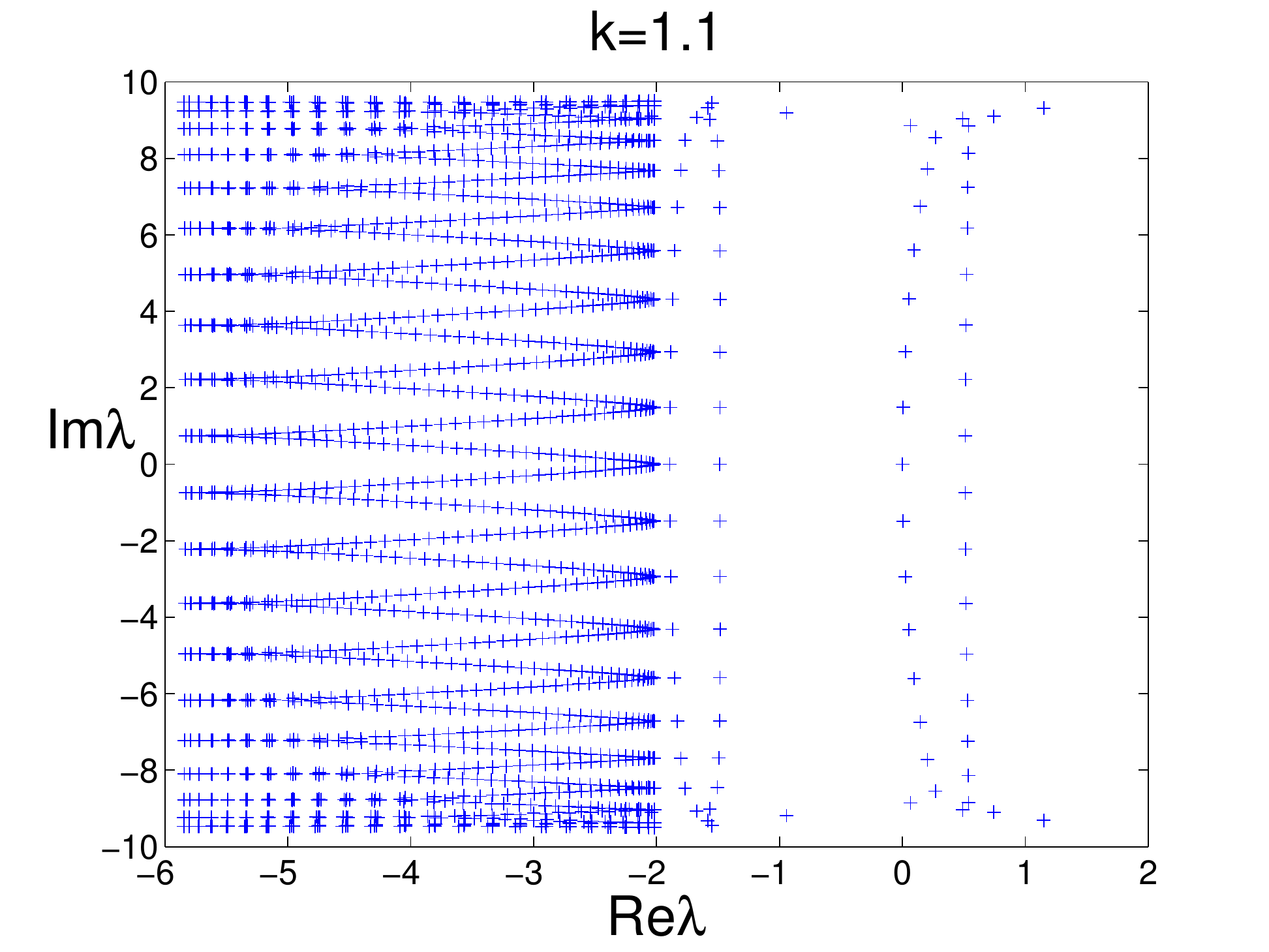}}}

\end{center}
\caption{Plot of the eigenvalues of the Jacobian associated with the linearization of Eq.~\eqref{tw1} about the traveling wave solution (blue pluses),
solved on $[-25,25]$ using $2001$ nodes and centered difference approximation. Here
$\mu=0.5$ and $k=2.25$, $1.5$ and $1.1$.}
\label{f500}
\end{figure}

Generally, this centered difference approximation is numerically unstable for the advection equation given by (\ref{tw1});
see \cite{morton}.
Here, we find that the nonlinear term does stabilize it for large enough values of $k$, but the instability in the numerical method is observed for smaller
values of $k$ even though the solutions of the traveling wave equation are stable,
according to  the forward difference approximation.
For example, in Fig.~\ref{f500} the eigenvalues obtained using the centered difference approximation of
Eq.~\eqref{tw11} are presented for $k=1.1$, $1.5$ and $2.25$ with $\mu=0.5$.
The real parts of the
eigenvalues mostly lie between $-5.86$ and $-2$ for $k=1.1$, between $-7.27$ and $-2$ for $k=1.5$ and between $-9.90$ and $-2$ for $k=2.25$,
which agree almost exactly with the continuum background theory based on  Eq.~\eqref{tw3}.
According to Fig.~\ref{f500}, however, with fixed $\mu=0.5$, the  centered difference approximation predicts that traveling waves become
unstable somewhere between $k=2.25$ and $k=1.5$, as some eigenvalues emerge with
positive real part due to the instabilities associated with the centered difference approximation. In the results presented in the manuscript, care has been
taken to avoid such spurious instabilities induced by the numerical scheme.

\end{document}